%% file: paper.tex
\documentclass[11pt,a4paper]{article}

\usepackage{jcappub}
\usepackage{amsfonts}
\usepackage{amsmath}
\usepackage{amssymb}
\usepackage{aas_macros}
\usepackage{graphicx}
\usepackage{color}
\usepackage{float}
\usepackage{bm}
\usepackage{natbib}
\usepackage{hyperref}
\usepackage{longtable}

\input{defs.tex}

\newcommand{\abs}[1]{\left\vert#1\right\vert}
\newcommand{\set}[1]{\left\{#1\right\}}

\newcommand{\zhat}{\mathbf{\hat{z}}}

\def\emph#1{\textit{#1}}

\def\comment#1{}
\definecolor{RedWine}{rgb}{0.743,0,0}
\definecolor{RoyalBlue}{rgb}{0.25,.41,.88}
\definecolor{ForestGreen}{rgb}{.13,.54,.13}
\definecolor{DeepPurple}{rgb}{.72,.18,1}
\definecolor{OrangeRed}{rgb}{1.0, 0.27, 0.0}

\allowdisplaybreaks

\title{Large-Scale-Structure Observables in General Relativity Validated at Second Order}

\author[a]{Antoine Villey,}
\author[b,c]{Yonadav Barry Ginat,}
\author[d,e]{Vincent Desjacques,}
\author[f,g]{Donghui Jeong}
\author[h]{and Fabian Schmidt}

\affiliation[a]{Centre de Physique Théorique, Ecole Polytechnique, 
91128 Palaiseau, France}
\affiliation[b]{Rudolf Peierls Centre for Theoretical Physics, University of Oxford, Parks Road, Oxford, OX1 3PU, United Kingdom}
\affiliation[c]{New College, Holywell Street, Oxford, OX1 3BN, United Kingdom}
\affiliation[d]{Physics department, Technion, 3200003 Haifa, Israel}
\affiliation[e]{Asher Space Research Institute, Technion, 3200003 Haifa, Israel}
\affiliation[f]{Department of Astronomy and Astrophysics and Institute for Gravitation and the Cosmos, The Pennsylvania State University, University Park, PA 16802, USA}
\affiliation[g]{School of Physics, Korea Institute for Advanced Study, Seoul 02455, Korea}
\affiliation[h]{Max-Planck-Institut f\"ur Astrophysik, Karl-Schwarzschild-Stra\ss e~1, 85748 Garching, Germany}

\emailAdd{antoine.villey@polytechnique.edu}
\emailAdd{yb.ginat@physics.ox.ac.uk}
\emailAdd{dvince@physics.technion.ac.il}
\emailAdd{djeong@psu.edu}
\emailAdd{fabians@mpa-garching.mpg.de}

\date{\today}

\label{firstpage}

\abstract{We present a second-order calculation of relativistic large-scale-structure observables in cosmological perturbation theory, specifically the ``cosmic rulers and clock'', which are the building-blocks of any other large-scale-structure observable, including galaxy number counts, on large scales. We calculate the scalar rulers (longitudinal perturbation and magnification) and the cosmic clock to second order, using a fully non-linear covariant definition of the observables. 
We validate our formul{\ae} on three non-trivial space-time metrics: two of them are null tests on metrics which are obtained by applying a gauge transformation to the background space-time, while the third is the ``separate universe'' curved background, for which we can also compute the observables exactly.
We then illustrate the results by evaluating the second-order observables in a simplified symmetric setup. On large scales, they are suppressed over the linear contributions by $\sim 10^{-4}$, while they become comparable to the linear contributions on mildly non-linear scales.
The results of this paper form a significant (and the most complicated) part of the relativistic galaxy number density at second order.}

\begin{document}

\maketitle

%
%


\section{Introduction}
\label{sec:intro}

The observed distribution of galaxies on the sky provides a plethora of cosmological information, both on the evolution of the Universe, large-scale structure (LSS), and on the formation mechanism of primordial fluctuations \cite{Laureijs:2011gra,Dore:2014cca,Aghamousa:2016zmz}. 
To interpret these observations however, we need to relate the observed galaxy sky positions (denoted by $\ntvh$) and redshifts (denoted by $\tilde{z}$) to the underlying space-time and matter distribution, a relation which is non-trivial even on very large scales: apart from galaxy bias, this includes selection effects (e.g.|the large-scale structure effects on observability) and the mapping from the galaxy rest frame to observed sky positions and redshifts \cite[see][for a detailed review of all these contributions]{Desjacques:2016bnm}.
On large scales, the latter, commonly known as ``relativistic effects'' as they are not captured by Newtonian physics, are not necessarily suppressed relative to the commonly included galaxy bias and redshift-space distortion contributions \cite[see][for a review of the linear-order relativistic effects]{JeongSchmidt2015}.


Since relativistic effects are only important on large scales, where perturbations are small, we can adopt a perturbative expansion in the space-time and matter sectors.
The formalism for studying relativistic effects at linear order was developed originally in refs.~\cite{Yoo:2009au,ChallinorLewis2011,Bonvin:2011bg,baldauf/etal:2011,Jeong:2011as}; these include the contributions from gravitational lensing (leading to magnification of light flux and distortions of the galaxy clustering), peculiar velocities, and wide-angle effects, which have all been incorporated \cite{Sachs:1967er,Dyer:1973zz,dyer/roeder:1974,weinberg:1976,Kaiser:1987qv,Sasaki:1987ad,Futamase:1989hba,Szalay:1997cc,bartelmann/schneider:2001,Suto:1999id,Matsubara:1999du,Matsubara:2000pr,Szapudi:2004gh,Papai:2008bd,Bonvin:2005ps,Raccanelli:2010hk,Yoo:2012se,Yoo:2013zga,Foglienietal2023,Wen:2024nhi,Bertacca:2012tp,Bertacca:2019wyg,Yoo:2013tc,Khek:2022dgw,Raccanelli:2015vla,Raccanelli:2023fle,Spezzati:2024ooc}.
At second order, however, many additional contributions exist, due to quadratic terms as well as post-Born terms in the solution of the geodesic equation. While the second-order expression for, e.g., the cosmological distance-redshift relation remains compact \cite{Umeh:2012pn,Ben-Dayan:2012lcv,Ben-Dayan:2014swa,Marozzi:2014kua}, the corresponding results for the observed galaxy over-density field $\dgobs$ are significantly more complex, typically resulting in hundreds of terms in expressions for $\dgobs$ \cite{YooZaldarriaga2014,Bertacca:2014dra,Bertaccaetal2014II,Bertaccaetal2015III,Fuentes:2019nel,MagiYoo2022}; a direct comparison between different expressions in the literature is highly non-trivial and has not yet been performed.
Nevertheless, given a proper definition of the density and observer's co-ordinates, a well-defined prediction for $\dgobs$ exists to which all calculations must converge. 

Given the complexity of the expressions, validation tests are highly desirable. One important such test is the invariance of $\dgobs$ under perturbative transformations of the space-time co-ordinates, so-called gauge transformations \cite{MagiYoo2022}. Other tests include symmetric setups where independent, simpler derivations are possible, such as a curved FLRW space-time.
The linear-order relativistic expression for $\dgobs$ has been shown to pass these tests \cite{Jeong:2011as}, but no such tests on the full second-order $\dgobs$ have been presented in the literature.

Clearly, approaches that simplify and make the calculation more transparent are highly desirable in this case. One option is to choose suitable co-ordinates, such as the geodesic light-cone gauge, which has been applied to the calculation of redshift and luminosity distance at linear and second order \cite{Gasperini:2011us,Fanizza:2013doa,Fanizza:2015swa,Fleury:2016htl,Fanizzaetal2022,Fanizzaetal2023,Fanizzaetal2023b}. An alternative route is by simplifying the observables themselves, employing the ``cosmic rulers'' and the ``cosmic clock'' introduced in refs.~\cite{SchmidtJeong2012,JeongSchmidt2014,JeongSchmidt2015}. The cosmic ruler and the cosmic clock refer to putative objects with in-principle knowable intrinsic shape and proper age, respectively. Cosmic rulers, in particular, provide an efficient and fully relativistic approach to derive key observables relevant for large-scale structure, such as magnification, shear, and the volume element corresponding to an observed solid angle and redshift interval. The central idea is to consider two space-time events near the source, separated by a known infinitesimal space-like distance in the source's rest frame. By mapping this separation to angular and redshift differences as measured by a distant observer, one can isolate the relevant relativistic distortions, be they shear/magnification, line of sight, or mixed between transverse and line-of-sight components. These rulers describe the distortion of shapes and time-keeping devices between the source's rest frame and the observer’s past light-cone, making them observable, gauge-invariant, and consistent with the equivalence principle.



The cosmic rulers can be classified uniquely, in terms of their transformation properties under rotations around the line of sight, into scalars, vectors, and tensors. 
In this paper, we calculate the scalar cosmic rulers and the cosmic clock, $\mathcal{T}$, at second order in cosmological perturbation theory; these pertain to the distortion along the line of sight, denoted by $\mathcal{C}$, the magnification $\mathcal{M}$, and the distinction between age as inferred from the observed redshift (via a fiducial FLRW background), and the proper age of the source, as measured locally in its rest frame.

Ref.~\cite{Ginatetal2021} established a fully covariant (relativistic) non-linear relation between the observed galaxy over-density, $\dgobs$, and the cosmic rulers and clocks (see \S\ref{sec:metricdet}). The ruler observables computed here are sufficient to obtain a second-order expression for the volume-element, an ingredient for the observed galaxy over-density field, while vector and tensor cosmic rulers are only needed explicitly at first order. Besides the volume element, $\dgobs$ requires (1) a boost from source's rest-frame to a fixed observed-redshift frame \cite{JeongSchmidt2015}, and (2) the galaxy bias expansion in the source's rest-frame \cite{Umehetal2019}. It is worth noting that, ranking by the number of independent contributions, the volume element is the most complicated ingredient of $\dgobs$.

The practical advantage of the approach proposed in ref.~\cite{Ginatetal2021} is that $\dgobs$ is broken up into many smaller gauge-invariant components, 
each of which is individually observable. We are then able to test the expression for each of the rulers individually---these expressions are much simpler than the full expression for $\dgobs$.

The calculation of the cosmic rulers and clock is kinematical
in the sense that we only need a given space-time $g_{\mu\nu}(x)$ which, following the discussion above, is assumed to be perturbatively close to an FLRW background. We neither need to make assumptions about the matter content of the Universe, nor about the validity of GR. Nevertheless, for all numerical results we will assume a $\Lambda$CDM background. 

The lengthy (yet manageable) calculation can be further divided into several steps, which are reflected in the sections of the paper:
\begin{itemize}
\item We define the cosmic clock $\mathcal{T}$ and, for computational purposes, introduce two different definitions of the cosmic rulers based on the pull-back of the spatial metric to the observer's light-cone (\S\ref{sec:frames}).
The first recovers the original proposal by \cite{SchmidtJeong2012}, while the second definition, following \cite{Ginatetal2021}, is more calculationally convenient. Both definitions are related by a unique, gauge-invariant, non-linear relation, and furthermore agree at linear order in perturbations.  
\item We then solve the null geodesic equation, to find the trajectory of light from the source to the observer (\S \ref{sec:geodesic equation}), which also yields the observed redshift (\S \ref{subsec:affine parameter}). We also derive the time-like geodesics of the source and the observer, and their proper times at the moments of emission and observation (\S \ref{sec:clock}). Given the solution for the null geodesic, all quantities of interest can be expressed in terms of the observed sky position and redshift (which we phrase more conveniently via an ``inferred position''). 
\item The redshift and the proper age of source and observer allow us to find an expression for the cosmic clock (\S \ref{sec:clock}), which is complete up to second order in perturbations. 
\item The cosmic clock and rulers are computed at second order in perturbations (\S \ref{sec:rulers}).
The resulting expressions for the cosmic rulers and clock are all written in terms of inferred co-ordinates. 
\item In \S \ref{sec: implementation} we describe how one can practically use the results of this paper, by providing a \verb"Mathematica" notebook where they are implemented. 
\end{itemize}
A key element of this work is a sequence of validation tests of our expressions for $\mathcal{T}$, $\mathcal{C}$ and $\mathcal{M}$, individually. We test the expressions for the following three distinct cases; the symbols refer to the perturbed FLRW metric defined in Eq.~\eqref{eqn: metric Poisson gauge}.
\begin{enumerate}
\item Spatially and temporally constant potential (scalar space-time perturbation) at second order: a constant potential can be absorbed by a global co-ordinate re-definition, and hence cannot affect the observable rulers and clocks. We verify explicitly that this is the case. This test validates all contributions that involve terms such as $\Phi, \Phi^2, \Psi, \Psi^2, \ldots$ as well as integrals thereof.
\item Gradient mode at second order: by performing a co-ordinate transformation up to $\mathcal{O}(\vx^2, \eta^2)$ on an unperturbed space-time, we obtain a non-trivial perturbed metric which, at linear order in perturbations, essentially corresponds to a  a pure-gradient mode $\Phi(\vx) = \v{A}\cdot \vx$. We show that all scalar rulers indeed remain zero for such a space-time perturbation. This complex test validates all contributions that involve spatial derivatives, such as $\partial_i\Phi$, $\partial_i\Phi\partial^i\Phi, \ldots$, of which there is a large number.
\item Curved FLRW background (separate universe): Thanks to the highly symmetric space-time, it is straightforward to compute the perturbation to the cosmic-ruler quantities that result when interpreting a curved FLRW space as a deviation from a flat FLRW background, with Euclidean spatial geometry. Conversely, in the small-curvature limit, a curved FLRW background can be expanded around the Euclidean background, resulting in a specific perturbed background with time-dependent potentials $\Phi(\eta), \Psi(\eta)$. Specialising the ruler perturbations to this perturbed background should yield agreement, order by order in curvature, with the curved background calculation. We show our second-order expressions also pass this test case, which probes in particular time derivatives of the potentials, or their time integrals.
\end{enumerate}
The \verb"Mathematica" notebook we provide with publication of this paper performs all three tests. We summarise the results and discuss them in \S \ref{sec: discussion}. 

Symbols and special notations used in this paper are listed in the table in Appendix \ref{appendix:symbol list}.  
Our perturbed FLRW metric is given by
\begin{equation}\label{eqn: metric Poisson gauge}
\begin{aligned}
\mathrm{d}s^2 &= a(\eta)^2\left[-(1+2\Psi)\mathrm{d}\eta^2 + 2\omega_i\mathrm{d}\eta \mathrm{d}x^i + \left(\delta_{ij} - 2\Phi \delta_{ij} + \frac{1}{2}h_{ij}\right)\mathrm{d}x^i\mathrm{d}x^j\right] \\
&\equiv g_{\mu\nu}\mathrm{d}x^\mu \mathrm{d}x^\nu \;,
\end{aligned}
\end{equation}
expressed in terms of the co-moving co-ordinates $x^i$ and the conformal time, which is related to the cosmic time $t$ by $\mathrm{d}t=a\mathrm{d}\eta$.
Here, the scalar potentials $\Psi$ and $\Phi$ include both first- and second-order contributions, while we take $\omega_i$ and $h_{ij}$ to be purely second order. 
That is, any primordial vector and tensor modes are treated only to linear order throughout.
Note that we do not require the metric to be in Poisson gauge, i.e.~$\omega_i$ and $h_{ij}$ need not necessarily be transverse and transverse-traceless at second order.

\section{Cosmic clock and ruler observables}
\label{sec:frames}

Before embarking on the calculations involved in the solution of the geodesic equations, let us first describe the two relevant co-ordinate systems for this work. Let $g_{\mu\nu}$ be the metric tensor, $u^\mu$ be the 4-velocity field of the source (where we assume that the source and the observer are co-moving with the galaxies) and $k^\mu$ be the wave 4-vector of the light travelling from the source to the observer. 

Suppose that we have a global co-ordinate system $x^\mu = (\eta,\vx)$ on a space-time manifold $\mathbb{M}$. A source emits light at $x^\mu_e = (\eta_e,\vx_e)$, which is received by the observer at $x_o^\mu=(\eta_o,\vx_o)$.
For convenience, we say that the observer
assigns to the source an \emph{inferred space-time position}
\begin{equation}\label{eqn: tilde co-ordinates definition}
\tilde x^\mu(\ntvh, \tilde z) =
(\tilde{\eta},\tilde{\vx}) =
(\overline{\eta}_o - \tilde{\chi}, \tilde\chi \ntvh)\;
\end{equation}
based on the observed redshift $\tilde{z}$ and sky position $\ntvh$ via a putative FLRW background described by the metric 
$a^2(\bar\eta) \left(-\mathrm{d}\bar\eta^2 + \gamma_{ij} \mathrm{d}x^i \mathrm{d}x^j\right)$.
The metric $\gamma_{ij}$ is non-trivial if one does not use Cartesian co-ordinates, even for a flat FLRW background. This defines the observer's {\it inferred co-moving co-ordinates} $\tilde x^i=\tilde\chi\tilde n^i$, where $\tilde\chi\equiv\chi(\tilde z)$ is the co-moving distance inferred from a measurement of $\tilde z$.

We reserve an over-line to denote the background relation between cosmic time and conformal time: $\overline{\eta}(t)$ is the background cosmic-time-to-conformal-time relation, and $\overline{t}(\eta)$ is its inverse. For simplicity, we keep the background scale factor $a$ as either a function of conformal time $\eta$, or of cosmic time, i.e. 
\begin{equation}\label{eqn: a bar definition}
a(t) = a(\overline{\eta}(t)).
\end{equation}
Throughout, we denote the proper time of co-moving observers by $\tau$, and the proper time of the observer at observation as $\tau_o$. We normalise the scale factor via the condition that the observer makes measurements at $a(\tau_o) = a(\overline{\eta}_o) = 1$ where $\overline{\eta}_o\equiv \overline{\eta}(\tau_o)$.
Additionally, derivatives with respect to $\eta$ will be denoted by a prime, i.e.~$\mathrm{d}f(\eta)/\mathrm{d}\eta = f'(\eta)$, while cosmic-time derivatives are denoted by a dot:~$\mathrm{d}f(t)/\mathrm{d}t = \dot{f}(t)$.

Note that $\tilde{x}^\mu$ are space-time co-ordinates on the observer's past light cone $\Sigma_{\tau_o}$ and, therefore, only three components of $\tilde{x}^\mu$ are independent, since $\eta_o - \tilde{x}^0 = \abs{\tilde{\vx}}$.
For the calculation in terms of global co-ordinates, we define the total displacements $\Delta x^\mu$ via \cite{JeongSchmidt2015}
\begin{equation}\label{eqn: total displacement definition}
\Delta x^\mu \equiv x_e^\mu - \tilde{x}^\mu\;.
\end{equation}
We also use the source's freely-falling reference frame, where we require the transformation between how lengths (and time-intervals) are measured in that frame, and how they would be interpreted by the observer, in the inferred co-ordinate system $\tilde{x}^\mu$. We do so by utilising the cosmic rulers and the cosmic clock introduced by refs.~\cite{SchmidtJeong2012,JeongSchmidt2014,JeongSchmidt2015}, whose non-linear definitions we give now. 

\subsection{Cosmic clock}

The cosmic clock is defined, non-linearly, as \cite{JeongSchmidt2014,Ginatetal2021}
\begin{equation}
\label{eq:Truler}
\mathcal{T} = \ln \left(\frac{\at(\overline{\eta}(\tau_e))}{\tilde{a}}\right) \;,
\end{equation}
where $\at(\overline{\eta}(\tau_e))$ is the background scale-factor-to-time relation evaluated at the proper time of the source at emission. We can conveniently separate $\mathcal{T}$ into two contributions, by introducing $a_e = a(\eta_e) = a\left(x^0_e\right)$, the \emph{co-ordinate} time at the space-time location of emission,
\begin{equation}
\mathcal{T} = \ln \left(\frac{\at(\tau)}{a\left(x^0_e\right)}\right) + \ln\left(\frac{a\left(x^0_e\right)}{\tilde{a}} \right) \equiv \ln \left(1+\Delta_{a,\tau} \right) + \ln\left(1+\Delta_{a,e} \right)
\label{eq:Tdef}
\end{equation}
where 
\begin{equation}
\label{eq:tildea}
\tilde a = \big(1+\tilde z\big)^{-1}
\end{equation}
is the scale factor inferred by the observer, and $\Delta_{a,\tau} \equiv \at(\tau)/a\left(x^0_e\right)-1$ and $\Delta_{a,e} \equiv a\left(x^0_e\right)/\tilde{a} - 1$ have been introduced to streamline subsequent calculations.
Note that only the sum of the two contributions is co-ordinate invariant and observable, and that $\mathcal T\to 0$ in the limit in which the source coincides with the observer; this reflects the fact that the scale factor is normalised such that $\at(\tau_o)\equiv 1$. 

A cautionary remark is in order here: at the conformal time co-ordinate $\eta_o$ of the observer's position, the scale factor $a(\eta_o)\neq1$ in general because the observer's proper time $\tau_o$ is different from the proper time read off from the conformal time co-ordinate $\tau_o\neq \overline{t}(\eta_o)$.
This will become fully clear once the source and observer geodesics are discussed in Section \S\ref{sec:geodesic equation} below.

\subsection{Cosmic rulers}

Ref.~\cite{SchmidtJeong2012} defined the cosmic rulers by considering the induced metric $g_s$ on constant-proper-time surfaces, intersecting the source's geodesic. By definition, these are orthogonal to the source 4-velocity $u^\mu$.
The mapping of infinitesimal distances in the source's rest frame (i.e.~on a fixed-proper-time-surface) to those estimated by a distant observer can be conveniently derived by defining suitable differential forms. These will help us to translate the physically intuitive picture of bridging together local, infinitesimal rulers without the need for an explicit ruler length. Concretely,
\begin{enumerate}
\item We introduce the embedding map $i:\Sigma_{\tau_o}\to\mathbb{M}$, which defines the pull-back $i^*$ of differential forms.
\end{enumerate}
The pull-back of the spatial metric for the source is key to the implementation of the local, infinitesimal ruler approach. We can proceed in two distinct ways. The first option generalises the approach taken by ref.~\cite{SchmidtJeong2012} at linear order in perturbations:
\begin{enumerate}
\addtocounter{enumi}{1}
\item We pull the spatial metric $g_s\equiv (g_{\mu\nu}+u_\mu u_\nu)\mathrm{d}x^\mu\otimes \mathrm{d}x^\nu$ back to $\Sigma_{\tau_o}$,
\begin{align}
i^*g_s &= \big(g_{\mu\nu}+u_\mu u_\nu\big) \frac{\partial x^\mu}{\partial \tilde x^i}\frac{\partial x^\nu}{\partial \tilde x^j} \mathrm{d}\tilde x^i\otimes \mathrm{d}\tilde x^j \\
&\equiv \tilde g_{ij}\, \mathrm{d}\tilde x^i\otimes \mathrm{d}\tilde x^j \nonumber \;,
\end{align}
where $\tilde x^i$ denote the observer's inferred co-moving co-ordinates and $\otimes$ is a tensor product.
\item We decompose $\mathrm{d}\tilde x^i\otimes \mathrm{d}\tilde x^j$ onto directions parallel and perpendicular to $\tilde n^i$, using the basis $(\ntvh,\tilde\ve_1,\tilde\ve_2)$, 
which is orthonormal with respect to the metric $\gamma_{ij}$: $\gamma_{ij}\tilde n^i\tilde n^j=1$, $\gamma_{ij}\tilde e_I^i\tilde e_J^j=\delta_{IJ}$ and $\gamma_{ij} \tilde{n}^i \tilde{e}_I^j = 0$ etc., where $I,J\in \set{1,2}$ and $\tilde e_{I}^i$ is the $i$th (contravariant) component of $\tilde\ve_I$ in the co-ordinate basis $\tilde\partial_i \equiv \partial/\partial\tilde x^i$.
Introducing the 1-forms $\mathrm{d}\tilde x^\parallel\equiv \tilde n_i \mathrm{d}\tilde x^i$ and $\mathrm{d}\tilde x^I\equiv \gamma_{ij}\tilde e_I^i \mathrm{d}\tilde x^j$,
we arrive at 
\begin{equation}
\label{eq:gs1}
i^*g_s = \tilde g_{\parallel\parallel}\, \mathrm{d}\tilde x^\parallel\otimes \mathrm{d}\tilde x^\parallel + 2 \tilde g_{\parallel I} \,\mathrm{d}\tilde x^\parallel\otimes \mathrm{d}\tilde x^I + \tilde g_{IJ}\, \mathrm{d}\tilde x^I\otimes \mathrm{d}\tilde x^J \;.
\end{equation}
Alternatively, one could also project $\mathrm{d}\tilde x^i\otimes \mathrm{d}\tilde x^j$ onto a complex (helicity) basis.

\item The components of the pulled-back metric tensor $i^* g_s$ are given by
\begin{align}
\label{eq:gpp}
\tilde g_{\parallel\parallel} &= \tilde g_{ij} \tilde n^i \tilde n^j \\
&\equiv \tilde a^2 \big(1 - \mathcal{C}\big)^2 \; ;\nonumber \\
\tilde g_{\parallel I} &= \tilde g_{ij} \tilde n^i \tilde e^j_I \nonumber \\
&\equiv -\tilde a^2 \mathcal{B}_I \; ;\nonumber \\
\tilde g_{IJ} &= \tilde g_{ij} \tilde e^i_I \tilde e^j_J \nonumber \\
&\equiv \tilde a^2\left(\delta_{IJ}-2\mathcal{A}_{IJ}\right) \nonumber \;,
\end{align}
where $\mathcal{A}_{IJ}$ is a $2\times 2$ symmetric matrix. The parametrisation (\ref{eq:gpp}) is chosen for consistency with ref.~\cite{SchmidtJeong2012}.
The rulers $(\mathcal{C},\mathcal{B}_I,\mathcal{A}_{IJ})$ can be used by the observer to measure deviations from the putative FLRW background on $\Sigma_{\tau_o}$.
\end{enumerate}
Alternatively, 
\begin{enumerate}
\addtocounter{enumi}{1}
\item We introduce an orthonormal tetrad $s_{\underline{\alpha}}=s_{\underline{\alpha}}^\mu \partial_\mu$ at the source such that the co-tetrad $s^{\underline{\alpha}}=s^{\underline{\alpha}}_\mu \mathrm{d}x^\mu$ satisfies $s_\mu^{\underline{0}}\equiv u_\mu$. Therefore, the spatial part of the metric, in the source co-moving frame, is given by $g_s = \delta_{\underline{ij}}\, s^{\underline{i}}\otimes s^{\underline{j}}$.
\item Given a null geodesic that connects the source and the observer, the co-tetrad fields $s^{\underline{i}}$ can be decomposed onto directions parallel and perpendicular to the photon wave-vector of that geodesic at the source, 
$k_s=k_s^\mu \partial_\mu$, before we pull back $g_s$. 
Concretely, we define 
$s^\parallel\equiv \hat{n}_{s\underline{i}}s^{\underline{i}}$ through the requirement $s^\parallel(k_s)=\sqrt{g_s(k_s,k_s)}$.
We introduce also $s^I\equiv \delta_{\underline{ij}}\, \hat s_I^{\underline{i}}\, s^{\underline{j}}$, where $I=1,2$ and $\hat s_I^{\underline{j}}$  are determined by $s^I(k) = 0$ and the conditions that the co-tetrad fields $s^\parallel$, $s^I$ are orthonormal.
The spatial metric becomes 
\begin{equation}\label{eqn: metric in terms of s's}
g_s = s^\parallel\otimes s^\parallel + \sum_{I} s^I\otimes s^I \;.
\end{equation}
The advantage of this decomposition with respect to $k_s$ lies in the fact that the parallel transport of the tetrad fields $(s_\parallel,s_1,s_2)$ along the photon geodesic can always be arranged to yield tetrad fields aligned with $(\ntvh,\tilde{\ve}_1,\tilde{\ve}_2)$ at the observer's position.
\item We pull the spatial metric back to $\Sigma_{\tau_o}$:
\begin{equation}
\label{eqn:gs2}
i^* g_s = \big(i^* s^\parallel\big)\otimes \big(i^* s^\parallel\big) + \sum_{J=1,2} \big(i^* s^J\big)\otimes \big(i^* s^J\big)\; ;
\end{equation}
and we decompose the 1-forms $(i^* s^\parallel)$ and $(i^* s^I)$ onto directions parallel and perpendicular to $\tilde n^i$. This defines the cosmic rulers $\mathfrak{C}$, $\mathfrak{D}_I$, $\hat{\mathfrak{D}}^I$  and $\mathfrak{A}^I_J$: 
\begin{align}
\label{eq:frak}
i^* s^\parallel & \equiv \tilde a \big[(1-\mathfrak{C})d \tilde{x}^\parallel - \mathfrak{D}_J d \tilde{x}^J\big] \\
i^* s^J & \equiv \tilde a \big[-\hat{\mathfrak{D}}^J d \tilde{x}^\parallel + \left(\delta^J_K - \mathfrak{A}^J_K\right)d \tilde{x}^K \big] \nonumber 
\end{align}
where again $J,K=1,2$ label the two independent directions transverse to $\ntvh$. We remark that the capitalised upper index $J$ in the second equation above can be lowered trivially---by the Euclidean metric $\delta_{IJ}$---so henceforth we will make no distinction between $\mathfrak{A}^J_K$ and $\mathfrak{A}_{JK}$, or $\hat{\mathfrak{D}}^J$ and $\hat{\mathfrak{D}}_J$.
\end{enumerate}
Substituting the relations (\ref{eq:frak}) into equation~\eqref{eqn:gs2} and comparing with the parametrisation (\ref{eq:gpp}), we find 
\begin{equation}\label{eqn:relation between fraktur and calligraphic rulers}
\begin{aligned}
\big(1-\mathcal{C}\big)^2 &= \big(1-\mathfrak{C}\big)^2 + \sum_J \hat{\mathfrak{D}}_J\hat{\mathfrak{D}}_J \\
\mathcal{B}_I &= \big(1-\mathfrak{C}\big) \mathfrak{D}_I + \sum_J \big(\delta_{IJ}-\mathfrak{A}_{IJ}\big) \hat{\mathfrak{D}}_J \\
\mathcal{A}_{IJ} &= \frac{1}{2}\left[\delta_{IJ} - \mathfrak{D}_I\mathfrak{D}_J - \sum_K \left(\delta_{IK} - \mathfrak{A}_{IK}\right)\left(\delta_{JK}-\mathfrak{A}_{JK}\right)\right]\;.
\end{aligned}
\end{equation}
These relations hold non-linearly, and all variables appearing here are individually gauge-invariant. To first order, we therefore find
\begin{equation}
\begin{aligned}
\mathcal{C} & = \mathfrak{C}, \\ 
\mathcal{B}_I & = \mathfrak{D}_I + \hat{\mathfrak{D}}_J \; ,\\ 
\mathcal{A}_{IJ} & = \mathfrak{A}_{IJ}\;,
\end{aligned}
\end{equation}
while the two sets of rulers differ at second order. In particular, $\mathfrak{A}_{JK}$ is not symmetric at second (and higher) order.

We further decompose $\mathcal{A}_{IJ}$ (respectively $\mathfrak{A}_{IJ}$) into its trace, $\mathcal{M} \equiv \tr \mathcal{A}_{IJ}$ (respectively $\mathfrak{M} \equiv \tr \mathfrak{A}_{IJ}$) and tensor parts, \emph{viz.}
\begin{equation}\label{eqn:A_IJ helicity decomposition}
\mathcal{A}_{IJ} \equiv \left(\begin{array}{cc}
\mathcal{M}/2 + \gamma_1 & \gamma_2 \\
\gamma_2 & \mathcal{M}/2 - \gamma_1
\end{array}\right) \;,
\end{equation}
and similarly for its 1-form counterpart. 
Let us remark here that while $\mathcal{A}_{IJ}$ is necessarily a symmetric matrix---because it is a reduced metric---$\mathfrak{A}_{IJ}$ is not necessarily symmetric (see \cite{Fanizzaetal2022} and the discussion in \S\ref{sec:rulers}). Finally, we introduce the symmetrised combination of $\mathfrak{D}_I$ and $\hat{\mathfrak{D}}_I$ via 
\begin{equation}\label{eqn:fraktur B definition}
\mathfrak{B}_I \equiv \mathfrak{D}_I + \hat{\mathfrak{D}}_I \;.
\end{equation}
Therefore, $\mathcal{B}_I=\mathfrak{B}_I$ at linear order.

The two implementations of the cosmic rulers presented above are completely equivalent. 
The advantage of the first implementation is the direct connection of the ``metric rulers'' $(\mathcal{C},\mathcal{B}_I,\mathcal{A}_{IJ})$ to the observables (see below). The advantage of the second implementation and its set of rulers $(\mathfrak{C},\mathfrak{B}_I,\mathfrak{A}_{IJ})$, the ``$1$-form rulers'', is simpler calculations, as will shortly be apparent from the volume form induced by $i^* g_s$ (see \S\ref{sec:metricdet}).

The ruler observables can be calculated in any chosen co-ordinate system.
For example, in the geodesic light-cone gauge, they become particularly simple, as, in the notation of \cite{Gasperini:2011us}, we have, essentially, 
$(i^* g_s)_{\parallel\parallel} = (\Upsilon + U)^2$,
$(i^* g_s)_{\parallel I} = -U_I$,
$(i^* g_s)_{IJ} = \gamma_{IJ}$.
In the following, we proceed directly in terms of an FLRW space-time perturbed by scalar perturbations at second order. 

\subsection{Relation to galaxy number counts and magnification}
\label{sec:metricdet}

The cosmic rulers defined above are directly related to a number of cosmological observables. In particular, ref.~\cite{Ginatetal2021} expressed the relativistic observed galaxy over-density field $\dgobs$ in terms of the cosmic rulers. Their equation (3.26) shows that, to determine the volume distortion contribution to $\dgobs$ at second order, only $\mathcal{T}$, $\mathcal{C}$ and $\mathcal{M}$ (or $\mathfrak{C}$ and $\mathfrak{M}$) are required at second order, \emph{viz.}
\begin{equation}\label{eqn:dgobs}
\frac{1+\dgobs}{1+\delta_{\rm g}^{\rm or}} = \left[ 1+ \mathcal{T}\frac{\Theta(\tilde{a})}{\tilde{H}} + \frac{\mathcal{T}^2}{2}\left(\left[\frac{\Theta(\tilde{a})}{\tilde{H}} \right]^2 + \frac{\mathrm{d}}{\mathrm{d} \ln \tilde{a}}\frac{\Theta(\tilde{a})}{\tilde{H}}\right)\right] \det \left(i^* g_s\right) \;,
\end{equation}
where $\Theta(\tilde{a})$ is defined as the rate of change (with respect to the source's proper time) of the logarithm of the volume $V_0$ of an infinitesimal ruler, 
\begin{equation}\label{eqn:ruler growth rate intrinsic}
\Theta(\tilde{a})\equiv \frac{1}{V_0}\frac{\mathrm{d}V_0}{\mathrm{d}\tau} \;,
\end{equation}
$\delta_{\rm g}^{\rm or}$ is the galaxy over-density on constant-observed-redshift hyper-surfaces, and the determinant $\det \left(i^*g_s\right)$ reads 
\begin{equation}\label{eqn:det i*gs}
\begin{aligned}
\det\!\big(i^* g_s\big) &= \tilde a^3 \abs{\det \left(\begin{array}{ccc}
		1-\mathfrak{C} & -\mathfrak{D}_1 & -\mathfrak{D}_{2} \\
		-\hat{\mathfrak{D}}_1 & 1-\mathfrak{A}_{11} & -\mathfrak{A}_{12} \\
		-\hat{\mathfrak{D}}_{2} & -\mathfrak{A}_{21} & 1-\mathfrak{A}_{22}
	\end{array}\right)} \\ &= \tilde a^3 \abs{\det \left(\begin{array}{ccc}
		(1-\mathcal{C})^2 & -\mathcal{B}_1 & -\mathcal{B}_{2} \\
		-\mathcal{B}_1 & 1-2\mathcal{A}_{11} & -2\mathcal{A}_{12} \\
		-\mathcal{B}_{2} & -2\mathcal{A}_{21} & 1-2\mathcal{A}_{22}
	\end{array}\right)}^{1/2}
\end{aligned}   
\end{equation}
in the co-ordinates $(\tilde x^\parallel,\tilde x^I)$.
This makes clear that $\mathcal{B}_I$ (or $\mathfrak{D}_I$, $\hat{\mathfrak{D}}_I$) only need to be known at first order to derive the second-order galaxy number counts (they do not appear in the boost from rest-frame to constant-observed-redshift frame, or the rest-frame bias expansion).

Another important example is the magnification $\mu$, which measures the change of an infinitesimal area element perpendicular to the photon wave-vector $k^\mu$. It is generally given by
\begin{align}
\mu^{-2} &= \det (\delta_{IJ} - 2\mathcal{A}_{IJ}) \\
&= (1-\mathcal{M})^2 - 4\left(\gamma_1^2 + \gamma_2^2\right) \nonumber \; .
\end{align}
At first order, this simplifies to $\mu=1+\tr \mathcal{A}_{IJ} \equiv 1 + \mathcal{M}$, and the relation given in \cite{JeongSchmidt2015}.

\section{Geodesic equation}
\label{sec:geodesic equation}

Having defined the observer's inferred co-ordinate system (on the past light-cone), the cosmic clock and the cosmic rulers, we now proceed to derive explicit expressions for $(\mathcal{T},\mathcal{C},\mathcal{B}_I,\mathcal{A}_{IJ})$ to second order in perturbations, after solving the geodesic equation for the  observer, the source, and the propagation of light from the source to the observer.

\subsection{Perturbed FLRW space-time}

Our physical metric is given in Eq.~\eqref{eqn: metric Poisson gauge}.
Null geodesics, in particular the light path from the source to the observer, can be obtained equivalently (and more simply) by integrating the geodesic equation for the conformally-transformed metric 
\begin{equation}
\label{eqn: conformal metric Poisson gauge}
\mathrm{d}s_{\rm conformal}^2 = -(1+2\Psi)\mathrm{d}\eta^2 + 2\omega_i\mathrm{d}\eta \mathrm{d}x^i + \left(\delta_{ij} - 2\Phi \delta_{ij} + \frac{1}{2}h_{ij}\right)\mathrm{d}x^i\mathrm{d}x^j \;.
\end{equation}
Without loss of generality, we can describe both source and observer geodesics as being described by a 4-velocity field
\begin{equation}\label{eqn: source 4 velocity}
u^\mu = a^{-1}(1-\Psi + \delta u,v^i) \;. 
\end{equation}
Here, 
\begin{equation}\label{eqn: delta u calculation}
\delta u = \frac{v^2 + 3\Psi^2}{2}
\end{equation}
follows from the normalisation condition: $u\cdot u = -1$. 


\subsection{Photon geodesic}

Let $\lambda$ be the affine parameter of the 4-trajectory $\xlc^\mu(\lambda)$ of light from the source to the observer, so that $k^\mu = \mathrm{d}\xlc^\mu/\mathrm{d}\lambda$ becomes its wave-vector. 
To match existing first-order calculations in the literature \cite{Jeong:2011as,SchmidtJeong2012,JeongSchmidt2014,JeongSchmidt2015}, we choose $\lambda$ to be the radial co-moving co-ordinate $\chi = |\vx|$ (whence $\mathrm{d}\xlc^0/\mathrm{d}\lambda < 0$). 
Furthermore, we fix $\chi=0$ at the observer's position. The geodesic equation is thus given by
\begin{equation}\label{eqn: light geodesic equation}
\frac{\mathrm{d}^2\xlc^\mu}{\mathrm{d}\chi^2} \equiv \frac{\mathrm{d}k^\mu}{\mathrm{d}\chi} =  -\Gamma^\mu_{\alpha\beta}k^\alpha k^\beta\;,
\end{equation}
where the Christoffel symbols of the conformally-transformed metric \eqref{eqn: conformal metric Poisson gauge} are listed to second order in Appendix \ref{appendix:Christoffel symbols}. One benefit of using the conformally equivalent metric \eqref{eqn: conformal metric Poisson gauge} is that the Christoffel symbols are zero at the zeroth (background) order, and only begin to contribute from the first (linear) order onward.
This allows us to solve equation \eqref{eqn: light geodesic equation} iteratively. 
At zeroth order, the solution is
\begin{equation}\label{eqn:zeroth-order geodesic}
x_{\textrm{lc},(0)}^{\mu}(\chi) = \left(\eta_0-\chi,\tilde{n}^i\chi\right) \;.
\end{equation}
We shall parametrise the higher-order contributions as
\begin{equation}\label{eqn: k mu definition}
k^\mu \equiv (-1 + \delta \nu,\tilde{n}^i + \delta n^i) \;.
\end{equation}
Up to second order in the perturbations, the geodesic equation is
\begin{equation} \label{eqn:geodesic to second order}
\frac{\mathrm{d}k^\mu_{[2]}}{\mathrm{d}\chi} 
= -\Gamma^{\mu,(1)}_{\alpha\beta}
k^\alpha_{[1]} k^\beta_{[1]} -\Gamma^{\mu,(2)}_{\alpha\beta}
k^\alpha_{(0)} k^\beta_{(0)} \;,
\end{equation}
where we denote by $Q^{[n]}$ or $Q_{[n]}$ the evaluation of a quantity $Q$ up to order $n$ included, whereas $Q^{(n)}$ or $Q_{(n)}$ denotes the order $n$ term in $Q$. 

To obtain the photon trajectory from the source to the observer, we must integrate the geodesic equation \eqref{eqn: light geodesic equation}, which is a set of 4 equations for $\delta \nu(\chi)$ and $\delta n^i(\chi)$. 
We describe the boundary conditions for these equations in \S\ref{subsec: source trajectory} and \S\ref{subsec: initial conditions}, before writing them down and formally integrating them in \S\ref{subsec:frequency shift}.


\subsection{Source and observer geodesics}
\label{subsec: source trajectory}


Both the source and the observer are assumed to be co-moving with the same time-like velocity field $u^\mu$, given by equation \eqref{eqn: source 4 velocity}. The final condition for the geodesic equation is that the source emitted its signal at $x^\mu_e$, which is received by the observer at $\tau_o$, with observed redshift $\tilde{z}$ and from a sky-direction $\ntvh$; the observer is situated at a global-co-ordinate position $x^\mu_o \equiv (\eta_0,\bm{0})$.

Let us start with the proper time of the source at emission, i.e.~the proper age of the source: the proper age increment is 
\begin{equation}
\mathrm{d}\tau = \sqrt{-g_{\mu\nu}\frac{\mathrm{d}x^\mu}{\mathrm{d}\eta}\frac{\mathrm{d}x^\nu}{\mathrm{d}\eta}} \mathrm{d}\eta \;,
\end{equation}
with 
\begin{align}
-g_{\mu\nu}\frac{\mathrm{d}x^\mu}{\mathrm{d}\eta}\frac{\mathrm{d}x^\nu}{\mathrm{d}\eta} &= a^2\left[(1+2\Psi) -\frac{2\omega_iv^i}{1-\Psi + \delta u} - \frac{v^2(1-2\Phi)}{(1-\Psi + \delta u)^2} - \frac{1}{2}h_{ij}\frac{v^iv^j}{(1-\Psi + \delta u)^2} \right]\nonumber \\
&\approx a^2\left(1+2\Psi - v^2 \right)
\label{eq:dtau2}
\end{align}
to second order.
Let $\vx_s(\eta) \equiv \vx_e + \delta \vx_s(\eta)$ be the source's geodesic, where $x_e^\mu \equiv (\eta_e,\vx_e)$ is the source's position (in the global co-ordinate system) at emission.
To first order, $\vx'_s = \vv(\vx_s,\eta) \approx \vv(\vx_e,\eta)$, whence 
\begin{equation}
\label{eq:delta_x_s}
\delta \vx_s(\eta_1) = \int_0^{\eta_1} \vv(\vx_e,\eta_2)\mathrm{d}\eta_2 - \int_0^{\eta_e} \vv(\vx_e,\eta_2)\mathrm{d}\eta_2 \;,
\end{equation}
assuming $0 \leq \eta_1 \leq \eta_e$. The second term on the right-hand side encapsulates the boundary condition $\vx_s(\eta_e) = \vx_e$, i.e.~the source is at the global co-ordinate position $\vx_e$ at the (conformal) time of emission.
Summarising,
\begin{equation}
\label{eq:tau_s}
\tau(\eta_e,\vx_e) = \int_0^{\eta_e} \left(1+ \Psi(\vx_e,\eta) + \partial_i\Psi \delta x^i_s(\eta) - \frac{1}{2}\Psi^2(\vx_e,\eta) -\frac{1}{2}v^2(\vx_e,\eta)\right)a(\eta)\mathrm{d}\eta
\end{equation}
is the source's proper time at emission to second order in perturbations.

Specialising the above to the observer, we can find the difference between $\tau_o$ and $t_0 \equiv \overline{t}(\eta_0)$. Defining $\delta \tau_o \equiv \tau_o - t_0$, we have
\begin{equation}\label{eqn: delta tau 0}
\delta \tau_o = \int_0^{\tilde{\eta}_0 + \Delta \eta^{(1)}_0} \left(\Psi(\bm{0} + \Delta \vx^{(1)}_0,\eta') + \partial_i\Psi \delta x^i_o(\eta') - \frac{1}{2}\Psi^2(\bm{0},\eta') -\frac{1}{2}v^2(\bm{0},\eta')\right)a(\eta')\mathrm{d}\eta' \; . 
\end{equation}
The displacement of the space-time co-ordinates of the observer is defined, as in equation \eqref{eqn: total displacement definition}, by $\Delta x^\mu_o \equiv x_o^\mu - \tilde{x}_o^\mu$; and $\delta x^i_o(\eta)$ is the (first-order) displacement of the observer's past trajectory relative to their spatial co-ordinate position $\vx_o = \bm{0}$. To first order, $\delta x^i_o(\eta)$ is given by 
\begin{equation}
\label{eq:delta_x_o}
\delta \vx_o(\eta_1) = \int_0^{\eta_1} \vv(\bm{0},\eta_2)\mathrm{d}\eta_2 - \int_0^{\eta_0} \vv(\bm{0},\eta_2)\mathrm{d}\eta_2 \;,
\end{equation}
where the second term ensures that $\delta \vx_o(\eta_0) = \bm{0}$, and we chose the spatial origin of co-ordinates as $\vx_o = \bm{0}$. The expression for $\delta\tau_o$ seems, \emph{prima facie}, to contain terms that have not yet been calculated, \emph{viz.} $\Delta x^\mu_{(1)}$, and whose values depend on $\delta \tau_o$ itself. However, they only appear at second order, and themselves are required only to first order, so in fact equation \eqref{eqn: delta tau 0} is not circular. Furthermore, this procedure has already been followed in refs.~\cite{SchmidtJeong2012,JeongSchmidt2015} to first order, so one may simply use their expressions for $\Delta x^\mu_{(1)}$ (or all other first-order quantities) here.

\subsection{Boundary conditions at observer}
\label{subsec: initial conditions}

The boundary conditions $\delta \nu_o$, $\delta n^i_o$ to the photon geodesic equation \eqref{eqn: light geodesic equation} are set by requiring, at all orders in perturbation theory, that photons have an observed in-coming direction $\ntvh$ on the sky, and a certain observer-frame frequency. The value of the latter is irrelevant, as only the ratio between emitted (in the source rest frame) and observed frequency matters, which is $1+\tilde z$. Hence, we can simply set the observed frequency to unity.
We can construct a normal Fermi tetrad by using the observer's four-velocity as the time-like component, $e_{\underline{0}}^\mu = u^\mu$, and orthogonal spatial unit vectors $e_{\underline{i}}^\mu$. In terms of this tetrad,
the boundary conditions are 
\begin{align}
\label{eqn:initial condition frequency}1 & = (g_{\mu\nu}e_{\underline{0}}^\mu k^\nu a^{-2})_o \\
\tilde{n}_i & = (g_{\mu\nu}e_{\underline{i}}^\mu k^\nu a^{-2})_o \;, \label{eqn:initial condition direction}
\end{align}
(the conversion from a tetrad index to a co-ordinate index for $\ntvh$ is trivial) 
where the components $e_{\underline{a}}^\mu$ are calculated explicitly in Appendix \ref{appendix:observer tetrad}. These boundary conditions thus become a system of linear equations in $\delta \nu_o$ and $\delta n^i_o$, which we can write as
\begin{equation}
A \left(\begin{array}{c}
	\delta \nu_o \\
	\delta n^i_o
\end{array}\right) = b \;,
\end{equation}
with solution $A^{-1}b$. 
The 4-vector $b$ starts at first order, while the $4\times 4$ matrix $A$ starts at zeroth order. Since $A^{-1}$ is multiplying a first-order perturbation, we need only evaluate it at first order. Explicitly, we have
\begin{equation}
A^{-1} = \left(\begin{array}{cc}
	1-\Psi & -v^j \\
	
	v^j & -\delta_j^k - \Phi\delta_j^k \\
\end{array}\right)\;,
\end{equation}
and 
\begin{equation}
b = \left(\begin{array}{c}
	(1+2\Psi)e_{\underline{0}}^0 + \omega_\parallel + (1-2\Phi)v_\parallel - a_o \\
	
	(1+2\Psi)e^0_{\underline{i}} - \omega_{\underline{i}} + (1-2\Phi)e_{\underline{i}}^k\tilde{n}_k + \frac{1}{2}h_{\underline{i}k}\tilde{n}^k - \tilde{n}_{\underline{i}}a_o
\end{array}\right)\;.
\end{equation}
Here and throughout, $\xi_\parallel \equiv \tilde{n}^i\xi_i$ for any 3-vector $\xi^i$.
Upon setting $\delta a_o \equiv a_o - 1$, where $a_o \equiv a(\eta_o)$ (recall that in general the observer's global co-ordinate $\eta_o$ is different from $\overline{\eta}(\tau_o)$), we obtain
\begin{align}
\label{eqn:delta nu at observer}\delta\nu_o = (A^{-1}b)_0 & = \Psi_o -\delta a_o + v_{o,\parallel} - 3\Psi_o^2  + \delta u_o -\Phi_o v_{o,\parallel} + \omega_{o,\parallel} - v_o^2 - v_{o,\parallel} \Psi_o \\ 
&\quad + \delta a_o(\Psi_o + v_{o,\parallel}) \nonumber\\
\delta_{ij}\delta n^j_o = (A^{-1}b)_i & = (\Psi_o + v_{o,\parallel} - \delta a_o)v_{o,i} - (1-\Phi) a_o e_{\underline{i}}^k\tilde{n}_k + \tilde{n}_i(1+\Phi_o) + (1 + \Phi_o)\tilde{n}_i \delta a_o \nonumber \\ &
\quad - \Big(a_o e_{\underline{i}}^0 + (\Phi_o + 2\Psi_o)v_{o,i} - \omega_{o,i} - 2\Phi_o^2 \tilde{n}_i + \frac{1}{2}h_{o,ik}\tilde{n}^k\Big)
\label{eqn:delta n at observer} \\ 
& = -v_i - v_i\delta a_o + \frac{1}{2}v_iv_\parallel - \frac{1}{4}h_i^k \tilde{n}_k + \Phi \tilde{n}_i + \frac{3}{2}\Phi^2 \tilde{n}_i + \delta a_o\tilde{n}_i(1+\Phi) \nonumber
\end{align} 
after substituting $e_{\underline{i}}^\mu$ from Appendix \ref{appendix:observer tetrad} to second order.

To parametrise the scale factor $a$ with the cosmic time $t$, we define $a(t) = a(\bar t(\tau) - \delta \tau)$ with $\bar t(\tau)\equiv \tau$ and $\delta \tau =\tau-t$ and arrive at the expansion
\begin{equation}\label{eqn:delta ln H}
\begin{aligned}
	\frac{a(t)}{\at(\tau)} & 
	= 1-\barH \delta \tau + \frac{\ddot{\at}(\tau)}{2\at(\tau)}\delta \tau^2 + \ldots \;,
\end{aligned}
\end{equation}
where $\barH \equiv H(\overline{\eta}(\tau))$, and $H$ is the background Hubble constant. This can be used to find an explicit expression for the perturbation $\delta a_o$ to the scale factor upon setting $t=t_o$ and $\tau=\tau_o$ in equation \eqref{eqn:delta ln H}, \emph{viz.}
\begin{equation}\label{eqn:delta a observer}
\delta a_o = H(\tau_o)(t_o - \tau_o) + \frac{\ddot{\at}(\tau_o)}{2}(t_o - \tau_o)^2 +\ldots \;.
\end{equation}

\subsection{Second-order shifts}
\label{subsec:frequency shift}
We solve the geodesic equation \eqref{eqn: light geodesic equation} perturbatively to second order. The linear (first) order perturbation solution has been presented in the literature, for example in ref.~\cite{SchmidtJeong2012}. To find the second-order solution, we have to collect all second-order quantities on the right-hand side of equation \eqref{eqn:geodesic to second order}, which requires $\xlc$ and $k^\alpha$ only to \emph{first order}, since the Christoffel symbols $\Gamma^\mu_{\alpha\beta}$ are already first order. We, however, need the Christoffel symbols to second order, including the linear-order terms evaluated along the first-order path $\xlc^{[1]}$. A line-of-sight integration then yields $k^\mu$ to second order.

A direct calculation for $\mu = 0$ yields 
\begin{align}
\frac{\mathrm{d}\delta \nu}{\mathrm{d}\chi} & = -\Psi' (1-2\Psi) \left(1-2\delta\nu^{(1)}(\chi)\right) +2\partial_i\Psi(1-2\Psi)\left[\tilde{n}^i\left(1-\delta\nu^{(1)}(\chi)\right) + \delta n^i_{(1)}(\chi)\right] \nonumber \\ &\quad - \frac{h'_\parallel}{4}
+ \partial_\parallel\omega_\parallel + \Phi'(1-2\Psi)\left(1+2\tilde{n}^i\delta n^{(1)}_i(\chi)\right)
\label{eqn: kernel for frequency shift} \;,
\end{align}
where we have shortened 
$\partial_\parallel = \tilde{n}^i\partial_i$.
All fields $\Psi$, $\Phi$, $\omega_i$, $h_{ij}$ and $v^i$ are evaluated at $x^{\mu}_{\textrm{lc},[1]} (\chi) = x_{\textrm{lc},(0)}^\mu(\chi) + x_{\textrm{lc},(1)}^\mu(\chi)$. This renders the right-hand-side of the above equation an explicit function $F^0_{[2]}(\chi)$ of the affine parameter, which may be integrated to yield, up to corrections at third order,
\begin{equation}\label{eqn: delta nu integral expression}
\delta \nu(\chi_e) - \delta \nu_o = \int_0^{\chi_e} F^0_{[2]}\left( x_{\textrm{lc},(0)}^\mu(\chi) + x_{\textrm{lc},(1)}^\mu(\chi)\right)\mathrm{d}\chi \;.
\end{equation}
Likewise, for $\mu=i$, the equation of motion for $\delta \nh$ evaluates to
\begin{align}
\frac{\mathrm{d} \delta n^i}{\mathrm{d}\chi} & =
- \Phi'\Big(2\tilde{n}^i+4\tilde{n}^i\Phi - 2\tilde{n}^i \delta \nu^{(1)}(\chi) + 2 \delta n^i_{(1)}(\chi)\Big) -\delta^{ij}\partial_j \Psi \Big(1+2\Phi - 2 \delta \nu^{(1)}(\chi)\Big) - \omega'^i \nonumber \\ & 
\qquad + \partial_{\parallel}\omega^i - \delta^{ij}\partial_j \omega_{\parallel}
+ \frac{1}{2} \tilde{n}^j h'^i_j - (1+2\Phi) \left(\delta^{ij}\partial_j\Phi - 2\partial_{\parallel} \Phi \tilde{n}^i\right) - \frac{1}{4}\Big(2\tilde{n}^j \partial_{\parallel} h^i_j - \delta^{ij}\partial_j h_{||}\Big) \nonumber \\ 
&\qquad \bigg. - 2 \delta^{ij}\partial_j\Phi \delta n_{\parallel}^{(1)}(\chi) 
+ 2 \tilde{n}^i \delta n^j_{(1)}(\chi) \partial_j \Phi 
+ 2 \delta n^i_{(1)}(\chi) \partial_{\parallel} \Phi \; .
\label{eqn:Geoe}
\end{align}
Defining the right-hand side, which is evaluated on the first-order light trajectory $x^\mu_{\textrm{lc},[1]}(\chi)$, as $F_{[2]}^i(\chi)$, we can recast the solution for $\mu=i$ into
\begin{equation}\label{eqn: delta n i integral expression}
\delta n^i(\chi_e) - \delta n^i_o = \int_0^{\chi_e} F_{[2]}^i\left( x_{\textrm{lc},(0)}^\mu(\chi) + x_{\textrm{lc},(1)}^\mu(\chi)\right) \mathrm{d}\chi \;.
\end{equation}
These solutions to the geodesic equation \eqref{eqn: light geodesic equation} will enable us to evaluate the change in frequency and direction along the null geodesic from the source to the observer. However, we have yet to relate it to the observed quantities $\tilde{z}$ and $\ntvh$; we will do so now. 

\section{Relating the inferred and the actual source positions}
\label{sec:source position}
Let $x_e^\mu \equiv \xlc^\mu(\chi_e)$ be the space-time position of the source, where $\chi_e$ is the affine parameter at emission. By equation \eqref{eqn:zeroth-order geodesic}, we thus have $x_e^\mu = x_{\textrm{lc},(0)}^\mu(\chi_e)$ at zeroth order. The solutions $\delta \nu$ and $\delta n^i$ in equations \eqref{eqn: delta nu integral expression} and \eqref{eqn: delta n i integral expression} are expressed, as yet, in terms of the global source co-ordinates $x^\mu_e$, which are not observable. The observer uses the measured redshift $\tilde{z}$ and incoming photon direction $\ntvh$ to assign to each source a position $\tilde{x}^\mu$ which is derived from the background FLRW metric expressions as in equation \eqref{eqn: tilde co-ordinates definition}. 
To relate $\tilde{x}^\mu$ and $x_e^\mu$, we first need the scale factor at emission, expressed in terms of the observed redshift $\tilde{z}$. 

\subsection{Observed redshift}
\label{subsec:affine parameter}
By definition, the observed redshift is
\begin{equation}
1+\tilde{z} = \frac{\big(k^{\rm ph}_\mu u^{\mu}\big)_e}{\big(k^{\rm ph}_\nu u^{\nu}\big)_o} \;,
\end{equation}
where $\big(k^{\rm ph}_\mu\big)_e$,  $\big(u^{\mu} \big) _e$, $\big(k^{\rm ph}_\mu\big)_o$ and $\big(u^{\mu}_g\big)_o$ are the physical (not the conformal) photon wave-vector and 4-velocity of source and observer, respectively. The boundary condition $\big(k^{\rm ph}_\nu u^{\nu}\big)_o = 1$ implies
\begin{equation}\label{eqn: redshift}
1+\tilde{z} = (k^{\rm ph}_\mu u^\mu)_e = \frac{1}{a_e}k_\mu u^{\mu}= \frac1{\tilde a} \;,
\end{equation}
where $a_e = a(\eta_e)$ is the background scale factor evaluated at the co-ordinate conformal time of emission (that is, at the source). 
The last equality, which defines the inferred scale factor at emission, follows from $\at(\tau_o)=1$.
Hence, to second order in perturbations, equation \eqref{eqn: redshift} becomes 
\begin{equation}\label{eqn: delta ln a}
\begin{aligned}
	\frac{a_e}{\tilde{a}} & = 1 + \delta u + \Psi - 2\Psi^2 -\delta \nu(1+\Psi) + v_\parallel(1-2\Phi) + v_i\delta n^i + \omega_i\tilde{n}^i \\
	& \equiv 1 + \Delta_{a,e} \;.
\end{aligned}
\end{equation}
At first order, $\Delta_{a,e}^{[1]}$ is equal to the quantity refs. \cite{SchmidtJeong2012,JeongSchmidt2014,JeongSchmidt2015} denoted by $\Delta \ln a$. 
To determine $\Delta_{a,e}$, and thus the redshift, to second order, all we need is $\delta n^i$ to first order (as it is dotted with a $\vv$, which is already first order), and $\delta \nu$ to second order, as given by equation \eqref{eqn: delta nu integral expression}. 

\subsection{Total displacements}
\label{subsec:Delta and delta}

We introduce the perturbation to the photon geodesic
\begin{equation}\label{eqn: lower-case delta x mu definition}
\delta x^\mu(\chi) \equiv \xlc^\mu(\chi) - x_{\textrm{lc},(0)}^\mu(\chi)\; ,
\end{equation}
such that
\begin{align}\label{eqn: total displacement mu}
\Delta x^\mu & = x_{\textrm{lc},(0)}^\mu(\chi_e) + \delta x^\mu(\chi_e) - \tilde{x}^\mu(\tilde{\chi}) \\ &
= (\eta_0-\tilde\chi,\tilde{n}^i \tilde\chi) + (-1,\tilde{n}^i) \delta\chi+ \delta x^\mu(\tilde\chi) + \frac{\mathrm{d}\delta x^\mu}{\mathrm{d}\tilde\chi}\delta\chi - \tilde x^\mu(\tilde\chi) \nonumber \\ &
= \delta x^\mu(\tilde{\chi}) + k^\mu(\tilde{\chi})\delta \chi \nonumber \;.
\end{align}
Here, 
\begin{equation}
\delta \chi \equiv \chi_e - \tilde{\chi} 
\end{equation}
and the second equality is valid to second order, because $x_{\textrm{lc},(0)}(\chi)$ is linear in $\chi$.

Following \cite{JeongSchmidt2015}, we require $a(\xlc^0(\chi_e)) = (1+\Delta_{a,e})\tilde{a}$, or
\begin{equation}
a\big(\tilde\eta+\Delta\eta\big) = \tilde{a}\times(1+\Delta_{a,e})\;,
\end{equation}
(since $\xlc^0(\chi_e)=\tilde\eta+\Delta\eta$)
to determine the total time shift $\Delta x^0=\Delta\eta$ at second order. From this we find
\begin{equation}\label{eqn: total displacement eta}
\Delta \eta = \frac{\Delta_{a,e}}{\tilde{\mathcal{H}}} - \frac{\tilde{a}''}{2\tilde{a}}\frac{\Delta_{a,e}^2}{\tilde{\mathcal{H}}^3} \;.
\end{equation}
Setting $\mu=0$ in equation \eqref{eqn: total displacement definition} and using $k^0(\tilde\chi) = -1 + \delta\nu(\tilde\chi)$, equation \eqref{eqn: total displacement eta} implies
\begin{align}\label{eqn:deltachi1}
\big[1-\delta\nu(\tilde{\chi})\big]\delta\chi  
&= \delta \eta(\tilde{\chi}) - \left(\frac{\Delta_{a,e}}{\tilde{\mathcal{H}}} - \frac{\tilde{a}''}{2\tilde{a}}\frac{\Delta_{a,e}^2}{\tilde{\mathcal{H}}^3}\right) \\
&= \int_0^{\tilde{\chi}}\delta \nu(\chi) \mathrm{d}\chi - \left(\frac{\Delta_{a,e}}{\tilde{\mathcal{H}}} - \frac{\tilde{a}''}{2\tilde{a}}\frac{\Delta_{a,e}^2}{\tilde{\mathcal{H}}^3}\right) \nonumber
\end{align}
at second order.


Likewise, on setting $\mu=i$ in equation \eqref{eqn: total displacement mu} and using $k^i(\tilde\chi) = \tilde{n}^i+\delta n^i(\tilde\chi)$ along with $\delta\chi$ given by equation \eqref{eqn:deltachi1}, the total spatial displacements are given by
\begin{align}
\Delta x^i & = \left[\int_0^{\tilde{\chi}}\delta \nu(\chi) \mathrm{d}\chi - \left(\frac{\Delta_{a,e}}{\tilde{\mathcal{H}}} - \frac{\tilde{a}''}{2\tilde{a}}\frac{\Delta_{a,e}^2}{\tilde{\mathcal{H}}^3}\right)\right]\frac{\tilde{n}^i + \delta n^i(\tilde{\chi})}{1-\delta\nu(\tilde{\chi})} + \int_0^{\tilde{\chi}}\delta n^i(\chi)\mathrm{d}\chi \label{eqn: total displacement x i}  \\ & 
= \left[\int_0^{\tilde{\chi}}\delta \nu(\chi) \mathrm{d}\chi - \left(\frac{\Delta_{a,e}}{\tilde{\mathcal{H}}} - \frac{\tilde{a}''}{2\tilde{a}}\frac{\Delta_{a,e}^2}{\tilde{\mathcal{H}}^3}\right)\right]
\Big[\big(1+\delta\nu^{(1)}(\tilde{\chi})\big)\tilde{n}^i + \delta n^i_{(1)}(\tilde{\chi})\Big]
%
%
+ \tilde{\chi}\delta n^i_o \nonumber \\ & \quad
+ \int_0^{\tilde{\chi}}\mathrm{d}\chi_e \int_0^{\chi_e}\mathrm{d}\chi~ \bigg[- \Phi'\Big(2\tilde{n}^i+4\tilde{n}^i\Phi - 2\tilde{n}^i \delta \nu^{(1)}(\chi) + 2 \delta n^i_{(1)}(\chi)\Big) - \omega'^i + \partial_{\parallel}\omega^i \nonumber \\ & \qquad
+ \frac{\tilde{n}^j h'^i_j}{2} - \delta^{ij}\partial_j \omega_{\parallel} -\delta^{ij}\partial_j \Psi \Big(1+2\Phi - 2 \delta \nu^{(1)}(\chi)\Big)
- \big(1+2\Phi\big) \big(\delta^{ij}\partial_j\Phi - 2\partial_{\parallel} \Phi \tilde{n}^i\big) \nonumber \\ & \qquad
+ 2 \delta n^i_{(1)}(\chi) \partial_{\parallel} \Phi - \frac{1}{4}\Big(2\tilde{n}^j \partial_{\parallel} h^i_j - \delta^{ij}\partial_j h_{||}\Big) - 2 \delta^{ij}\partial_j\Phi \delta n_{\parallel}^{(1)}(\chi) 
+ 2 \tilde{n}^i \delta n^j_{(1)}(\chi) \partial_j \Phi 
\bigg] \nonumber \;.
\end{align}
At this point, we have all the ingredients to write down the cosmic clock and cosmic ruler observables at second order in perturbations. 

\section{Cosmic clock at second order}
\label{sec:clock}

To compute $\mathcal{T}$, we need to know the proper age (i.e.~the proper time) of the source at emission, which we already found in \S \ref{subsec: source trajectory}, in equation \eqref{eq:tau_s}.
All that is left to do now is to relate the source's space-time co-ordinates $x_e^\mu = \xlc^\mu(\chi_e) = (\eta_e,\vx_e)$ to the inferred co-ordinates $\tilde{\chi}$ and $\ntvh$, and then insert equation \eqref{eq:tau_s} into the argument of the function $\at$.
The last stage is carried out through the Taylor expansion 
\begin{equation}\label{eqn: delta ln alpha}
\begin{aligned}
	\frac{\at(\tau)}{a(t)} & \equiv 1+ \Delta_{a,\tau} \\ &
	= 1 + H(t)(\tau - t) + \frac{a''(t)}{2a(t)}(\tau-t)^2 + \ldots ,
\end{aligned}
\end{equation}
where $t = \overline{t}(\eta) = \int_0^\eta a(\eta')\mathrm{d}\eta'$. As a result, 
\begin{align}\label{eqn: delta tau definition}
H \delta \tau &\equiv H(t)(\tau - t) \\ 
&= H\int_0^{\eta_e} \left(\Psi(\vx_e,\eta') + \partial_i\Psi \delta x^i_s(\eta') - \frac{1}{2}\Psi^2(\vx_e,\eta') -\frac{1}{2}v^2(\vx_e,\eta')\right)a(\eta')\mathrm{d}\eta' \nonumber \;.
\end{align}
Since this expression is already first order, it needs to be evaluated at a co-ordinate $x^\mu_e$ 
which is correct only to first order, i.e.~at $\eta_e = \tilde{\eta} + \Delta \eta^{(1)}$ and $\vx_{e} = \tilde{\vx} + \Delta \vx^{(1)}$ (cf.~equation \eqref{eqn: total displacement definition}), where the expressions for $\Delta x_\mu^{(1)}$ are first-order ones (which, as mentioned above, can be found in the literature \cite[e.g.][]{JeongSchmidt2015}).
Thus, to the appropriate order,
\begin{equation}\label{eqn: delta tau calculated}
\delta \tau = \int_0^{\tilde{\eta} + \Delta \eta^{(1)}} \left[\Psi(\tilde{\vx} + \Delta \vx^{(1)},\eta') + \partial_i\Psi \delta x^i_s(\eta') - \frac{1}{2}\Psi^2(\tilde{\vx},\eta') -\frac{1}{2}v^2(\tilde{\vx},\eta')\right]a(\eta')\mathrm{d}\eta',
\end{equation}
where $\Delta \vx^{(1)}$ is evaluated at $(\tilde \vx,\tilde \eta)$, i.e.~it is a constant with respect to the $\eta'$ integration. 
On using equation \eqref{eqn:delta ln H}, this implies
\begin{equation}\label{eqn: delta ln alpha calculated}
\begin{aligned}
1+\Delta_{a,\tau} & = \left(1-\barH \delta \tau + \frac{\ddot{\at}(\tau)}{2\at(\tau)}\delta \tau^2\right)^{-1} + \ldots \\ &
= 1+ \barH \delta \tau + \left({\barH}^2- \frac{\ddot{\at}(\tau)}{2\at(\tau)}\right)\delta \tau^2 + \ldots \; ,
\end{aligned}
\end{equation}
where $\delta \tau$ is now given by equation \eqref{eqn: delta tau calculated}. 
Substituting equations \eqref{eqn: delta ln alpha calculated} and \eqref{eqn: delta ln a} for the first and second summand of equation \eqref{eq:Tdef}, respectively, we arrive at
\begin{align}\label{eqn:cosmic clock second order}
\mathcal{T} & = \barH \delta \tau + \frac{1}{2}\left(\tilde{H}^2- \frac{\ddot{\at}(\tau)}{\at(\tau)}\right)\delta \tau^2 \\ &
\quad + \delta u + \Psi(x) - 2\Psi^2 -\delta \nu(1+\Psi) + v_\parallel(x)(1-2\Phi) + v_i\delta n^i + \omega_i\tilde{n}^i - \frac{1}{2}\left[\Psi -\delta \nu + v_\parallel \right]^2 \nonumber \\ &
= \tilde{H} \int_0^{\tilde{\eta}} \left[\Psi(\tilde{\vx},\eta') + \partial_i\Psi(\tilde{\vx},\eta')\left(\delta x_{s}^i + \Delta x^i_{(1)}\right) - \frac{1}{2}\Big(\Psi^2(\tilde{\vx},\eta') + v^2(\tilde{\vx},\eta')\Big)\right]a(\eta')\mathrm{d}\eta'  \label{eqn:cosmic clock second order explicit} \\ &
\quad + \left[\frac{(\tilde{\mathcal{H}}' - \mathcal{H}^2)}{\tilde{a}^2}\left(\tilde{a}\Delta \eta^{(1)} + \int_0^{\tilde{\eta}} \Psi(\tilde{\vx},\eta')a(\eta')\mathrm{d}\eta'\right) \right]\int_0^{\tilde{\eta}} \Psi(\tilde{\vx},\eta')a(\eta')\mathrm{d}\eta' \nonumber \\ &
\quad + \tilde{\mathcal{H}}\Delta \eta^{(1)}\Psi(\tilde{x})
+ \frac{1}{2}\left(\tilde{H}^2- \frac{\ddot{\at}(\tau)}{\at(\tau)}\right)\left(\int_0^{\tilde{\eta}} \Psi(\tilde{\vx},\eta')a(\eta')\mathrm{d}\eta'\right)^2 - \frac{1}{2}\Big(\delta \nu^2 + v_\parallel^2\Big) + v_i\delta n^i + \omega_i\tilde{n}^i \nonumber \\ &
\quad + \delta u + \Psi + \partial^{\mu}\Psi\Delta x^{(1)}_\mu - \frac{5}{2}\Psi^2 -\delta \nu(1-v_\parallel) + v_\parallel(1-2\Phi-\Psi) + \partial^\mu v_{\parallel}\Delta x^{(1)}_{\mu} \nonumber
\end{align}
to second order in perturbations, where the second equality follows from expanding some of the quantities in equation \eqref{eqn:cosmic clock second order} and substituting in equation \eqref{eqn: delta tau calculated}.

Notice that $\mathcal{T}$ vanishes at the observer's position, where $\Delta_{a,e} = \delta a_o$ and
\begin{equation}
\left(\Delta_{a,\tau}\right)_o = H(\tau_o) \left(\tau_o-t_o\right) + \left(\tilde{H}_0^2- \frac{\ddot{\at}(\tau_o)}{2\at(\tau_o)}\right)\left(\tau_o-t_o\right)^2 \;.
\end{equation}
Therefore, by equation \eqref{eqn:delta a observer} we get $\mathrm{e}^{\mathcal{T}} = \left[1+ \left(\Delta a_{a,e}\right)_o\right]\left[1+ \left(\Delta a_{a,\tau}\right)_o\right] = 1$ to second order.

Appendix \ref{appendix:T} gives an expression for $\mathcal{T}$ with the conformal Hubble parameter.


\section{Cosmic rulers at second order}
\label{sec:rulers}

While the metric rulers $(\mathcal{C},\mathcal{B}_I,\mathcal{A}_{IJ})$ defined by \cite{SchmidtJeong2012,JeongSchmidt2014,JeongSchmidt2015} are more closely related to observables, the $1$-form rulers $(\mathfrak{C},\mathfrak{B}_I,\mathfrak{A}_{IJ})$ are calculationally more convenient. Therefore, we adopt the second definition outlined in \S\ref{sec:frames} for practical computations, but the two are convertible via equations \eqref{eqn:relation between fraktur and calligraphic rulers}. Let us now repeat some of the steps involved in their construction in more detail. Concretely, we define an orthonormal tetrad $s_{\underline{\alpha}}^\mu$ at the source, in exact analogy to the observer tetrad $e_{\underline{\alpha}}^\mu$. On setting $s_{\underline{0}}^\mu= u^\mu$, the spatial metric $g_s$ for the source is given by $g_s=s^{\underline{i}}\otimes s^{\underline{i}}$, where the tetrad indices $\underline{i}$ are raised and lowered by the canonical metric $\delta_{\underline{ij}}$.

The crucial step in the derivation of $(\mathfrak{C},\mathfrak{B}_I,\mathfrak{A}_{IJ})$ consists in projecting the ``shadow'' of the photon 4-trajectory on the spatial hyper-surface in the source rest frame. For this purpose, we introduce a new co-tetrad field $s^\parallel\equiv \hat n_{s\underline{i}}\, s^{\underline{i}}$ and constrain $\hat n_{s\underline{i}}$ from the requirement $s^\parallel(k_s)=\sqrt{g_s(k_s,k_s)}$, i.e.
\begin{equation}
\hat n_{s\underline{i}} s_\mu^{\underline{i}} k_s^\mu = \sqrt{(k_s^\mu s_\mu^{\underline{j}})(k_{s\nu} s_{\underline{j}}^\nu)} \;,
\end{equation}
where $g_s(k_s,k_s)$ is proportional to the proper-length increment along the light beam, as measured in the source's rest-frame.
The particular choice 
\begin{equation}\label{eqn: n_s hat definition}
\hat n_{s\underline{i}} = \frac{k_{s\mu} s^\mu_{\underline{i}}}{\sqrt{(k_s^\mu s_\mu^{\underline{j}})(k_{s\nu} s_{\underline{j}}^\nu)}}
\end{equation}
implies the relation
\begin{equation}
s^\parallel(s_\parallel) = s^\parallel(\hat n_s^{\underline{i}} e_{\underline{i}})= \hat n_s^{\underline{i}} \hat n_{s\underline{i}} = 1\;,
\end{equation}
which constrains the tetrad field $s_\parallel=\hat n_s^{\underline{i}}s_{\underline{i}}$ along the ``direction'' of $s^\parallel$.
Next, we introduce two additional co-tetrad fields $s^I = \hat e_{I\underline{j}}s^{\underline{j}}$ and fix $\hat e_{I\underline{j}}$ from the requirement 
\begin{equation}
s^\parallel(s_I) = 0 = \hat n_{s\underline{i}} \hat e_I^{\underline{i}}\; ,
\end{equation}
where $s_I$ is the tetrad field dual to $s^I$. Likewise, the orthonormality condition $s^I(s_J)\equiv \delta_J^I$ implies
\begin{equation}
\hat e_{I\underline{i}}\,\hat e_J^{\underline{i}} = \delta_{IJ} \;.
\end{equation}
The relation 
\begin{equation}
\delta^{\underline{i}}_{\underline{j}} = \hat n_s^{\underline{i}}\hat n_{s\underline{j}}+\hat e_1^{\underline{i}}\hat e_{1\underline{j}}+\hat e_2^{\underline{i}}\hat e_{2\underline{j}}
\end{equation}
enables us to write the spatial metric $g_s$ for the source as
\begin{equation}
g_s = s^\parallel\otimes s^\parallel + \sum_I s^I\otimes s^I \;,
\end{equation}
which is equation \eqref{eqn: metric in terms of s's}.
Furthermore, it allows us to define a projection operator
\begin{equation}
P_s^{\underline{ij}} = \delta^{\underline{ij}} - \hat n_s^{\underline{i}}\hat n_s^{\underline{j}} = \hat e_1^{\underline{i}}\hat e_1^{\underline{j}}+\hat e_2^{\underline{i}}\hat e_2^{\underline{j}}\; ,
\end{equation}
which project onto the plane transverse to $\hat n_s^{\underline{i}}$. Incidentally, one could also define tetrad elements ``perpendicular'' to $\hat n_s^{\underline{i}}$, as $s_{\perp,\underline{i}}^\mu \equiv P_s^{ij}s_{\underline{j}}^\mu$. Likewise, the matrix
\begin{equation}
P^{ij} \equiv \gamma^{ij} - \tilde{n}^i \tilde{n}^j
\end{equation}
defines a projection operator at the source, such that $\tilde{x}_\perp^i \equiv P^i_j\tilde{x}^j$ is the component of $\tilde\vx$ in the plane transverse to $\ntvh$.

Pulling back $g_s$ onto the observer's past light cone $\Sigma_{\tau_o}$ and using $i^* \mathrm{d}x^\mu = \frac{\partial x^\mu}{\partial\tilde x^i}\mathrm{d}\tilde x^i$ leads to equation \eqref{eqn:gs2}, where 
\begin{equation}
\label{eqn:spullback}
\begin{aligned}
	i^* s^\parallel &= \hat n_{s\underline{i}} s_\mu^{\underline{i}} \frac{\partial x^\mu}{\partial \tilde x^k}\mathrm{d}\tilde x^k
	=\hat n_{s\underline{i}} s_\mu^{\underline{i}} \frac{\partial x^\mu}{\partial \tilde x^k}\left(\tilde n^k \mathrm{d}\tilde x^\parallel + \sum_I \tilde e_I^k \mathrm{d}\tilde x^I\right)\;,\\
	i^* s^I &= \hat e_{I\underline{i}}  s_\mu^{\underline{i}} \frac{\partial x^\mu}{\partial \tilde x^k}\mathrm{d}\tilde x^k
	=\hat e_{I\underline{i}} s_\mu^{\underline{i}} \frac{\partial x^\mu}{\partial \tilde x^k}\left(\tilde n^k \mathrm{d}\tilde x^\parallel + \sum_J \tilde e_J^k \mathrm{d}\tilde x^J\right) \;; 
\end{aligned}
\end{equation}
here, all the quantities in the right-hand side are now functions of the inferred co-ordinates $\tilde x^i$. 
These give the $1$-form cosmic rulers, because from the definition (\ref{eq:frak}), one obtains the relations
\begin{equation}
\label{eq:stetradproj}
\begin{aligned}
	\tilde a \big(1-\mathfrak{C}\big) &= \hat n_{s\underline{i}} s_\mu^{\underline{i}} \frac{\partial x^\mu}{\partial \tilde x^k}\tilde n^k  \\
	\tilde a \mathfrak{D}_I &= -\hat n_{s\underline{i}} s_\mu^{\underline{i}} \frac{\partial x^\mu}{\partial \tilde x^k}\tilde e_I^k  \\
	\tilde a \hat{\mathfrak{D}}_I &= -\hat e_{I\underline{i}} s_\mu^{\underline{i}} \frac{\partial x^\mu}{\partial \tilde x^k}\tilde n^k  \\
	\tilde a \big(\delta_{IJ}-\mathfrak{A}_{IJ}\big) &=\hat e_{I\underline{i}} s_\mu^{\underline{i}} \frac{\partial x^\mu}{\partial \tilde x^k}\tilde e_J^k \;,
\end{aligned}
\end{equation}
which are valid at any order in perturbations.
The 9 degrees of freedom encoded in the functions $\mathfrak{C}$, $\mathfrak{D}_I$, $\hat{\mathfrak{D}}_I$ and $\mathfrak{A}_{IJ}$ emerge from the asymmetric nature of the $3\times 3$ matrix $s_\mu^{\underline{i}}\frac{\partial x^\mu}{\partial\tilde x^k}$. The 3 degrees of freedom of the anti-symmetric component are the 3 Euler angles characterising the rotation between a set of rulers defined at the source and the set defined by the observer.\footnote{The observer could, in principle, reconstruct these angles from the measured position on the sky and knowledge of the perturbations along the line of sight.} In the Born approximation (valid at first order), this rotation reduces to the identity. 
We show that $\hat{\mathfrak{D}}^{(1)}_I =  \mathfrak{D}_I^{(1)}$ in this limit (at first order), in Appendix \ref{appendix:DI and hat DI}. 

We can now proceed to compute $\mathfrak{C}$ and $\mathfrak{M} \equiv \delta_I^J \mathfrak{A}^I_J$ at second order from the expressions given in \S\ref{sec:frames}. Our definition \eqref{eqn: total displacement definition} of the total displacements $\Delta x^\mu$ implies
\begin{equation}
\frac{\partial x^i}{\partial \tilde{x}^k} = \delta^i_k + \frac{\partial \Delta x^i}{\partial \tilde{x}^k}\;,
\end{equation}
and also
\begin{equation}
\frac{\partial x^0}{\partial \tilde{x}^k} = -\tilde{n}_k + \frac{\partial \Delta x^0}{\partial \tilde{x}^k}\; ,
\end{equation}
upon using the light-cone condition. To compute $\mathfrak{C}$, we need
\begin{equation}
g_{\mu\nu} \frac{\partial x^\mu}{\partial \tilde x^k}s^{\nu\underline{i}}\tilde{n}^k = g_{j\nu} \left(\tilde{n}^k\delta^j_k + \tilde{n}^k\frac{\partial \Delta x^j}{\partial \tilde{x}^k} \right)s^{\nu\underline{i}} + g_{0\nu} \left(-1 + \tilde{n}^k\frac{\partial \Delta x^0}{\partial \tilde{x}^k} \right)s^{\nu\underline{i}}\; ,
\end{equation}
which, on inserting the metric \eqref{eqn: metric Poisson gauge}, becomes
\begin{align}
g_{\mu\nu} \frac{\partial x^\mu}{\partial \tilde x^k}s^{\nu\underline{i}}\tilde{n}^k & = a^2\left[\omega_j s^{0\underline{i}} + \left(\delta_{jk} - 2\Phi \delta_{jk} + \frac{1}{2}h_{jk}\right)s^{k\underline{i}}\right] \left(\tilde{n}^j + \frac{\partial \Delta x^j}{\partial \tilde{\chi}} \right) \\ & \quad
+ a^2\Big[-\big(1+2\Psi\big)s^{0\underline{i}} + \Omega_K s^{k\underline{i}}\Big] \left(-1 + \frac{\partial \Delta x^0}{\partial \tilde{\chi}} \right) \nonumber \;.
\end{align}
Upon substituting the expressions for the tetrad $s^{\mu\underline{i}}$ (which are the same as in Appendix \ref{appendix:observer tetrad}, but evaluated at the source instead of the observer), writing $a = \tilde{a}\left(1+\Delta_{a,e}\right)$ and keeping terms only to second order, the contraction of the previous expression with $\hat n_{s\underline{i}}$ yields
\begin{align}
\label{eqn:C 2nd order} 1-\mathfrak{C}  = (1+\Delta_{a,e})&\left[ \nh_s^iv_i +  (\Psi - \Phi)v_\parallel + \frac{h_\parallel}{4} + \frac{v_\parallel^2}{2} - v_\parallel \frac{\partial \Delta \eta^{[1]}}{\partial \tilde{\chi}} \right. \\ &
\left. + \left(1-\Phi - \frac{\Phi^2}{2}\right)\left(\nh_s^i\tilde{n}_i + \nh_s^i\frac{\partial \Delta x_i}{\partial \tilde{\chi}}\right)\right] \nonumber\;.
\end{align}
Likewise, the same procedure applied to $\mathfrak{M}$ eventually gives 
\begin{align}
\label{eqn:M 2nd order} 2 - \mathfrak{M}  = (1+\Delta_{a,e})&\left[ P_s^{ij}P_{ij}\left(1-\Phi - \frac{\Phi^2}{2}\right) + \frac{P^{ij}h_{ji}}{4} + \frac{v^iP^j_{i}v_{j}}{2} \right. \\
& 
\left. - v_jP^{ij}\frac{\partial \Delta \eta^{[1]}}{\partial \tilde{x}_\perp^i} + \left(1-\Phi\right)P^i_{s,j}\frac{\partial \Delta x^j}{\partial \tilde{x}_\perp^i}\right] \nonumber\;.
\end{align}
The second-order expressions for $\mathcal{C}$ and $\mathcal{M}$ follow directly from substituting equations \eqref{eqn:C 2nd order} and \eqref{eqn:M 2nd order}, along with the first-order expressions for $\mathcal{B}_I$ given in refs.~\cite{SchmidtJeong2012,JeongSchmidt2015}, into the relations \eqref{eqn:relation between fraktur and calligraphic rulers} between the different sets of rulers.
The observant reader may notice that equations \eqref{eqn:C 2nd order} and \eqref{eqn:M 2nd order}, when truncated at first order, are not identical to, e.g., equation (51) of ref.~\cite{JeongSchmidt2015}. We explain why there is no disagreement in Appendix \ref{appendix:ns vs n}.

\section{Tests and Implementation}
\label{sec: implementation}

\subsection{Test Cases}


The expressions for $\mathcal{T}$, $\mathcal{M}$, $\mathcal{C}$ are quite lengthy and involved. Furthermore, there are subtleties in the calculation, such as the boundary conditions at source and observer. For this reason, we devise tests that we subject our results for $\mathcal{T}$, $\mathcal{M}$, $\mathcal{C}$ to.
There are two kinds of tests: \emph{null} tests and \emph{consistency} tests. Null tests employ a metric obtained by a co-ordinate transformation from an unperturbed background space-time; since the results have to be independent of co-ordinates, we expect all observables to yield null values (or their expected values in an unperturbed universe). In the second kind, a consistency test, we map a known solution, namely a curved FLRW space-time, to a non-trivial perturbation about a Euclidean FLRW space-time, and compare the results to the expected results for a curved universe, based on the definition of the ruler observables.
For each case, we calculate the cosmic rulers using the formul\ae\ derived above, and then compare them with the expected values. 
The calculations are summarised in \S \ref{appendix:test constant potential} and in Appendix \ref{appendix:tests}.


For reasons of convenience, we test the 1-form scalar rulers, $\mathfrak{C}$ and $\mathfrak{M}$, as their expressions are simpler.
For corresponding test results for the rulers $\mathcal{C}$ and $\mathcal{M}$ (as well as all others) at linear order, see ref.~\cite{SchmidtJeong2012}.

\subsubsection{Null test: constant potential}
\label{appendix:test constant potential}
This test considers constant, non-zero Bardeen potentials $\Psi = \Phi = \textrm{const}$ on top of an Einstein--de-Sitter (EdS) background space-time. It is straightforward to see by a co-ordinate transformation ($\vx \mapsto \vx\sqrt{1-2\Phi}$, $\eta \mapsto \eta\sqrt{1+2\Psi}$) that the full space-time is in fact an EdS space-time, so we expect that all cosmic ruler perturbations vanish at all orders in cosmological perturbations. We remark, that for consistency, the background must be EdS, for otherwise constant-in-time Bardeen potentials might not solve the Einstein equations. 

The primary effect of such a metric enters through gravitational-redshift terms in the geodesic equation, leading to corrections to the line-of-sight length in $\mathfrak{C}$ and $\mathfrak{M}$, and to gravitational redshift (or Sachs--Wolfe effects) in $\mathcal{T}$.
Additionally, the perturbations do not vanish at the observer’s position, so this test checks the cancellation of observer terms, which is a prerequisite for gauge invariance. This is the simplest of the three tests, and we include it in the main text for illustration; the other two tests, discussed below, are more involved and are described in Appendix \ref{appendix:tests}, in \S\S \ref{appendix:test gradient mode}--\ref{appendix:test separate universe}, respectively. 

Let us start with the cosmic clock $\mathcal{T}$. 
By equation \eqref{eqn: kernel for frequency shift}, $F^0_{[2]}(\chi) = 0$, whence $\delta\nu(\chi) = \delta\nu_o$, and likewise $\delta n^i(\chi) = \delta n^i_o$. Equation \eqref{eqn:delta nu at observer}, gives
\begin{equation}\label{eqn:delta nu observer constant potential test case}
\delta \nu_o = \Psi - \delta a_o(1-\Psi) - \frac{3}{2}\Psi^2 \;,
\end{equation}
whence equation \eqref{eqn: delta ln a} yields
\begin{equation}
\Delta_{a,e} = -\frac{\Psi^2}{2} + \Psi -\delta\nu_o(1+\Psi)\;,
\end{equation}
while
\begin{equation}
\Delta_{a,\tau} = Ht\left(\Psi - \frac{\Psi^2}{2}\right) + \frac{\ddot{a}(t)}{2a(t)}t^2\left(\Psi - \frac{\Psi^2}{2}\right)^2 \;.
\end{equation}
In EdS, we have $H(t) = 2/(3t)$, and $\ddot{a}/a = -2/(9t^2)$, so that
\begin{equation}
\Delta_{a,\tau} = \frac{2}{3}\left(\Psi - \frac{\Psi^2}{2}\right) - \frac{1}{9}\left(\Psi - \frac{\Psi^2}{2}\right)^2 = \frac{2}{3}\Psi - \left(\frac{2}{3}\Psi\right)^2 \;,
\end{equation}
and $\delta a_o = -\frac{2}{3}\Psi + \frac{8}{9}\Psi^2$, where we have used $\barH = H(1-\Psi)$ \cite{Dai:2015jaa}. Hence, equation \eqref{eqn:delta nu observer constant potential test case} becomes
\begin{equation}
\delta \nu_o = \frac{5}{3}\Psi - \frac{55}{18}\Psi^2\;.
\end{equation}
The other component of $\mathcal{T}$ is $\Delta_{a,e}$, which, in this special case, reduces to
\begin{equation}
\Delta_{a,e} = \Psi - \frac{\Psi^2}{2} - \delta \nu_o(1+\Psi) = - \frac{2}{3}\Psi + \frac{8}{9}\Psi^2
\end{equation}
to second order in the perturbations. Therefore, we find
\begin{equation}
\exp \mathcal{T} = (1+\Delta_{a,\tau})(1+\Delta_{a,e}) = 1 + \frac{4}{9}\Psi^2 + \left(-\frac{2}{3}\Psi + \frac{8}{9}\Psi^2\right) \left(\frac{2}{3}\Psi - \frac{4}{9}\Psi^2\right) = 1
\end{equation}
whence $\mathcal{T} = 0$ to second order; that is, the cosmic clock is unaffected by a constant potential, as it should be.

Let us now move on to the cosmic rulers.
The change in $k^i$ is proportional to $\tilde{n}^i$ since there is no other direction in the constant potential case. As a result, $n_s^i = \tilde{n}^i$ and we find 
\begin{equation}
\delta n^i = \delta n^i_o = \left(\frac{31 \Psi ^2}{18}+\frac{\Psi }{3}\right)\tilde{n}^i\; ,
\end{equation}
and
\begin{equation}
\frac{\partial \delta \chi}{\partial \tilde{\chi}} = \frac{4 \Psi }{3}-\frac{4 \Psi ^2}{9} \;.
\end{equation}
This gives $\Delta x_\parallel = \tilde{\chi}\left(\frac{31 \Psi ^2}{18}+\frac{5 \Psi }{3}\right)$, whence
\begin{equation}
\mathfrak{C} = 1- (1+\Delta_{a,e})\sqrt{1-2\Psi}\left(1+\frac{\partial \Delta x_\parallel}{\partial \tilde{\chi}}\right) = O(\Psi^3) \;.
\end{equation}
Furthermore, $P_s^{ij}\frac{\partial \Delta x_i}{\partial x_\perp^j} = \frac{2}{\tilde{\chi}}\Delta x_\parallel$ because the rest of the terms vanish by symmetry, whence
\begin{equation}
\mathfrak{M} = 2 - (1+\Delta_{a,e})\sqrt{1-2\Psi}\left(2+\frac{2}{\tilde{\chi}}\Delta x_\parallel\right) = O(\Psi^3) \;.
\end{equation}
All of this demonstrates that the cosmic clock and 1-form rulers are invariant under a constant potential shift, up to including second order. 

\subsubsection{Null test: pure gradient mode at second order}

We now construct a specific second-order gauge-mode which corresponds to a generalisation of what in linear theory would be a pure gradient mode: $\Phi, \Psi \propto \vk \cdot \vx$, for some constant vector $\vk$. We obtain the perturbed space-time  by performing a general second-order large gauge-transformation on an unperturbed FLRW background, which is at most quadratic in the space-time position (see Appendix \ref{appendix:test gradient mode} for details). The components of the resultant metric are:
\begin{align}
\Psi & = \psi_0 \vk \cdot \vx + \left(\frac{5}{6} + 3a_1\right)\psi_0^2(\vk \cdot \vx)^2 +3a_2 \psi_0^2k^2x^2 + \left(5a_3 - \frac{1}{18}\right)\psi_0^2k^2\eta^2 \nonumber \\
\Phi & = \psi_0 \vk \cdot \vx - \left(2a_1 -\frac{77}{54}+\frac{5}{3}b_1 + \frac{2}{3}b_4
\right)\psi_0^2(\vk \cdot \vx)^2 - \left(2a_2 +\frac{25}{27} + \frac{5}{3}b_3 + \frac{1}{3}b_4\right)\psi_0^2 k^2 x^2 \nonumber \\ & \quad 
-\left(2a_3-\frac{1}{54} + b_2 + \frac{1}{3}b_5\right)\psi_0^2k^2\eta^2 \label{eqn:metric pure gradient mode} \\
\frac{\omega_i}{\psi_0^2} & = \left(-\frac{11}{9} -2a_1 +2b_5\right)(\vk \cdot \vx)\eta k_i + \left(\frac{5}{9} - 2a_2 +2b_2\right)k^2\eta x_i \nonumber \\
\frac{h^{ij}}{\psi_0^2} & = \left(-\frac{2}{9} + 4b_5\right)\left(k^ik^j - \frac{\delta^{ij}}{3}k^2\right)\eta^2 + \left( \frac{50}{9} + 4b_4\right)\left(k^ik^j - \frac{\delta^{ij}}{3}k^2\right)x^2 \nonumber \\ & 
+\left( \frac{50}{9} + 8b_3\right)\left(x^ix^j - \frac{\delta^{ij}}{3}x^2\right)k^2  
+ \vk \cdot \vx\left(4b_1+4b_4-\frac{50}{9} \right) \left[k^ix^j + x^ik^j - \frac{2\delta^{ij}}{3}\left(\vk\cdot\vx\right) \right]. \nonumber
\end{align}
The coefficients $a_{1,2,3}$ and $b_{1,\ldots,5}$ are parameters, which we keep free;\footnote{They could be constrained by requiring that the metric be in Poisson gauge, or that $h_{ij}$ vanish at $\eta = 0$. We will not require these restrictions here, as $h_{ij}$ and $\omega^i$ are considered second-order anyway.} the three components of $\vk$ are also kept free. This metric describes an FLRW space-time for \emph{any} values of the parameters $a_i$, $b_i$, whence we expect all cosmic rulers to vanish to second order in $\psi_0$.\footnote{This metric differs from the one described in \cite[][equations (5.16--5.19)]{MirbabayiZaldarriaga2015}; there, e.g.~the Kretschmann and Ricci scalars do not assume their FLRW values for every value of the parameters.}

We perform this test in Appendix \ref{appendix:test gradient mode}. 
Given the complexity of the metric, the fact that our ruler perturbations indeed yield zero on this metric constitutes a non-trivial confirmation of the expressions for $\mathcal{T}$, $\mathcal{C}$ and $\mathcal{M}$. 
Among other things, it checks velocity-dependent contributions to the cosmic rulers and corrections to the direction of light propagation due to the special spatial direction---namely $\vk$---inherent in the metric. It also tests the intricate cancellations between derivative terms at linear order in $\vk$, and, because the metric is time-dependent at second order, effects such as the integrated Sachs–Wolfe effect.

\subsubsection{Constant-curvature (separate universe) test case}

While the previous two test cases were null tests, that dealt with a flat FLRW metric expressed in non-standard co-ordinate systems, here we consider a curved FLRW space-time, on scales smaller than its radius of curvature, which is an example of a consistency test.
This case is equivalent to an isotropic long mode via the separate-universe picture, where the local over-density mimics a non-zero spatial curvature and induces a slightly different local Hubble parameter \cite{Dai:2015jaa}. Unlike the other two test cases, we do not expect $\mathcal{T}$, $\mathcal{C}$ or $\mathcal{M}$ to vanish in this case. Therefore, we first compute them from their definitions, and then compute them again using our second-order expressions, in order to compare the two results. 

Under a stereographic projection, a 3-sphere of curvature $K$ (or radius of curvature $1/\sqrt{K}$) can be described by the 3-metric
\begin{equation}
\mathrm{d}s_3^2 
= \frac{\delta_{ij}\mathrm{d}x^i\mathrm{d}x^j}{\left(1+\frac14K|{\bf x}|\right)^2}
\simeq 
\left(
1 - \frac12 K|{\bf x}|^2 + \frac3{16} K^2|{\bf x}|^4
\right) \delta_{ij}\mathrm{d}x^i\mathrm{d}x^j
\end{equation}
where $|\vx|^2 = \delta_{ij} x^i x^j$ and, in the second equality, the metric has been expanded to second order in $K$ in order to identify
\be
\Phi \supseteq \frac14 K|{\bf x}|^2 - \frac{3}{32} K^2|{\bf x}|^4 
= \frac14 K \chi^2 - \frac{3}{32} K^2\chi^4 \;.
\label{eq:Phi_const_curbaure}
\ee
While the co-moving distance-to-redshift relation $\chi(z)$ remains the same, the presence of non-zero spatial curvature alter the co-moving angular diameter distance according to
\be
d_A(z) = \sin_K\chi(z)
=
\left\{
\begin{array}{ll}
\chi(z) & K=0 \\
K^{-1/2} \sin\big[K^{1/2}\chi(z)\big] & K>0 \\
(-K)^{-1/2} \sinh\big[(-K)^{1/2}\chi(z)\big] & K<0 \\
\end{array}
\right. \;.
\ee
The full metric is 
\begin{equation}\label{eqn: curved space metric with cosmic time}
\mathrm{d}s^2 = -\mathrm{d}t^2 + a_K^2 \frac{\delta_{ij}\mathrm{d}x^i\mathrm{d}x^j}{\left(1+\frac14K|{\bf x}|\right)^2} \;.
\end{equation}
where the dependence of the scale factor $a_K(t)$ on $K$ arises through the Friedmann equations.

In this paper, the fiducial background adopted by the observer is always a flat FLRW metric. Therefore, a curved-space FLRW space-time differs from the background, leading to non-zero deviations at both first and second orders.
The reasons are twofold: first of all, the time-dependence of $a$ is not the same as in the flat case, so when we convert equation $\eqref{eqn: curved space metric with cosmic time}$ from cosmic time to conformal time via $\mathrm{d}t=a_K\mathrm{d}\eta$, this will create a non-zero $\Psi$; secondly, the spatial slice is itself curved, manifesting in a non-zero $\Phi$. 
Another option for defining the conformal time is to \emph{define} conformal time with the background scale factor, denoted $a_0$, i.e.~by setting $\mathrm{d}t=a_0\mathrm{d}\eta$. In this case, one obtains $\Psi = 0$. 
We consider both options in this paper, when performing this test in Appendix \ref{appendix:test separate universe}, as they represent algebraically distinct tests of our expressions for the cosmic rulers. 

As remarked above, also need to compute the expected ruler observables directly in a curved background. As an example, in the afore-mentioned case of $\mathrm{d}t=a_0\mathrm{d}\eta$, we have
\begin{equation}
\exp \mathcal{T} = \frac{a_0(t)a_K(\tau_o)}{a_K(t)}. 
\end{equation}
(Recall that the background scale-factor is normalised to unity at $\tau_o$, i.e.~$a_0(\tau_o) = 1$, so $a_K(\tau_o) \neq 1$ in general.) $\mathfrak{C}$ and $\mathfrak{M}$ are calculated in Appendix \ref{appnedix:test case curvature confomal time} (equations \eqref{eqn:constant curvature C expected} and \eqref{eqn:constant curvature M expected}). 
This separate-universe test probes both the space-dependent terms involving derivatives in the formul\ae\ for the cosmic rulers and clock, as well as the time-dependent terms arising from the difference between the functions $a_0$ and $a_K$.  

\subsection{Numerical results}

Since the computation of the rulers is a challenging numerical task for a viable $\Lambda$CDM cosmology, we shall plot, as an example, the cosmic rulers for the simplest possible space-time: an EdS background where $\Phi = \Psi = \varphi_0 \Re\exp(\mathrm{i}\vk\cdot \vx)$, with the remaining metric components set to zero, and the velocity field given by the first-order relations in FLRW. For simplicity, our calculation will assume that the observer knows what the actual background is, although the observer could also use a background defined by local measurements \cite{Ginatetal2021}.

\begin{figure}
\centering
\includegraphics[width=0.49\textwidth]{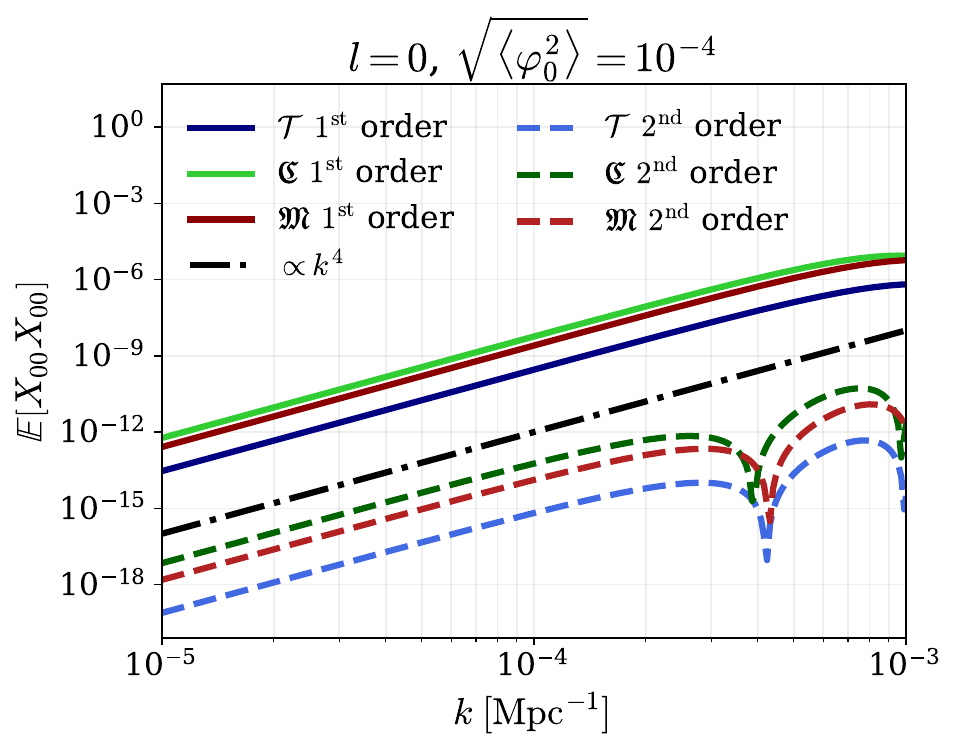}
\includegraphics[width=0.49\textwidth]{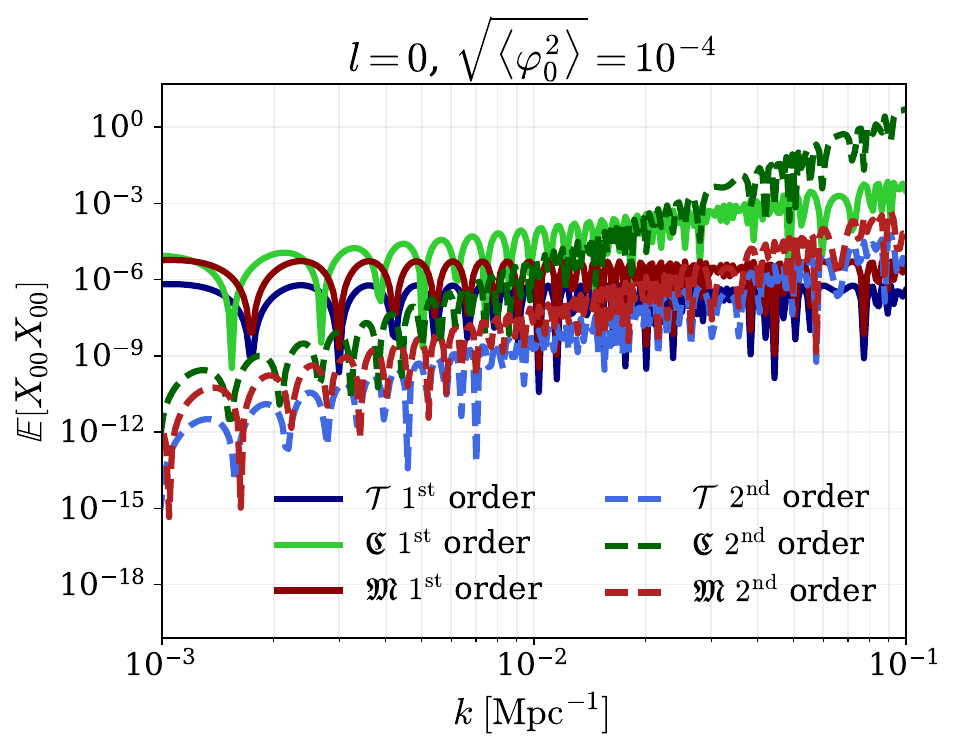}
\caption{Plots of $\mathbb{E}(X_{l0}X_{l0})$, for $X \in\set{\mathcal{T},\mathfrak{C},\mathfrak{M}}$, for potentials described by a pure sinusoidal mode. Here, $l=0$, $\tilde{z} = 1$ and $\sigma(k) = 10^{-4}$. The \emph{left} panel shows low $k$ and the \emph{right} panel shows larger values of $k$, where the second-order correction starts to dominate over the first order. Dashed lines show the pure first-order contribution in the top row, and the pure second-order part is plotted as full lines.}
\label{fig:ps_exponential l=0}
\end{figure}

\begin{figure}
\centering
\includegraphics[width=0.49\textwidth]{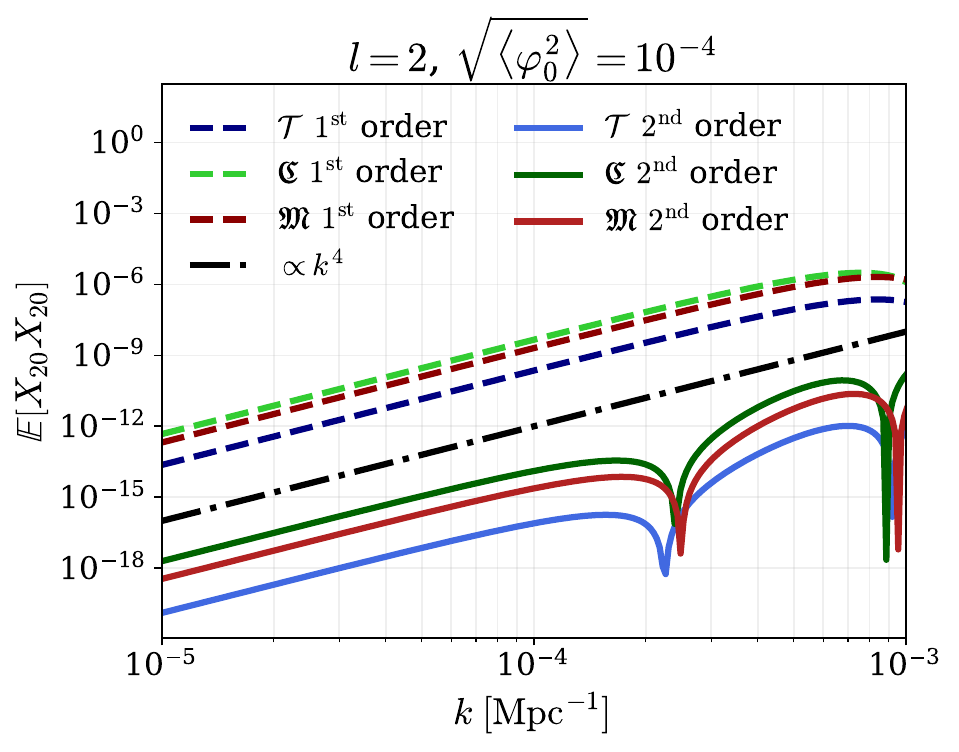}
\includegraphics[width=0.49\textwidth]{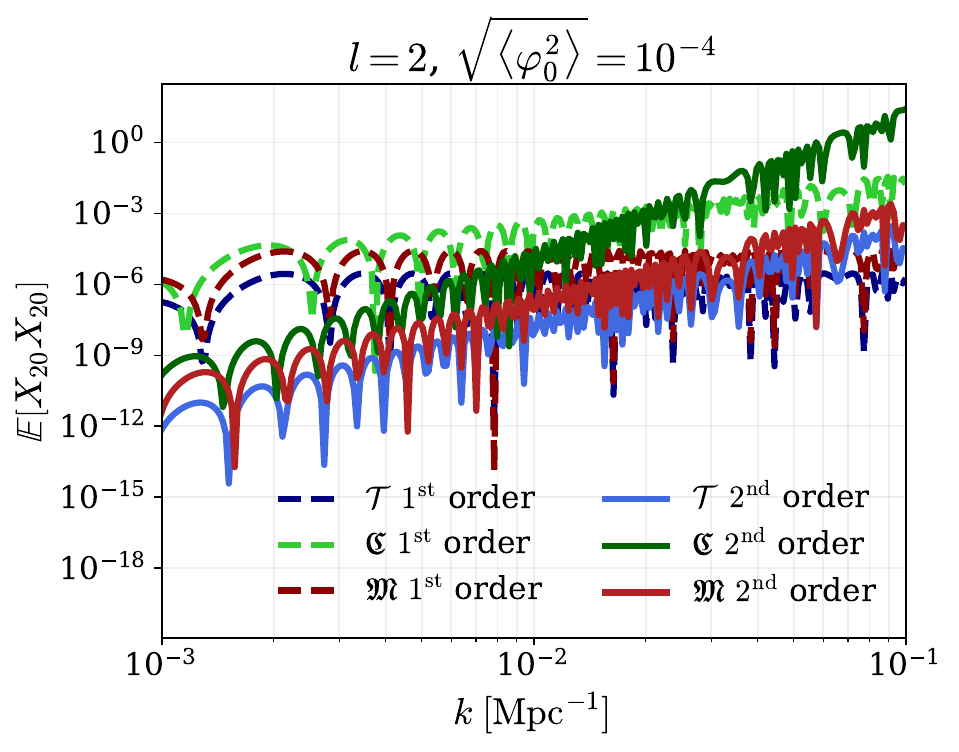}
\caption{Plots of $\mathbb{E}(X_{l0}X_{l0})$, for $X \in\set{\mathcal{T},\mathfrak{C},\mathfrak{M}}$, for potentials described by a pure sinusoidal mode. Here, $l=2$, $\tilde{z} = 1$ and $\sigma(k) = 10^{-4}$. The \emph{left} panel shows low $k$ and the \emph{right} panel shows larger values of $k$, where the second-order correction starts to dominate over the first order. Dashed lines show the pure first-order contribution in the top row, and the pure second-order part is plotted as full lines.}
\label{fig:ps_exponential l=2}
\end{figure}

We compute the cosmic clock $\mathcal{T}$ and $1$-form rulers $\mathfrak{C}$ and $\mathfrak{M}$ as functions of the inferred co-ordinates from equations \eqref{eqn:cosmic clock second order explicit}, \eqref{eqn:C 2nd order} and \eqref{eqn:M 2nd order}---using the \verb"Mathematica" file provided. 
We then decompose each ruler, collectively denoted by $X$, in spherical harmonics, \emph{viz.}
\begin{equation}
X(z, \ntvh) \equiv \sum_{l=0}^{\infty} \sum_{m=-l}^{l} X_{lm}(z) Y_{lm}(\ntvh) \;.
\end{equation}
As a proxy for the correlation function 
\begin{equation}
\langle X_{lm}(z) X_{l'm'}(z) \rangle = \delta_{ll'} \delta_{mm'} C_{l}(z) 
\end{equation}
we calculate the average $\mathbb{E}(X_{l0}X_{l0})$ over $\varphi_0$, of $X_{l0}X_{l0}$, for a normally-distributed amplitude $\varphi_0$, with standard deviation $\langle\varphi_0^2\rangle^{1/2} =10^{-4}$; this amplitude is motivated by the typical amplitude of potential perturbations in the Universe.
The results are shown in figures \ref{fig:ps_exponential l=0}--\ref{fig:ps_exponential l=2} for the monopole ($\ell=0$) and the quadrupole ($\ell=2$), respectively.
For the particular choice made here, the $l=1$, dipole, term vanishes because both $\Phi$ and $\Psi$ are even functions of $\vx$. Similarly, if $\vk$ is aligned with the $\zhat$-axis, there is no dependence on the azimuthal number and the $m\neq 0$ terms vanish as well. Therefore, the only non-vanishing multipole moments are those with $m=0$. 

At low $k$, $\mathbb{E}(X_{l0}X_{l0}) \propto k^4$, as expected from the arguments emphasised in this paper---the constant and linear terms (in $\vk$) do not contribute to the cosmic rulers, as demonstrated by the two corresponding test cases.
At the level of the large-scale structure observables, this behaviour ensures that the monopole of these observables does not diverge in the limit $k\to 0$ \cite[see][]{Ginatetal2021,CastorinaDiDio2021,Desjacques:2020zue,Foglieni:2023xca}.
The second-order contributions become comparable to the first-order ones for $k\gtrsim 0.1\ {\rm Mpc}^{-1}$ in all three rulers. For the specific background and perturbations considered here, these are largest for $\mathfrak{C}$. Adding up contributions from other modes is likely to reduce the amplitude of the oscillations.

We remark that $\mathbb{E}(X_{l0}X_{l0})$ is not sufficient to compute $C_l$, because the full expression for $C_l$ (for cosmological perturbations which are Gaussian random fields) would couple two wave-vectors, with a kernel prescribed by the full expressions for the cosmic rulers. Instead, $\mathbb{E}(X_{l0}X_{l0})$ simply selects the single-wave-vector contributions, and should be viewed as a qualitative proxy for the amplitude of the ruler observable at scale $\vk$. 


\subsection{Mathematica notebook}
\label{subsec:mathematica notebook}
Additionally, we provide with this paper a \verb"Mathematica" notebook that evaluates the expressions for the rulers and the clock, with three special cases corresponding to the three test cases.\footnote{Link here: \url{https://github.com/byg3/GR_effects}, presently containing the test cases; the general notebook will be released after publication.}

The file requires the metric components and the velocity field as inputs, as functions of $(\eta,\vx)$, and the background scale factor $\at$, which should be normalised to unity at the observer’s position (to all orders in cosmological perturbation theory), i.e.~$\at(\tau_o) = 1$. 
At present, the integrations are performed analytically, and while they may not be tractable for a realistic metric, they still yield integral expressions for the rulers. The test cases described in this section are included in the notebook and can be computed directly there.

To use the notebook, input the velocity field, the perturbations of the metric, and the background function in the section titled \verb"Input perturbations, velocity field and" \verb"background function". To do so, it is necessary to uncomment the first part of the section. Evaluating the entire notebook will go through the calculations in this paper, and the final results for $\mathcal{T}$, $\mathfrak{C}$ and $\mathfrak{M}$ should appear at the end.
To run the three test cases, simply uncomment the corresponding part of the \verb"Input..." section and then run the notebook, rather than inserting a new metric. 
In the \verb"Mathematica" notebook, we systematically write functions of $\chi$ and $\tilde{n}^i$ when calculating $\mathfrak{M}$ and $\mathfrak{C}$. Derivatives with respect to $\tilde{x}^i$ can be evaluated using the identity
\begin{equation}
\frac{\partial}{\partial\tilde{x}^i} = \tilde{n}_i \frac{\partial}{\partial \tilde{\chi}} + \frac{1}{\tilde{\chi}} P^k_i \frac{\partial}{\partial \tilde{n}^k} \;,
\end{equation}
which simplifies to
\begin{equation}
\frac{\partial}{\partial \tilde{x}_\perp^i} = P^{ij}\frac{1}{\tilde{\chi}}\frac{\partial}{\partial \tilde{n}^j}
\end{equation}
when the derivative is taken with respect to $\tilde x^i_\perp$.

\section{Conclusions}
\label{sec: discussion}



\subsection{Summary}

In this paper, we have outlined a calculation of the scalar cosmic rulers at second order in cosmological perturbation theory. This calculation was performed by solving the geodesic equation, and re-expressing the solution in terms of the observer's inferred co-ordinate system, $\tilde{x}^\mu$, i.e.~in terms of $\tilde{z}$ and $\ntvh$. Then, we tested our expressions in three non-trivial configurations, where the results were known beforehand based on physical principles, such as the equivalence principle. 

The first two tests we performed were null tests: an FLRW metric (in fact, an EdS metric) in non-standard co-ordinates. Being observable, and hence independent of the choice of co-ordinate system, the cosmic rulers must vanish in these test cases. 
We confirmed that indeed, both for the constant potential test case, and for the metric \eqref{eqn:metric pure gradient mode}, all of the terms in equations \eqref{eqn:cosmic clock second order explicit}, (\ref{eqn:C 2nd order}--\ref{eqn:M 2nd order}) are non-vanishing but cancel among themselves to give the desired result. 

The third test we performed was an FLRW space-time with non-zero $\Omega_{K}$ and $\Omega_m$. This, again, is an FLRW metric, and therefore the required values for $\mathcal{T}$, $\mathfrak{C}$ and $\mathfrak{M}$ are known in principle. We confirmed that, when these are expanded to second order in $\Omega_K$, and then expressed in terms of the inferred co-ordinates of an observer using a \emph{flat} FLRW space-time as a background, they match the perturbative expressions in equations \eqref{eqn:cosmic clock second order explicit}, (\ref{eqn:C 2nd order}--\ref{eqn:M 2nd order}). The expressions in Appendix \ref{appendix:test separate universe} provide an illustration for how complicated these become at second order, even for such a simple space-time. 
Passing all of these checks lends credence to our expressions for the scalar cosmic rulers. 

Furthermore, we have estimated the importance of second-order contributions for a single plane wave perturbation. While a thorough investigation of relativistic second-order contributions in a realistic cosmological model---which would be relevant, for example, to galaxy-bispectrum constraints on inflation \cite[see][for recent results]{Cabass:2022wjy,Cabass:2022ymb,DAmico:2022gki,Cagliari:2025rqe}---is beyond the scope of this paper, we expect the order of magnitude to be preserved: second-order contributions become comparable to first-order ones on mildly non-linear scales.

Finally, following equation \eqref{eqn:dgobs}, equations \eqref{eqn:cosmic clock second order explicit}, (\ref{eqn:C 2nd order}--\ref{eqn:M 2nd order}) constitute a key, and the most complex, component of the second-order expression for the observed relativistic galaxy over-density $\dgobs$. 
The remaining part, computing the galaxy density contrast $\delta_{\rm g}^{\rm or}$ in the constant-observed-redshift gauge, can likewise be validated using the test cases presented here, at it is separately observable. Moreover, it can be further broken down into the bias expansion in the galaxy rest-frame, and the boost from rest-frame to constant-observed redshift frame. We leave these developments to future work.


\subsection{Relation to other works}


A comprehensive comparison of our results with second-order GR calculations available in the literature is a challenging task, because the latter mainly focus on the observed number counts $\dgobs$
and not on the individual ruler components. We hope that our work will motivate direct calculations of the cosmic rulers themselves, as these should allow for a simpler comparison between different calculations.
In what follows, we briefly review a few related studies and contrast them to our work.

The authors of refs.~\cite{YooZaldarriaga2014,Bertacca:2014dra,Bertaccaetal2014II,DiDio:2014lka} computed $\dgobs$ to second order for a $\Lambda$CDM model including all general-relativistic effects (this was extended by \cite{Bertaccaetal2015III} to flux-limited surveys) albeit without explicit validation on test cases. 
By contrast, ref.~\cite{MagiYoo2022} explicitly validated their second-order expressions by comparing their gauge-transformation properties using two independent methods: a first check based on their expressions in terms of metric perturbations, and a second based on their non-linear relations.
Our null tests can be seen as special cases of (large) gauge transformations as considered by ref.~\cite{MagiYoo2022}. In addition, we propose tests with non-vanishing results such as the separate-universe test case.

Ref.~\cite{Nielsen:2016ldx} focused on the dominant terms proportional to $(\partial_i\partial_j \Phi)^2/(aH)^4$ and presented a formalism to compute leading contributions at any perturbative order (including explicit third-order expressions). 
In principle, this allows for the computation of one-loop corrections to the correlation function and power spectrum of observed number counts. The dominant contributions stem from density fluctuations, redshift-space distortions (Newtonian terms), and gravitational lensing. 
The other, sub-dominant terms, require a careful treatment, which could be carried out with the formalism introduced in this work.

Finally, a self-consistent prediction of galaxy number counts should also include relativistic corrections to the galaxy perturbative bias expansion at second order. However, these do not yield new bias parameters, but instead modify the existing bias parameters through GR corrections to the matter density evolution~\cite{Umehetal2019}.\footnote{See however \cite{Yoo2023} for possible complications for relativistic galaxy bias models.}



\subsection{Future work}


There are several pathways for extending the results of this paper. The first is to calculate the other, non-scalar, cosmic ruler observables at second order, namely $\mathcal{B}_I$ and the rest of $\mathcal{A}_{IJ}$. These are relevant, for example, for lensing. Another direction is to use the result for the second-order galaxy density to derive relativistic predictions for galaxy bispectrum estimators on the full sky 
\cite{Montandonetal2025}; these results are relevant for primordial non-Gaussianity measurements, e.g.~using SPHEREx \cite{Dore:2014cca}. 
Other possible generalisations of the treatment of cosmic rulers presented here include accounting for a biased fiducial cosmology via the Alcock--Paczynski effect \cite{Alcock:1979mp}, and for magnification bias (for e.g.~flux-limited surveys of emission line galaxies). The latter is immediate, since we have already computed $\mathcal{M}$ at second order here.

Furthermore, given the results in figures \ref{fig:ps_exponential l=0}--\ref{fig:ps_exponential l=2}, it would be interesting to compare the relativistic second-order contributions with their Newtonian counterparts as functions of scale, that is, by keeping only the leading $(k/aH)^4$ and sub-leading $(k/aH)^3$ contributions. We plan to perform such a comparison in a future work.




\acknowledgments
A.V.'s work was conducted as a research internship at Oxford with Y.B.G.~as part of an Oxford--\'{E}cole-Polytechnique partnership. F.S.~thanks
Julian Adamek,
Pierre Bechaz
Ruth Durrer,
Francesca Lepori,
Roy Maartens,
and Matteo Magi for helpful discussions during the 2025 MIAPbP focus workshop on GR corrections. Y.B.G.~is grateful for the kind hospitality of the Max Planck Institute for Astrophysics, where some early work on this research was done. D.J.~and F.S.~thank the hospitality of the Technion during a workshop organised by V.D.~in December 2022. This work was supported by a Leverhulme Trust International Professorship Grant (No. LIP-2020-014). The work of A.V.~and Y.B.G.~was partly supported by a Simons Investigator Award to A.A.~Schekochihin. V.D.~acknowledges support by the Israel Science Foundation (grants no. 2562/20). 
D.J.~was supported by NSF grants (AST-2307026, AST-2407298) at Pennsylvania State University and by KIAS Individual Grant PG088301 at Korea Institute for Advanced Study.

\appendix

\section{Christoffel Symbols}
\label{appendix:Christoffel symbols}
The non-zero Christoffel symbols of the metric \eqref{eqn: conformal metric Poisson gauge} are
\begin{align}
	\Gamma_{00}^0 & = \Psi'(1-2\Psi)\; , \\ 
	\Gamma_{0i}^0 & = \partial_i\Psi(1-2\Psi)\; , \\ 
	\Gamma_{ij}^0 & = \frac{1}{4}h'_{ij} -\frac{\partial_i\omega_j + \partial_j\omega_i}{2}- \Phi'(1-2\Psi)\delta_{ij}\;,  \\ 
	\Gamma_{00}^i & = \partial_i\Psi(1+2\Phi) + \omega'_i \;, \\ 
	\Gamma_{0j}^i & = \frac{\partial_j\omega_i - \partial_i \omega_j}{2} + \frac{1}{4}h'^i_j - \Phi'(1+2\Phi)\delta^i_j \;, \\ 
	\Gamma_{jk}^i & = \left(1+2\Phi\right)\left(\partial_i\Phi\delta_{jk}-\partial_j\Phi\delta^i_k - \partial_k\Phi\delta^i_j\right) + \frac{1}{4}\left(\partial_k h_{ij} + \partial_j h_{ki} - \partial_i h_{jk}\right)\;,
\end{align}
where, as usual, a prime denotes a derivative with respect to conformal time.

\section{Observer's Orthonormal Tetrad}
\label{appendix:observer tetrad}
As in \cite{SchmidtJeong2012} we define the observer's orthonormal tetrad $e_{\underline{a}}^\mu$ by 
\begin{equation}
	g_{\mu\nu}e^\mu_{\underline{a}}e^\nu_{\underline{b}} = \eta_{\underline{a}\underline{b}} \;,
\end{equation}
with
\begin{equation}
	e^\mu_{\underline{0}} = u_o^\mu = \frac{1}{a}(1-\Psi+\delta u, v^i)\;,
\end{equation}
the observer's 4-velocity, and $\delta u$ is defined in equation \eqref{eqn: source 4 velocity}.

The requirement $0 = g_{\mu\nu}e_{\underline{0}}^\mu e_{\underline{i}}^\nu$, implies that
\begin{equation}
	e_{\underline{i}}^0 = \frac{(1-\Phi)v_i + \omega_i}{a(1+\Psi + \delta u -2\Psi^2)} \;.
\end{equation}
The other components of $e_{\underline{i}}^\mu$ are found from the equations $\delta_{\underline{ij}} = g_{\mu\nu}e^\mu_{\underline{i}}e^\nu_{\underline{j}}$, which yield
\begin{equation}
	a^{-2}\delta_{\underline{ij}} = -v_iv_j+(1-2\Phi)\delta_{kl}e_{\underline{i}}^k e_{\underline{j}}^l + \frac{1}{2}h_{ij} \;.
\end{equation}
This equation is solved by
\begin{equation}
	e_{\underline{i}}^k = \frac{1}{a\sqrt{1-2\Phi}}\left(\delta_i^k - \frac{1}{4}h_i^k + \frac{1}{2}v_iv^k\right) \;.
\end{equation}

\section{Another derivation of $\cal T$}
\label{appendix:T}

We can also calculate equation (\ref{eq:Truler}) to second order in the perturbations as follows. Expanding $a(\tau)$ around $\bar t(\tilde z)$, we find
\begin{equation}
	\mathcal{T} = \frac{\tilde\cH}{\tilde a}\big(\tau-\bar t(\tilde z)\big) + \frac{1}{2\tilde a^2}\left(\tilde{\cH}' - \tilde\cH^2\right)\big(\tau-\bar t(\tilde z)\big)^2 + \dots
\end{equation}
where $\tau-\bar t(\tilde z)$ is the perturbation to the source proper time on slices of constant observed redshift. Using the background time-redshift relation
\begin{equation}
	\bar t(\tilde z) = \int_0^{\tilde\eta}\! a(\eta')\,\mathrm{d}\eta' =
	\int_{\tilde z}^\infty\!\!\frac{\mathrm{d} z'}{(1+z') H(z')} \;,
\end{equation}
this becomes
\begin{align}
	\tau(\tilde z,\vx) - \bar t(\tilde z) &= \tilde a\big(1+\Psi\big) \Delta \eta + \frac{1}{2}\tilde a \tilde\cH \Delta \eta^2+\int_0^{\tilde\eta}\! a(\eta') \Psi\,\mathrm{d}\eta' - \frac{1}{2}\int_0^{\tilde\eta}\!\big(\Psi^2+v^2\big)\,\mathrm{d}\eta' \;.
\end{align}
The perturbation $\Delta \eta$ to the time co-ordinate follows from expanding $a_e=a(\tilde\eta+\Delta \eta)$ in equation (\ref{eqn: delta ln a}) at second order in $\Delta \eta$,
\begin{equation}
	a_e \approx \tilde a \left[1 + \cH \Delta \eta + \frac{1}{2}\Big(\tilde \cH'+\tilde\cH^2\Big)\Delta \eta^2\right] \;.
\end{equation}
Solving for $\Delta \eta$ yields
\begin{equation}
	\begin{aligned}
		\tilde\cH\Delta \eta &= \delta u + \Psi - 2\Psi^2 -\delta \nu(1+\Psi) + v_\parallel(1-2\Phi) + v_i\delta n^i + \omega_i\tilde{n}^i \\
		& \quad - \frac{1}{2} \left(1+\frac{\tilde\cH'}{\tilde\cH^2}\right)
		\Big(\Psi - \delta\nu + v_\parallel\Big)^2
	\end{aligned}
\end{equation}
at second order in the perturbations. Therefore,
\begin{equation}\label{eqn:cosmic clock conformal Hubble}
	\begin{aligned}
		\mathcal T &= \bigg. \delta u + \Psi - 2\Psi^{(1)2} -\delta \nu(1+\Psi^{(1)}) + v_\parallel(1-2\Phi^{(1)}) + v_i\delta n^i + \omega_i\tilde{n}^i
		+ \Psi^{(1)} \tilde \cH \Delta \eta^{(1)}\\
		& \quad
		-\frac{1}{2}\Big(\Psi^{(1)}-\delta\nu^{(1)}+v_\parallel^{(1)}\Big)^2
		+ \frac{\tilde\cH}{\tilde a}\int_0^{\tilde\eta}\! a(\eta')\,\Psi\,\mathrm{d}\eta'
		-\frac{\tilde\cH}{2\tilde a}\int_0^{\tilde\eta}\! a(\eta') \Big(\Psi^{(1) \,2}+v^{(1)\, 2}\Big)\,\mathrm{d}\eta' \\
		&\quad +\frac{1}{\tilde a}\Big(\tilde\cH' -\tilde\cH^2\Big)\Delta \eta^{(1)}
		\int_0^{\tilde\eta}\! a(\eta') \Psi^{(1)}\,\mathrm{d}\eta'
		+ \frac{1}{2\tilde a^2}\Big(\tilde\cH' -\tilde\cH^2\Big)
		\left(\int_0^{\tilde\eta}\! a(\eta') \Psi^{(1)}\,\mathrm{d}\eta'\right)^2 \;,
	\end{aligned}
\end{equation}
where it is understood that $\Psi=\Psi(\vx,\eta) = \Psi(\tilde\vx,\tilde\eta)+\partial_\mu\Psi\, \Delta x^\mu$ and $v_\parallel=v_\parallel(\vx,\eta) = v_\parallel(\tilde\vx,\tilde\eta) + \partial_\mu v_\parallel \, \Delta x^\mu$.
It can be verified straightforwardly that this expression is equal to that in equation \eqref{eqn:cosmic clock second order explicit}.
\section{Vector 1-form rulers at first order}
\label{appendix:DI and hat DI}
At first order, equation \eqref{eq:stetradproj} implies that 
\begin{equation}
	\mathfrak{D}_I = -\hat{n}_{s\underline{i}}^{[0]}\left(s_\mu^{\underline{i}} \frac{\partial x^\mu}{\partial \tilde{x}^k}\right)^{[1]}\tilde{e}_I^k -\hat{n}_{s\underline{i}}^{(1)}\left(s_\mu^{\underline{i}} \frac{\partial x^\mu}{\partial \tilde{x}^k}\right)^{[0]}\tilde{e}_I^k,
\end{equation}
and 
\begin{equation}
	\hat{\mathfrak{D}}_I = -\hat{e}_{I\underline{i}}^{[0]}\left(s_\mu^{\underline{i}} \frac{\partial x^\mu}{\partial \tilde{x}^k}\right)^{[1]}\tilde{n}^k -\hat{e}_{I\underline{i}}^{(1)}\left(s_\mu^{\underline{i}} \frac{\partial x^\mu}{\partial \tilde{x}^k}\right)^{[0]}\tilde{n}^k.
\end{equation}
The first two terms on both equations are equal, because at zeroth order, $\hat{e}_{I\underline{i}} = \tilde{e}_{I\underline{i}}$ and $\hat{n}_{s}^k = \tilde{n}^k$, and also the matrix $\left(s_\mu^{\underline{i}} \frac{\partial x^\mu}{\partial \tilde{x}^k}\right)^{[1]}$ is symmetric. 

Comparing the second terms, we find that $s_\mu^{\underline{i}} \frac{\partial x^\mu}{\partial \tilde{x}^k} = \delta^{\underline{i}}_k$, which turns the second terms into $-n_{s\underline{i}}^{(1)}\tilde{e}^k_I \delta^{\underline{i}}_k$ and $-\hat{e}_{I\underline{i}}\tilde{n}^k\delta^{\underline{i}}_k$, respectively. These are equal by the exchange symmetry between the observer and the source. Together, we find that at first order $\hat{\mathfrak{D}}_I = \mathfrak{D}_I$.

\section{Comparison of ruler expressions with first-order results in the literature}
\label{appendix:ns vs n}

If we keep only the first-order terms in equation \eqref{eqn:C 2nd order}, we find
\begin{equation}
	\mathcal{C} = \mathfrak{C} = \tilde{n}_i \delta n^i_{s} - v_{||} +\Phi - \frac{\partial \Delta x_{||}}{\partial \tilde{\chi}} - \Delta \ln(a) \;.
\end{equation}
This is similar to the equation (51) of ref.~\cite{JeongSchmidt2015}, under the assumption (as we make here) that the cosmic ruler does not evolve with time. However, we have an extra term $\tilde{n}_i\delta n^i_{s}$ where $\delta n^i_{s} \equiv n^i_{s} - \tilde{n}^i$---an apparent discrepancy. Since both $n^i_{s}$ and $\tilde{n}^i$ are unit vectors, their first-order difference should be orthogonal to each one of them at first order.
Indeed, the photon's 4-momentum is 
\begin{equation}
	k_i = g_{0 0} (-1 + \delta \nu) s_i^0 + g_{0 j} (-1 + \delta \nu) s_i^j + g_{j 0} (\tilde{n}^j+\delta n^j) s_i^0 + g_{j k} (\tilde{n}^j+\delta n^j) s_i^k 
\end{equation}
where all the quantities are evaluated at the source position. Recalling that $s_i^0$ and $s_i^k - \delta_i^k$ are purely first-order quantities, we can write (at first order)
\begin{equation}
	k^i = \tilde{n}^i + \delta n^i
\end{equation}
where $\delta n^i$ is a first-order quantity, whence
\begin{align}
	n^i_{s} 
	&= \frac{\tilde{n}^i + \delta n^i}{\sqrt{1 + 2\tilde{n}^j \delta n_j}} \\
	&= \tilde{n}^i + \delta n^i - \tilde{n}^i \tilde{n}^j \delta n_j \;.
\end{align}
Therefore, we obtain
\begin{equation}
	\delta n^i_{s}\tilde{n}_i = 0 \;,
\end{equation}
thereby resolving the discrepancy. 
The same issue arises with $\mathcal{M}$ and is resolved in the same manner. 

\section{Test Cases}
\label{appendix:tests}

The two test cases below were described in \S \ref{sec: implementation}, and are now laid out in detail. Ultimately, the evaluation of the rulers was done with the \verb"Mathematica" notebook published together with this work, as described below. We encourage readers who implement the equations given in this paper to also apply the test cases to their implementation. 
\subsection{Pure Gradient Mode}
\label{appendix:test gradient mode}
At second order a gradient mode $\Psi = \Phi = \psi_0 \vk \cdot \vx$ is not a pure gauge mode, as it induces a second-order quadrupole. However, one can correct for this, and we do so by starting with an EdS space-time 
\begin{equation}
	\mathrm{d}s^2 = a^2(\eta')\left(-\mathrm{d}\eta'^2 + \delta_{ij}\mathrm{d}x'^i\mathrm{d}x'^j\right)
\end{equation}
with $a(\eta') = \eta'^2/\eta_0^2$. Next, we perform the following co-ordinate transformation---the most general quadratic transformation in $x^\mu$: ${x^\mu}' \mapsto x^\mu$, where
\begin{align}
	\eta' & = \eta\left[1 + \frac{1}{3}\psi_0\vk \cdot \vx + \psi_0^2\left(a_1 (\vk \cdot \vx)^2 + a_2 x^2 k^2 + a_3 k^2 \eta^2\right)\right]. \\
	\vx' & = \vx\left(1-\frac{5}{3}\psi_0 \vk \cdot \vx\right) + \frac{5}{6}\psi_0 x^2 \vk + \psi_0\frac{\eta^2}{6}\vk \nonumber \\ &
	+ \psi_0^2\left[ b_1 (\vk \cdot \vx)^2\vx + b_2\eta^2k^2\vx  + b_3 k^2 x^2 \vx + b_4 x^2(\vk \cdot \vx)\vk + b_5 \eta^2 (\vk \cdot \vx) \vk\right] \;.
\end{align}
The coefficients $a_{1,2,3}$ and $b_{1,\ldots,5}$ are parameters, which we keep free.\footnote{They could be constrained by requiring that the metric be in Poisson gauge, i.e.~that $\omega_i$ be transverse and $h_{ij}$ be transverse-traceless. We will not require these restrictions here, but continue to assume that $h_{ij}$ and $\omega^i$ are second order in the perturbations.} Thus, the second-order velocity-field is 
\begin{equation}
	\vv = -\psi_0\frac{\eta}{3}\vk - \psi_0^2\left[\left(\frac{5}{9} + 2b_2\right)k^2\eta \vx + \left(2b_5-\frac{1}{3}\right)\eta(\vk\cdot \vx)\vk \right] \;.
\end{equation}
This transformation turns the metric into
\begin{align}
	\Psi & = \psi_0 \vk \cdot \vx + \left(\frac{5}{6} + 3a_1\right)\psi_0^2(\vk \cdot \vx)^2 +3a_2 \psi_0^2k^2x^2 + \left(5a_3 - \frac{1}{18}\right)\psi_0^2k^2\eta^2 \nonumber \\
	\nonumber\Phi & = \psi_0 \vk \cdot \vx - \left(2a_1 -\frac{77}{54}+\frac{5}{3}b_1 + \frac{2}{3}b_4
	\right)\psi_0^2(\vk \cdot \vx)^2 - \left(2a_2 +\frac{25}{27} + \frac{5}{3}b_3 + \frac{1}{3}b_4\right)\psi_0^2 k^2 x^2 \nonumber \\ &
	-\left(2a_3-\frac{1}{54} + b_2 + \frac{1}{3}b_5\right)\psi_0^2k^2\eta^2 \\
	\frac{\omega_i}{\psi_0^2} & = \left(-\frac{11}{9} -2a_1 +2b_5\right)(\vk \cdot \vx)\eta k_i + \left(\frac{5}{9} - 2a_2 +2b_2\right)k^2\eta x_i \nonumber \\
	\frac{h^{ij}}{\psi_0^2} & = \left(-\frac{2}{9} + 4b_5\right)\left(k^ik^j - \frac{\delta^{ij}}{3}k^2\right)\eta^2 + \left( \frac{50}{9} + 4b_4\right)\left(k^ik^j - \frac{\delta^{ij}}{3}k^2\right)x^2 \nonumber \\ & 
	+\left( \frac{50}{9} + 8b_3\right)\left(x^ix^j - \frac{\delta^{ij}}{3}x^2\right)k^2  
	+ \left(4b_1+4b_4-\frac{50}{9} \right) \left(\vk \cdot \vx\right) \left[k^ix^j + x^ik^j - \frac{2\delta^{ij}}{3}\left(\vk\cdot\vx\right) \right] \nonumber \;.
\end{align}
Checking the vanishing of the rulers up to order $\psi_0^2$ is computationally intensive and cannot be done by hand. Therefore, we have implemented this test in a \verb"Mathematica" notebook (see \S \ref{subsec:mathematica notebook}). The main steps are listed below.

We begin with the the frequency-shift to order $\psi_0^2$. Equation \eqref{eqn: kernel for frequency shift} returns
\begin{equation}
	\begin{aligned}
		\delta \nu & = -\frac{\psi_0}{3}  (\eta_0-6 \tilde{\chi} ) \vk\cdot\ntvh -\frac{1}{9}\psi_0^2\Big[(9a_1+2) \tilde{\chi}  (2 \eta_0-7 \tilde{\chi} ) (\vk \cdot \ntvh)^2\Big]\\ & 
		\quad +\frac{1}{9}\psi_0^2 k^2 \Big[-6 \eta_0 \tilde{\chi}  (3a_2+21 a_3-2)+21 \tilde{\chi}^2 (3 a_2+3 a_3-1)+(63a_3-1) \eta_0^2\Big] \;.
	\end{aligned}
\end{equation}
Besides, the perturbation $\delta a_o = a_o -1$ is given by
\begin{equation}
	\delta a_o = -2a_3\psi_0^2k^2\eta_0^2 \;,
\end{equation}
and the first-order total displacements are
\begin{equation}
	\Delta x^i_{(1)} = \frac{1}{3} \psi_0 \left(5 \tilde{n}^i \tilde{\chi}^2 k_\parallel+\eta_0 k^i \tilde{\chi} -3 k^i \tilde{\chi}^2\right),
\end{equation}
and
\begin{equation}
	\Delta\eta_{(1)} = \frac{1}{3} \psi _0 \tilde{\chi}  (\tilde{\chi}-\eta _0) k_\parallel \;.
\end{equation}
These yield, from equation \eqref{eqn: delta ln a},
\begin{equation}
	\begin{aligned}
		\Delta\ln a &= -\frac{2}{3}\psi_0 \tilde{\chi}\vk \cdot \ntvh
		- \psi_0^2 \bigg[\left(2 a_1+\frac{7}{9}\right) \tilde{\chi}^2 (\vk\cdot \ntvh)^2 \\
		&\qquad + \frac{2}{9}k^2 \left(\tilde{\chi}  (9 a_2\tilde{\chi} +\eta_0-3 \tilde{\chi})+9 a_3 (\eta_0-\tilde{\chi})^2\right)\bigg]\;, 
	\end{aligned}
\end{equation}
while, for $\Delta_{a,\tau}$ (defined in equation \eqref{eqn: delta ln alpha calculated}), we obtain
\begin{equation}
	\begin{aligned}
		\Delta_{a,\tau} &= \frac{2}{3}\psi_0 \tilde{\chi} \vk \cdot \ntvh + \psi_0^2 \bigg[\left(2 a_1+\frac{11}{9}\right) \tilde{\chi}^2 (\vk\cdot\ntvh)^2 \\
		&\qquad +\frac{2}{9}k^2 \left(\tilde{\chi}  (9a_2 \tilde{\chi} +\eta_0-3 \tilde{\chi} )+9 a_3 (\eta_0-\tilde{\chi} )^2\right)\bigg]
	\end{aligned}   
\end{equation}
with $\tilde{\eta} = \eta_0 - \tilde{\chi}$. Together, these expressions imply that $\exp \mathcal{T} = 1 + O(\psi_0^3)$, as expected from a space-time that was derived from an FLRW universe by a co-ordinate transformation. 

Additionally, we have 
\begin{equation}
	\begin{aligned}
		\delta \chi^{(2)} & = \frac{2\psi_0^2\tilde{\chi} }{27}  \bigg\{18 a_3 \big(81 \tau_o^2-9 \tau_o \tilde{\chi} +\tilde{\chi} ^2\big) k^2+\tilde{\chi}  \Big[26 \tilde{n}_1 k_1 \tilde{\chi}  \big(\tilde{n}_2 k_2+\tilde{n}_3 k_3\big) \\ & 
		\quad +9 \tau_o \big(k_2^2+k_3^2\big) +k_1^2 \Big(\tilde{\chi}  \big(7-13 \tilde{n}_2^2-13 \tilde{n}_3^2\big)+9 \tau_o\Big) \\ & 
		\quad +\tilde{\chi}  \Big(\big(13 \tilde{n}_2^2-6\big) k_2^2+26 \tilde{n}_2 \tilde{n}_3 k_2 k_3+\big(13 \tilde{n}_3^2-6\big) k_3^2\Big)\Big]\bigg\} 
	\end{aligned}
\end{equation}
and consequently, the total displacements are found to be
\begin{align}
	\nonumber \Delta x_1^{(2)} & = -\frac{\psi_0 ^2\tilde{\chi}}{9} \left\{k^2 \tilde{n}_1 \left[-9 \tau_o^2 (90 a_3-9 b_2-9 b_5+1)+\tilde{\chi} ^2 \left(-9 b_1 \left(\tilde{n}_2^2+\tilde{n}_3^2-1\right)+9 b_2+9 b_3 \right.\right.\right.\\ & 
	\nonumber \quad \left.\left.+9 b_4+9 b_5+50 \tilde{n}_2^2+50 \tilde{n}_3^2-11\right)-9 \tau_o \tilde{\chi}  (6 b_2+6 b_5+1)\right]-\tilde{\chi} ^2 \left[\tilde{n}_2 k_2 \left(k_1 \left(2 (9 b_1-50) \tilde{n}_3^2 \right.\right.\right. \\ &
	\nonumber \quad \left.\left.-18 b_1-9 b_4-9 b_5+76\right)+2 (50-9 b_1) \tilde{n}_1 \tilde{n}_3 k_3\right)+\tilde{n}_1 \left(k_2^2 \left((50-9 b_1) \tilde{n}_3^2+9 b_1+9 b_4 \right.\right.\\ & 
	\nonumber \quad \left.\left.+9 b_5-26\right)+k_3^2 \left(2 (50-9 b_1) \tilde{n}_3^2+9 b_1+9 b_4+9 b_5-26\right)\right)+\tilde{n}_3 k_1 k_3 \left(2 (9 b_1-50) \tilde{n}_3^2 \right.\\ & 
	\nonumber \quad \left.-18 b_1-9 b_4-9 b_5+76\right)-(9 b_1-50) \tilde{n}_2^2 \left(\tilde{n}_1 \left(2 k_2^2+k_3^2\right)-2 \tilde{n}_3 k_1 k_3\right) \\ & 
	\nonumber \quad \left.+2 (9 b_1-50) \tilde{n}_2^3 k_1 k_2\right]+9 (9 b_5-1) \tau_o^2 \left[-\tilde{n}_1 \left(k_2^2+k_3^2\right)+\tilde{n}_2 k_1 k_2+\tilde{n}_3 k_1 k_3\right] \\ & \quad
	\left. -6 (9 b_5-1) \tau_o \tilde{\chi}  \left[-\tilde{n}_1 \left(k_2^2+k_3^2\right)+\tilde{n}_2 k_1 k_2+\tilde{n}_3 k_1 k_3\right]\right\} \;, \\  
	\nonumber \Delta x_2^{(2)} & = -\frac{\psi_0 ^2\tilde{\chi}}{9} \left\{-9 \tau_o^2 \left[(90 a_3 -9b_2)k^2 \tilde{n}_2 + (1-9 b_5) k_2 (\ntvh \cdot \vk)\right]-\tilde{\chi} ^2 \left[k^2 \tilde{n}_2 \left(9 b_1 \left(\tilde{n}_2^2+\tilde{n}_3^2-1\right) \right.\right.\right.\\ & 
	\nonumber \quad \left.-9 b_2-9 b_3-50 \tilde{n}_2^2-50 \tilde{n}_3^2+35\right)+\tilde{n}_2 k_2 \left(k_2 \left(2 (50-9 b_1) \tilde{n}_2^2+(50-9 b_1) \tilde{n}_3^2+9 b_1-9 b_4 \right.\right.\\ & 
	\nonumber \quad \left.\left.-9 b_5-74\right)+2 (50-9 b_1) \tilde{n}_1 \tilde{n}_2 k_1\right)+\tilde{n}_3 k_3 \left(-2 (9 b_1-50) \tilde{n}_2 (\tilde{n}_1 k_1+\tilde{n}_2 k_2) \right.\\ & 
	\nonumber \quad \left.\left.-3 k_2 (3 b_4+3 b_5+8)\right)-\left((9 b_1-50) \tilde{n}_2 k_3^2 \left(\tilde{n}_2^2+2 \tilde{n}_3^2-1\right)\right)-3 \tilde{n}_1 k_1 k_2 (3 b_4+3 b_5+8)\right] \\ & \quad 
	\left. -3 \tau_o \tilde{\chi}  \left[(18 b_2+5) k^2 \tilde{n}_2+2 (9 b_5-1) k_2 (\ntvh \cdot \vk)\right]\right\} \;, \\
	\nonumber \Delta x_3^{(2)} & = -\frac{\psi_0 ^2\tilde{\chi}}{9} \left\{-9 \tau_o^2 \left[(90 a_3 -9b_2)k^2 \tilde{n}_3+(1-9 b_5) k_3 (\ntvh \cdot \vk)\right]-\tilde{\chi} ^2 \left[k^2 \tilde{n}_3 \left(9 b_1 \left(\tilde{n}_2^2+\tilde{n}_3^2-1\right) \right.\right.\right.\\ & 
	\nonumber \quad \left. -9 b_2-9 b_3-50 \tilde{n}_2^2-50 \tilde{n}_3^2+35\right)-k_3 \left(2 (9 b_1-50) \tilde{n}_3^2+9 b_4+9 b_5+24\right) (\tilde{n}_1 k_1+\tilde{n}_2 k_2) \\ & 
	\nonumber \quad +\tilde{n}_3 k_3^2 \left((50-9 b_1) \tilde{n}_2^2+2 (50-9 b_1) \tilde{n}_3^2+9 b_1-9 b_4-9 b_5-74\right) \\ & 
	\nonumber \quad \left.-(9 b_1-50) \tilde{n}_3 k_2 \left(2 \tilde{n}_1 \tilde{n}_2 k_1+k_2 \left(2 \tilde{n}_2^2+\tilde{n}_3^2-1\right)\right)\right] \\ & \quad
	\left. -3 \tau_o \tilde{\chi}  \left[(18 b_2+5) k^2 \tilde{n}_3+2 (9 b_5-1) k_3 (\ntvh \cdot \vk)\right]\right\} \;, \\ 
	\nonumber \Delta \eta^{[2]} & = -\frac{\psi_0}{9} (3 \tau_o-\tilde{\chi} ) \left\{\tilde{\chi}  \left[k^2 \psi_0  (9 a_1 \tilde{\chi} +9 a_2 \tilde{\chi} +3 \tau_o+\tilde{\chi} )-(9 a_1+4) k^2 \tilde{\chi}  \psi_0  \left(\tilde{n}_2^2+\tilde{n}_3^2\right) \right.\right.\\ & 
	\nonumber \quad +(9 a_1+4) \tilde{\chi}  \psi_0  \left(2 \tilde{n}_1 k_1 (\tilde{n}_2 k_2+\tilde{n}_3 k_3)+k_2^2 \left(2 \tilde{n}_2^2+\tilde{n}_3^2-1\right) +k_3^2 \left(\tilde{n}_2^2+2 \tilde{n}_3^2-1\right) \right.\\ & \quad 
	\left.\left.\left.+2 \tilde{n}_2 \tilde{n}_3 k_2 k_3\right)+3 \ntvh\cdot \vk\right]+9 a_3 k^2 \psi_0  \left[27 \tau_o^2-6 \tau_o \tilde{\chi} +\tilde{\chi} ^2\right]\right\} \;.
\end{align}
Inserting these formul\ae\ and their derivatives (with respect to $\tilde{\chi}$ and $\ntvh$) into equations \eqref{eqn:C 2nd order} and \eqref{eqn:M 2nd order} yields, miraculously, that $\mathfrak{C} = \mathcal{O}(\psi_0^3)$ and $\mathfrak{M} = \mathcal{O}(\psi_0^3)$. Furthermore, $\mathcal{M}^{[2]}$ and $\mathcal{C}^{[2]}$ only depend on the first-order non-scalar rulers, which satisfy $\mathfrak{D}_i = \hat{\mathfrak{D}}_i = \mathcal{B}_i/2$. Ref.~\cite{SchmidtJeong2012} already tested the expression for $\mathcal{B}_i^{[1]}$ for a pure gradient mode, that is, $\mathcal{B}_i^{[1]} = \mathcal{O}(\psi_0^2)$; and similarly $\mathfrak{A}_{IJ}^{[1]} = \mathcal{O}(\psi_0^2)$. Therefore, by equation \eqref{eqn:relation between fraktur and calligraphic rulers}, $\mathcal{M} = \mathcal{C} = \mathcal{O}(\psi_0^3)$. 

Some of these calculations were performed with the aid of the \verb"xPand" package\footnote{Documentation available on \url{http://www.xact.es/xPand/}.} for \verb"Mathematica" \cite{Pitrou_2013}.


\subsection{Separate Universe (Spatial Curvature Perturbations)}
\label{appendix:test separate universe}

For this test case, we use an FLRW universe with constant spatial curvature, with the understanding that its expansion history differs from that of the fiducial background used for comparison when evaluating the cosmic rulers---that is, the fiducial background is assumed to be a flat space-time (\S \ref{appendix: curved space correct expansion history}). Fixing a positive spatial curvature does not completely determine the space-time metric; to fully specify it, one must also provide the expansion history, i.e., the conformal time as a function of cosmic time.
In this Appendix, we will consider two different definitions of the conformal time:
\begin{enumerate}
	\item \S \ref{appendix: curved space correct expansion history}, where we use an EdS universe without curvature to define the relation between the global conformal time co-ordinate and the global cosmic time co-ordinate. This approach is used to test the consistency of $\mathcal{T}$. 
	\item \S \ref{appnedix:test case curvature confomal time}, where the conformal time is defined using the curved expansion history. Here, we test $\mathcal{T}$ as well as $\mathfrak{C}$ and $\mathfrak{M}$. The relevant \verb"Mathematica" notebook is published with this paper (see \S \ref{subsec:mathematica notebook}). 
\end{enumerate}
Since the purpose of the test cases is primarily mathematical---to verify the validity of the second-order formul\ae\ for the cosmic rulers---we adopt both approaches here. 

\subsubsection{Curved background: cosmic time}
\label{appendix: curved space correct expansion history}



Let us consider the case where the space-time is simply an FLRW metric with non-vanishing $\Omega_m$ and $\Omega_K$, such that the expansion history differs from that of the fiducial, flat universe with the same value of $\Omega_m$. The metric is 
\begin{equation}
	\mathrm{d}s^2 = -\mathrm{d}t^2 + a_K^2 \frac{\delta_{ij}dx^idx^j}{\left(1+\frac14K|{\bf x}|\right)^2} \;.
\end{equation}
Here, $a_K$ denotes the true scale factor, to be contrasted with the scale factor $a_0$ of an assumed flat background. To compare the expansion histories of these two backgrounds, we expand the metric of the curved space-time to second order in $K|\vx|^2$ and write 
\begin{equation}
	\mathrm{d}s^2 = -\mathrm{d}t^2 + a_0^2\left(
	1 - \frac12 K|{\bf x}|^2 + \frac3{16} K^2|{\bf x}|^4
	\right) \frac{a_K^2}{a_0^2} \delta_{ij}\mathrm{d}x^i\mathrm{d}x^j \;.
\end{equation}
We perform the test case twice: in this section, using cosmic time, which does not allow a fully explicit test;\footnote{This is not fully explicit because the metric itself involves the ratio $a_K/a_o$, which in turn depends on $\mathcal{T}$.} 
and in the next section, using conformal time as defined in the curved space-time.

In this space-time, we choose to normalise $\at(\tau_o)=1$, i.e. $a_0(\tau_o) =1$, because cosmic time coincides with proper time (age). Therefore, $a_K(\tau_o) \neq 1$, and we have
\begin{equation}
	\exp \mathcal{T} = \frac{a_0(t)a_K(\tau_o)}{a_K(t)} \;.
\end{equation}
Here $a_K(t)$ is given implicitly in terms of the eccentric anomaly $\theta$ by 
\begin{equation}\label{eqn:curved universe scale factor solution}
	\begin{aligned}
		a_K(\theta) & = A(1-\cos \theta)\;, \\ 
		H_0 t(\theta) & = B(\theta - \sin\theta)\;,
	\end{aligned}
\end{equation}
where $B = \frac{1-\Omega_K }{2 \abs{\Omega_K}^{3/2}}$. 
The ratio $A/B$ is determined by requiring that at early times, this should match a matter-dominated universe with the same $\Omega_m$, \emph{viz}.~$A=\left(B/\sqrt{2}\right)^{2/3}$. 

Solving Kepler's equation for $\theta(t)$ we find 
\begin{equation}\label{eqn:eccentric anomaly solution second order}
	\begin{aligned}
		\theta & = \abs{\Omega_K}^{3/2} \left[\frac{H_0t}{5}-\left(\frac{2}{3}\right)^{2/3} \sqrt[3]{H_0t}\right]+2^{2/3} \sqrt{\abs{\Omega_K}} \sqrt[3]{3H_0 t} \\ & \quad 
		+ \abs{\Omega_K}^{5/2} \left[\left(\frac{2}{3}\right)^{5/3} \sqrt[3]{H_0t}+\frac{1}{175} \sqrt[3]{2}(3H_0t)^{5/3}-\frac{H_0t}{5}\right] + \mathcal{O}\big(\Omega_K^3\big)
	\end{aligned}
\end{equation}
Therefore, to second order in $\Omega_K$ the cosmic clock is given by 
\begin{equation}
	\begin{aligned}
		\exp \mathcal{T} = 1 + \frac{\Omega_K}{1-a_0} + \frac{\left[\sqrt[3]{6} \left(45 \sqrt[3]{2} (H_0t)^{2/3}-91 \sqrt[3]{3}\right)(H_0t)^{2/3}+122\right] \Omega_K^2}{1050} + \mathcal{O}\big(\Omega_K^3\big) \;.
	\end{aligned}
\end{equation}
Let us now check that this equation is satisfied by the second-order expression \eqref{eqn:cosmic clock second order} for the cosmic clock. To this end, we define the conformal time with the background scale-factor, i.e. $\mathrm{d}t = a_0\mathrm{d}\eta$, whence $\Psi = 0$, $h_{ij} = 0$, $\omega_i = 0$, $\vv = 0$ and 
\begin{equation}
	\Phi = \mathcal{T} - \mathcal{T}^2 - \delta a_{K,o} + 2\delta a_{K,o} \mathcal{T} -\frac{\delta a_{K,o}^2}{2} + \frac{K}{4}x^2 - \frac{K}{2}\mathcal{T}x^2 + \frac{K}{2}x^2\delta a_{K,o} - \frac{3}{32}K^2x^4,
\end{equation}
where $\delta a_{K,o} \equiv a_K(\tau_o) - 1$. 
Therefore, $\delta \tau = 0$ and $\mathcal{T} = -\delta \nu - \delta \nu^2/2$. 

By equation \eqref{eqn: kernel for frequency shift}, 
\begin{equation}
	F_{[2]}^0 = \Phi'\left(1+2\tilde{n}_i\delta n^i_{(1)}\right)\; ;
\end{equation}
and the first-order shift in the photon momentum is given by \cite{SchmidtJeong2012}
\begin{equation}
	\delta n^i_{(1)} = \left(2\mathcal{T} + \frac{K}{4}x^2 - \delta a_{K,o}\right)\tilde{n}^i.
\end{equation}
Therefore, we have 
\begin{equation}
	F_{[2]}^0 = \mathcal{T}'(1+2\mathcal{T})
\end{equation}
and, since $\delta \nu_o = 0$ from equation \eqref{eqn:delta nu at observer},
\begin{equation}\label{eqn:constant curvature test case delta nu}
	\delta \nu = \int_0^{\chi_e} \mathcal{T}'(1+2\mathcal{T}) \mathrm{d}\chi = \int_{\eta_0}^{\eta_e} \frac{\mathcal{T}'(1+2\mathcal{T})}{-1+\delta \nu^{(1)}} \mathrm{d}\eta \approx -\int_{\eta_0}^{\eta_e} \mathcal{T}'(1+\mathcal{T}) \mathrm{d}\eta = -\mathcal{T} - \frac{\mathcal{T}^2}{2} \;.
\end{equation}
The second equality follows from the fact that $F_{[2]}^0$ depends on $\chi$ only through $\eta = x^0_{\rm lc}$. This immediately gives $\delta \nu^{(1)} = -\mathcal{T}^{(1)}$, which was used in the third equality. Hence equation \eqref{eqn:cosmic clock second order} satisfies this test case since 
\begin{equation}
	\mathcal{T} \underset{\eqref{eqn:cosmic clock second order}}{=} -\delta \nu - \frac{\delta \nu^2}{2} \underset{\eqref{eqn:constant curvature test case delta nu}}{=} \mathcal{T}
\end{equation}
as required. 

\subsubsection{Curved background: conformal time}
\label{appnedix:test case curvature confomal time}

Here we perform the test using the conformal time $\eta$ defined in the actual space-time, i.e. via $\mathrm{d}t = a_K\mathrm{d}\eta$. 
The metric is 
\begin{equation}\label{curvedmet}
	\mathrm{d}s^2 = a_K^2(\eta) \left[- \mathrm{d}\eta^2 + \frac{\delta_{ij}\mathrm{d}x^i\mathrm{d}x^j}{\left(1 + \frac{1}{4}K x^2\right)^2}\right]\;,
\end{equation}
where $a_K$ is the scale-factor in the curved space-time. 
The calculation consists of three parts: first, we compute the values that the rulers should take in a curved FLRW space-time, based on their non-linear definitions, and expand them to second order in the curvature $K$; then, we calculate the analogous results using the perturbative approach developed in this paper; and finally, we compare the two by matching both the expansions and the parameterisations appropriately.

\paragraph{Calculation of the perturbations of the metric.}

Before going any further, let us express the metric \eqref{curvedmet} in the form of equation \eqref{eqn: metric Poisson gauge}. To do so, we need to define the conformal time, as above, and express $a_K$ in terms of the conformal time $\eta$, defined this time via $\mathrm{d}t = a_K\mathrm{d}\eta$. 
The solution is as above
\begin{equation}
	\begin{aligned}
		a_K(t) &= \frac{a_o \Omega_o}{2(\Omega_m-1)}\big(1-\cos\theta\big)\;, \\
		H_ot &= \frac{\Omega_o}{2(\Omega_0-1)^{\frac{3}{2}}}\big(\theta-\sin\theta\big) \;,
	\end{aligned}
\end{equation}
where $a_o$ and $H_o$ are the modified scale factor, and the Hubble constant at $\eta_o$. The conformal time is
\begin{equation}
	\eta(\theta) = \int_0^\theta \frac{\mathrm{d}\theta'}{a_o H_o(\Omega_0-1)^{\frac{1}{2}}} 
	= \frac{\theta}{a_oH_o (\Omega_m-1)^{\frac{1}{2}}} \;.
\end{equation}
From Friedmann's equation $\Omega_m + \Omega_K = 1$, we have 
\begin{equation}
	a_K(\eta) = \frac{a_o H_o^2 \left(1 + \frac{K}{H_o^2}\right)}{2K}\left[1-\cos(a_o \sqrt{K}\eta)\right] \;.
	\label{aKcom}
\end{equation}
The Hubble parameter $H_o$ is related to $\eta_o$ via 
\begin{equation}
	H_o^2 = \frac{2K}{1-\cos\left(a_o \sqrt{K}\eta_o\right)} - K
\end{equation}
since $a_K(\eta_o) = a_o$. 
Inserting that expression back into \eqref{aKcom}, we get
\begin{equation}
	\frac{a_K(\eta)}{a_o}= \frac{1-\cos\left(a_o \sqrt{K}\eta\right)}{1-\cos\left(a_o \sqrt{K}\eta_o\right)} \;.
	\label{ak}
\end{equation}
Without loss of generality, we will hereafter take $\eta_o$ to be independent of $K$. 

The dependence of $a_o$ on $K$ is determined by the initial condition $a_0(\overline{\eta}(\tau_o)) = 1$, where $a_0(\eta) = \left(\eta/\eta_o\right)^2$ is the background scale factor. Since we decided to fix $\eta_o$, we need to express this condition in terms of $\eta_o$. We have: 
\begin{equation}
	\tau_o = \int_0^{\eta_o}a_K(\eta) d\eta \;,
\end{equation}
whence by equation \eqref{ak},  
\begin{equation}
	\tau_o = \frac{1}{\cos \left(a_o
		\eta_o \sqrt{K}\right)-1}\left[\frac{\sin \left(a_o  \eta_o 
		\sqrt{K}\right)}{\sqrt{K}}-a_o  \eta_o \right] \;.
\end{equation}
Using $\overline{\eta}(t) = \left(3t\eta_o^2\right)^{\frac{1}{3}}$, this yields
\begin{equation}
	\eta_o = 3 \left[\frac{1}{2} a_o \eta_o  \csc
	^2\left(\frac{1}{2} a_o  \eta_o  \sqrt{K}\right)-\frac{\cot
		\left(\frac{1}{2} a_o  \eta_o 
		\sqrt{K}\right)}{\sqrt{K}}\right] \;,
	\label{etaocurved}
\end{equation}
which we need only solve to second order in $K$. Expanding this equation at second order in $K$, we find\footnote{We remark that $a_o$ is bounded near $0$, which follow from the fact that $a_K$ simply converges towards the function $a_0$, as $K\to 0$.}
\begin{equation}
	1 = \frac{(a_o  \eta_o)^{2/3}}{\eta_o^{2/3}}+\frac{19 a_o ^4 \eta_o^{10/3} K^2
		(a_o  \eta_o)^{2/3}}{28350}+\frac{1}{45} a_o
	^2 \eta_o^{4/3} K (a_o  \eta_o)^{2/3} \;.
\end{equation}
Let $a_K(\eta_o) \equiv 1 + \delta_K a_o$ where $\delta_K a_o = \mathcal{O}(\Omega_K)$ is small relative to unity.\footnote{$\delta_K a_o$ is not to be confused with $\delta a_o$ in the main body of the paper. There, we had $\delta a_o = a_0(\eta_o) - 1$ (which is actually zero here with the initial conditions we chose) whereas here, $\delta_K a_o = a_K(\eta_o) - 1$.} Substituting this definition into the previous equation and expanding to second order in $\delta_K a_o$, we obtain
\begin{equation}
	\begin{aligned}
		0 &= \frac{(\delta_K a_o)^2 \left(209 \eta_o ^4 K^2+1800 \eta_o ^2
			K-4050\right)}{36450} \\ &\quad +\frac{\delta_K a_o\left(19 \eta_o ^4 K^2+360
			\eta_o ^2 K+4050\right)}{6075} 
		+\frac{19 \eta_o ^4 K^2}{28350}+\frac{\eta_o ^2 K}{45} \;.
	\end{aligned}
\end{equation}
By solving this equation for $\delta_K a_o$ at second order in $\Omega_K$, we determine the explicit dependence of $a_K$ on $K$ at second order, to be
\begin{equation}\label{eqn:a k of eta curvate test case}
	a_K(\eta) = \left(1-\frac{\eta_o ^2 K}{30}+\frac{3 \eta_o ^4 K^2}{1400}\right) \frac{1-\cos\left[\left(1-\frac{\eta_o ^2 K}{30}+\frac{3 \eta_o ^4 K^2}{1400}\right) \sqrt{K}\eta\right]}{1-\cos\left[\left(1-\frac{\eta_o ^2 K}{30}+\frac{3 \eta_o ^4 K^2}{1400}\right) \sqrt{K}\eta_o\right]} \;.
\end{equation}

Given that the function $a_K(\eta)$ is known explicitly, we can expand the metric \eqref{curvedmet} at second order in $K$ to find, finally, 
\begin{equation}
	\label{eqn:curvedmet1}
	\mathrm{d}s^2 = a_0^2(\eta)\left[-(1+2\Psi)\mathrm{d}\eta^2 +(1- 2\Phi) \delta_{ij} \mathrm{d}x^i \mathrm{d}x^j \right] \;,
\end{equation}
with 
\begin{align}
	\Psi &= \frac{\eta ^4 K^2}{160}-\frac{1}{360} \eta ^2 \eta_o^2
	K^2-\frac{13 \eta_o^4 K^2}{16800}-\frac{\eta ^2
		K}{12}+\frac{\eta_o^2 K}{20} \;,\label{eqn:constant curvature psi}\\
	\Phi &= -\frac{\eta ^4 K^2}{160}+\frac{1}{360} \eta ^2 \eta_o^2K^2+\frac{13 \eta_o^4 K^2}{16800}-\frac{3}{32} K^2
	x^4-\frac{1}{24} \eta ^2 K^2 x^2 \; ,\label{eqn:constant curvature phi} \\
	&\quad +\frac{1}{40} \eta_o^2 K^2 x^2+\frac{\eta ^2 K}{12}-\frac{\eta_o^2 K}{20}+\frac{K x^2}{4}\nonumber 
\end{align}
for the Bardeen's potentials.


\paragraph{Calculation of the expected value of the cosmic rulers.}

Unlike the other test cases, in this case, the value of the cosmic rulers is nonzero. We will calculate the expected values for the different rulers based on their definitions, starting with $\mathcal{T}$, which is defined by
\begin{equation}
	\mathcal{T}(x) = \ln\left[\frac{a_0(\overline{\eta}(\tau))}{a_0(\Tilde{x}^0)}\right] = \ln\left[a_0(\overline{\eta}(\tau))\frac{a_K(\eta_o)}{a_K(x^0)}\right] \;,
\end{equation}
where $\tilde{a} = a_0(\Tilde{x}^0)$. 
Note that, since this space-time is an FLRW space-time, $\tau$ is simply the cosmic time, $t$.
Using the expressions of $\overline{\eta}$, $\tau$, $a_0$, and $a_K(\eta)$ given equation \eqref{eqn:a k of eta curvate test case}, we proceed to calculate $\mathcal{T}$ at second order in $K$. The result is
\begin{equation}\label{eqn: curvature T expected}
	\mathcal{T}(x^\mu) = \frac{K^2}{16800} \Big(11 \eta^4-56 \eta ^2 \eta_o^2+45 \eta_o^4\Big)+\frac{K}{20} \big(\eta^2-\eta_o^2\big)\;.
\end{equation}
Notice that $\mathcal{T}$ vanishes at $x^\mu = x^\mu_o$ by definition.

We now turn to the calculation of the expected values of $\mathfrak{M}$ and $\mathfrak{C}$. We begin by solving the geodesic equation for a photon in a curved FLRW universe. 
In this paper, we choose the radial co-moving co-ordinate $\chi$ as the affine parameter (cf.~\S \ref{sec:geodesic equation}), but here we must specify that the latter refers to the $\chi$ of the flat fiducial background, not the $\chi$ of the actual FLRW space-time.  
To compare the values of the rulers found in this section with those calculated using the perturbative approach presented in this article, we need to establish a second-order relation between these two parametrisations. 
To distinguish between the background's $\chi$ and that of the FLRW space-time considered here, we will denote the latter by $\chi_K$. 

With the metric (\ref{eqn:curvedmet1}), the only non-zero Christoffel symbols for the conformally re-scaled metric---equation \eqref{curvedmet} divided by $a_K^2$---are
\begin{equation}
	\Gamma^i_{jk} = \frac{K}{2\left(1+\frac{1}{4}Kx^2\right)}\left(\delta_{jk}x^i - \delta^i_j x_k - \delta^i_k x_j\right) \;.
\end{equation}
Substituting this expression into the photon geodesic equation, the photon 4-trajectory is
\begin{align}
	x^0(\chi_K) &= \eta_o - \chi_K\; ,
	\label{eta2curved}\\
	x^k(\chi_K) &= \tilde{n}^k \left(\chi_K +\frac{1}{12}K \chi_K^3 + \frac{1}{120}K^2\chi_K^5\right)\; ,
	\label{x2curved}
\end{align}
at second order in $\chi_K^2K$. 
The orthonormal tetrad $s_{\underline{\alpha}}^\mu$ defined at the source is given by 
\begin{align}
	s_{\underline{0}}^\mu &= a_K^{-1}(1, \bm{0})\;, \\
	s_{\underline{j}}^\mu &= a_K^{-1}\left(0, \delta_j^i \left[1+\frac{Kx^2}{4}\right]\right)\;.
\end{align}
Since this space-time is isotropic, we can set $\nvh_{s} = \ntvh$ in what follows. We now proceed to calculate the cosmic rulers using equation \eqref{eqn:spullback}. For $\mathfrak{C}$ and $\mathfrak{B}_i$, we need
\begin{equation}
	i^*s^{||} = g_{\mu \nu}(x^\beta) \frac{\partial x^\mu}{\partial \tilde{x}^k} \mathrm{d}\tilde{x}^k s_i^\nu \tilde{n}^i =  \frac{a_K(x^\beta)}{(1+\frac{1}{4}Kx^2)}\tilde{n}_i \frac{\partial x^i}{\partial \tilde{x}^k} \mathrm{d}\tilde{x}^k \;.
\end{equation}
On writing $x(\chi_K) = \chi_K +\frac{K \chi_K^3}{12} + \frac{K^2\chi_K^5}{120}$, the term $\tilde{n}_i \frac{\partial x^i}{\partial \tilde{x}^k}$ involving the Jacobian becomes
\begin{align}
	\tilde{n}_i \frac{\partial x^i}{\partial \tilde{x}^k} &= \tilde{n}_i \left( \tilde{n}_k \frac{\partial}{\partial \tilde{\chi}} + \frac{1}{\tilde{\chi}} P^j_k \frac{\partial}{\partial \tilde{n}^j}  \right) \left[ \tilde{n}^i x(\chi_K) \right] \\
	&= \tilde{n}_k \frac{\mathrm{d} x}{\mathrm{d}\tilde{\chi}_K}
\end{align}
and, consequently, 
\begin{equation}
	i^*s^{||} = \frac{a_K(x^\beta)}{(1+\frac{1}{4}Kx^2)} \frac{\mathrm{d} x}{\mathrm{d}\tilde{\chi}_K} \mathrm{d}\tilde{x}_{||} \;.
\end{equation}
Using the definition of the 1-form rulers \eqref{eq:frak}, we read off
\begin{align}
	\mathfrak{B}_i &= 0 \;, \\
	1 - \mathfrak{C} &= \frac{a_K(x^0(\chi_K))}{a_0(\eta_o-\Tilde{\chi}_K)(1+\frac{1}{4}Kx(\chi_K)^2)} \frac{\mathrm{d}x(\chi_K)}{\mathrm{d}\Tilde{\chi}_K} \;.
	\label{Ccurvedtran}
\end{align}
To express the cosmic rulers as functions of $\Tilde{\chi}_K$ defined via the observed redshift,\footnote{Notice, that this is not $\tilde{\chi}$, because the affine parameter is $\chi_K$, not $\chi$.} we use the relation
\begin{equation}
	\frac{1}{a_0(\eta_o - \Tilde{\chi}_K)} \equiv \frac{a_K(\eta_o)}{a_K(x^0(\chi_K))}\;,
\end{equation}
which can be solved for $\chi_K$ order by order. At second order, the solution is
\begin{equation}
	\begin{aligned}
		\chi_K & = \frac{K^2}{23040} \big(\tilde{\chi}_K-\eta_o\big) \Big(63\tilde{\chi}_K^4-252 \eta_o \tilde{\chi}_K^3+284 \eta_o^2
		\tilde{\chi}_K^2-64 \eta_o^3 \tilde{\chi}_K\Big)
		\\ & \quad 
		+\frac{K}{24} \big(\tilde{\chi}_K-\eta_o\big) \big(\tilde{\chi}_K^2-2 \eta_o \tilde{\chi}_K\big)+\tilde{\chi}_K \;.
	\end{aligned}
\end{equation}
Using the second-order solution to the geodesic equation, equations \eqref{eta2curved} and \eqref{x2curved}, we eventually obtain
\begin{equation}\label{eqn:constant curvature C expected}
	\begin{aligned}
		1-\mathfrak{C} & = \frac{K^2}{89600} \Big(192 \eta_o^4+875 \tilde{\chi}_K^4-3500 \eta_o \Tilde{\chi}_K^3+4480 \eta_o^2 \Tilde{\chi}_K^2-1960 \eta_o^3 \Tilde{\chi}_K\Big)
		\\ & \quad 
		+\frac{K}{40} \Big(2 \eta_o^2+5 \Tilde{\chi}_K^2-10\eta_o \Tilde{\chi}_K\Big)\;.
	\end{aligned}
\end{equation}
Likewise, the starting point for the derivation of $\mathfrak{M}$ is 
\begin{equation}
	i^*s^J 
	= g_{\mu \nu}(x^\beta) \frac{\partial x^\mu}{\partial \tilde{x}^k} \mathrm{d}\tilde{x}^k s_j^\nu P^{ij} \hat{e}_{J, i} =  \frac{a_K(x^\beta)}{1+\frac{1}{4}Kx^2} P^{ij}\frac{\partial x_j}{\partial \tilde{x}^k} \mathrm{d}\tilde{x}^k\hat{e}_{J, i} \;.
\end{equation}
The computation of the projection $P^{ij} \frac{\partial x_j}{\partial \tilde{x}^k}$ gives 
\begin{align}
	P^{ij} \frac{\partial x_j}{\partial \tilde{x}^k} &= P^{ij} \left( \tilde{n}_k \frac{\partial}{\partial \tilde{\chi}} + \frac{1}{\tilde{\chi}} P^l_k \frac{\partial}{\partial \tilde{n}_l}  \right) \left[ \tilde{n}_j x(\chi_K) \right] \\
	&= P^i_k \frac{x(\chi_K)}{\tilde{\chi}_K} \;.
\end{align}
As a result, we have
\begin{equation}
	i^*s^J = \frac{a_K(x^\beta)}{1+\frac{1}{4}Kx^2} \frac{x(\chi_K)}{\tilde{\chi}_K} \mathrm{d}\tilde{x}^i_\perp \hat{e}_{J, i} \;.
\end{equation}
Using the definition \eqref{eq:frak} of the rulers, we read off
\begin{equation}
	2- \mathfrak{M} = \frac{2a_K(x^0(\chi_K))}{a_0(\eta_o-\Tilde{\chi}_K)(1+\frac{1}{4}K\chi_K^2)} \frac{x(\chi_K)}{\Tilde{\chi}_K}
	\label{Mcurvedtran}
\end{equation}
After expressing $\chi_K$ as a function of $\tilde{\chi}_K$ and substituting the solution to the geodesic equation, we eventually obtain
\begin{equation}\label{eqn:constant curvature M expected}
	\begin{aligned}
		\mathfrak{M} & = \frac{K^2}{403200} \Big(11025 \tilde{\chi}_K^4-1728 \eta_o^4-51975 \eta_o \tilde{\chi}_K^3+24080 \eta_o^2 \tilde{\chi}_K^2+8820 \eta_o^3 \tilde{\chi}_K\Big) \\ & \quad 
		-\frac{K}{20} \Big(2 \eta_o^2-5 \tilde{\chi}_K^2-5 \eta_o \tilde{\chi}_K\Big) \;.
	\end{aligned}
\end{equation}

\paragraph{Calculation of the rulers with a perturbative approach}
We now turn to the calculation of $\mathcal{T}$ using the formula in equation \eqref{eqn:cosmic clock conformal Hubble}. 

First, let us check that equation \eqref{eqn:cosmic clock second order} gives the correct result at first order. At this order in perturbations, $\mathcal{T}$ reduces to 
\begin{equation}
	\label{eqn:T11}
	\mathcal{T}^{(1)} = -\delta \nu + \Psi + \frac{\mathcal{H}(\overline{\eta}(\tau))}{a_0(\overline{\eta}(\tau))} \int_0^{x^0} a_0(\eta)\Psi(\eta) \mathrm{d}\eta\;,
\end{equation}
with $v^i=v_\parallel=0$ owing to the homogeneity and isotropy of this space-time.
We use \eqref{eqn: kernel for frequency shift} to calculate $\delta\nu$. We get
\begin{equation}
	\frac{\mathrm{d} \delta \nu^{(1)}}{\mathrm{d}\chi} = - \Psi'|_{\eta = \eta_o - \chi} + \Phi'_{x^i = \tilde{n}^i \chi, \eta = \eta_o - \chi}\;,
\end{equation}
in which the arguments of the Bardeen's potentials involve the zeroth order trajectory since $\Phi$ and $\Psi$ are already first-order quantities. 
Substituting expressions \eqref{eqn:constant curvature psi} and \eqref{eqn:constant curvature phi} for the Bardeen's potentials and integrating the resulting equation with the initial condition \eqref{eqn:delta nu at observer} yields
\begin{equation}
	\delta \nu^{(1)}(\chi) = -\frac{1}{30}\eta_o ^2 K+\frac{1}{3}\eta_o  K \chi-\frac{1}{6}K \chi^2
	\label{deltavu1}
\end{equation}
or, since $\chi = \eta_o - x^0$ (this is a background co-ordinate relation),
\begin{equation}
	\delta \nu^{(1)} = \frac{1}{3} \eta_o K \big(\eta_o-x^0\big)-\frac{1}{6} K \big(\eta_o-x^0\big)^2-\frac{1}{30}\eta_o^2 K \;.
\end{equation}
In equation \eqref{eqn:T11}, the integral involves the first-order quantity $\Psi$, so we can just take the zeroth-order value  of $x^0$ and $\overline{\eta}(\tau)=x^0$. Using the expression for $\delta \nu^{(1)}$ and the zeroth-order value of $\overline{\eta}(\tau)$, we finally have at first-order: 
\begin{equation}
	\mathcal{T}^{(1)}(x^\mu) = \frac{K}{20}\Big(\big(x^0\big) ^2-\eta_o^2\Big) \;,
\end{equation}
which matches the expected result. 

At second order, the cosmic clock follows from equation \eqref{eqn:cosmic clock second order} upon setting $v_i=\omega_i=0$,
\begin{equation}
	\begin{aligned}
		\mathcal{T} & = - \delta \nu + \Psi - \Psi^2 - \frac{\delta \nu^2}{2} 
		+ \frac{\mathcal{H}(\overline{\eta}(\tau))}{a(\overline{\eta}(\tau))} \left[ \int_0^{x^0} a(\eta') \left(\Psi (\Vec{x}, \eta')   - \frac{ \Psi^2(\Vec{x}, \eta')}{2} \right) \mathrm{d}\eta' \right] \\ & \quad
		+ \left[\frac{\mathcal{H}(\overline{\eta}(\tau))^2}{a(\overline{\eta}(\tau))^2}- \frac{1}{2} \frac{a''(\overline{\eta}(\tau))}{a(\overline{\eta}(\tau))^2}\right] \left[\int_0^{x^0} a(\eta')\Psi (\Vec{x}, \eta') \mathrm{d}\eta'\right]^2 \;.
		\label{TsecondK}
	\end{aligned}
\end{equation}
Here again, we calculate $\delta \nu$ to second order with equation \eqref{eqn: kernel for frequency shift}. We have
\begin{equation}
	\frac{\mathrm{d} \delta \nu^{(2)}}{\mathrm{d}\chi} = -\Psi' \Big(1-2\Psi - 2\delta \nu^{(1)}\Big) + \Phi' \Big( 1 - 2 \Psi + \delta_{ij}\tilde{n}^i(\delta n^j)^{(1)}\Big)\;,
	\label{diffdeltanu}
\end{equation}
where $\Psi$ and $\Phi$ are evaluated on the first-order trajectory (in the post-Born approximation). To integrate this equation, we thus need to calculate the photon 4-trajectory at first order. For this purpose, we need
\begin{equation}
	\eta^{(1)}(\chi) = \int_0^\chi \left(-1 + \delta \nu^{(1)}(\chi')\right)\mathrm{d}\chi' \;.
\end{equation}
On inserting the first-order value of $\delta \nu$ from \eqref{deltavu1} under the integral sign and the initial condition $ \eta^{(1)}(0) = \eta_o$, $\eta^{(1)}(\chi)$ becomes
\begin{equation}
	\eta^{(1)}(\chi) = \eta_o +\chi  \left(-\frac{1}{30}\eta_o ^2 K-1\right)+\frac{1}{6} \eta_o  K \chi ^2-\frac{1}{18}K \chi ^3 \;.
\end{equation}
Likewise, we use the first-order equation (15) from \cite{JeongSchmidt2015} to find $x^{(1)}$. In our case, the former translates into
\begin{equation}
	\label{eqn:ddndchi}
	\frac{\mathrm{d}}{\mathrm{d}\chi} \Big[\big(\delta n^i\big)^{(1)} - 2 \Phi \tilde{n}^i\Big] = -\partial_i \Psi - \partial_i \Phi \;,
\end{equation}
in which zeroth-order values of the photon 4-trajectory are used in the argument of $\Phi$ and $\Psi$. The initial condition is given by equation (20) of the same paper and, in our case, simplifies to
\begin{equation}
	(\delta n_o^i)^{(1)} = \Phi(\eta_o, 0) \tilde{n}^i \;.
\end{equation}
On integrating equation \eqref{eqn:ddndchi} and expanding the result at first-order in $\eta_o^2K$ etc., we obtain 
\begin{equation}
	(\delta n^i)^{(1)}(\chi) = \frac{1}{60} K \big(2 \eta_o ^2 \tilde{n}^i-20 \eta_o  \tilde{n}^i \chi +25 \tilde{n}^i \chi ^2\big)\;.
\end{equation}
A second integration over $\chi$, with the initial condition $(x^i)^{(1)}(0) = 0$, yields
\begin{equation}
	(x^i)^{(1)}(\chi) = \frac{K}{180} \Big(6 \eta_o^2 \tilde{n}^i\chi -30 \eta_o \tilde{n}^i \chi ^2+25\tilde{n}^i \chi ^3 \Big)+\tilde{n}^i\chi 
\end{equation}
and, since $\tilde{n}^i$ is of norm unity,
\begin{equation}
	x^{(1)}(\chi) = \frac{K}{180} \Big(6 \eta_o ^2 \chi -30 \eta_o  \chi ^2+25 \chi ^3\Big)+\chi \;,
\end{equation}
which is our final expression for the first-order $x^{(1)}$.

Knowing the first-order trajectory of the photon and using the initial condition \eqref{eqn: kernel for frequency shift}, we can integrate equation \eqref{diffdeltanu} to find 
\begin{equation}
	\begin{aligned}
		\delta \nu^{(2)}(\chi) & = \frac{K^2}{75600}\Big(78 \eta_o ^4+420 \eta_o ^3 \chi -9450 \eta_o ^2 \chi ^2+10220 \eta_o  \chi ^3-2555 \chi ^4\Big)\\ & \quad
		+\frac{K}{30} \Big(-\eta_o^2+10 \eta_o  \chi -5 \chi ^2\Big) \;.
		\label{deltanu2c}
	\end{aligned}
\end{equation}
To find $\delta \nu$ at emission, we need to get the first-order $\chi$ at emission, which satisfies $\eta^{(1)}(\chi_e) = x^0$. Here, we want to express quantities as functions of $x$, not $\Tilde{x}$, which is why we do not apply the method presented in \S\ref{sec:geodesic equation}, as it requires of the redshift and thus $\Tilde{x}$. 
Using the expression we derived for $\eta^{(1)}(\chi)$, the condition $\eta^{(1)}(\chi_e) = x^0$ becomes
\begin{equation}
	\eta_o +\chi_e  \left(-\frac{1}{30}\eta_o ^2 K-1\right)+\frac{1}{6} \eta_o  K \chi_e ^2-\frac{1}{18}K \chi_e ^3 = x^0.
\end{equation}
Again, this equation is complicated to solve but we can simplify it by recalling that we only need a first-order expression in $K$ for $\chi_e$. Expanding the latter around its zeroth order value $\eta_o - x^0$, we find
\begin{equation}
	\chi_e = \eta_o +\frac{1}{90} K \Big(7 \eta_o ^3+5 \big(x^0\big)^3-12 \eta_o ^2 x^0\Big)-x^0\;.
\end{equation}
On inserting this result into equation \eqref{deltanu2c}, we obtain the second-order frequency shift $\delta \nu^{(2)}$ at emission, 
\begin{equation}
	\delta \nu^{[2]} = \frac{K^2}{25200} \left[-429 \eta_o ^4-385 \big(x^0\big)^4+840 \eta_o ^2 \big(x^0\big)^2\right]+\frac{K}{30}\left[4 \eta_o ^2-5 \big(x^0\big)^2\right] \;.
\end{equation}
The last ingredient we need is the first-order value of $\overline{\eta}(\tau)$. Using $\overline{\eta}(t) = \left(3t\eta_o^2\right)^{\frac{1}{3}}$ as well as equation \eqref{eq:tau_s}, we get 
\begin{equation}
	\overline{\eta}(\tau) = \frac{1}{60} K x^0 \left[\eta_o ^2-(x^0)^2\right]+x \;.
\end{equation}
After combining all the partial results into equation \eqref{TsecondK}, we ultimately obtain the expression
\begin{equation}
	\mathcal{T}(x) = \frac{K^2}{16800}\Big[11 (x^0) ^4-56 \big(x^0\big) ^2 \eta_o^2+45 \eta_o^4\Big]+\frac{K}{20} \Big[\left(x^0\right) ^2-\eta_o^2\Big]
\end{equation}
for the second-order cosmic clock, which matches the expected result equation \eqref{eqn: curvature T expected}! 

We can now calculate $\mathfrak{M}$ and $\mathfrak{C}$ using equation \eqref{eqn:C 2nd order} and \eqref{eqn:M 2nd order}. For the test case considered here, we get
\begin{equation}
	1-\mathfrak{C} = \left(1 - \delta \nu + \Psi - \frac{\Psi^2}{2} - \Psi \delta \nu \right) \left( 1- \Phi- \frac{\Phi^2}{2}+ \tilde{n}_k \big(1- \Phi\big) \frac{\partial \Delta x^k}{\partial \tilde{\chi}} \right) 
	\label{Ccurved}
\end{equation}
for $\mathfrak{C}$, and 
\begin{equation}
	2- \mathfrak{M} = \left(1 - \delta \nu + \Psi - \frac{\Psi^2}{2} - \Psi \delta \nu \right) \left(2 -  2\Phi - \Phi^2 + P^{lj} \frac{\partial \Delta x_j}{\partial \tilde{x}^l}\big(1-\Phi\big) \right)
	\label{Mcurved}
\end{equation}
for $\mathfrak{M}$.
To evaluate the second bracket of both equations, we first need to calculate the photon 4-trajectory using the results of section \ref{sec:geodesic equation}. For this purpose, we iterate the first-order result obtained at the beginning of this section. We shall skip the details of this calculation here since it follows the first-order derivation quite closely (but goes to second order). We find 
\begin{align}
	\eta^{[2]}(\chi) & = \eta_o+K^2 \left(\frac{13 \eta_o^4 \chi }{12600}+\frac{\eta_o^3 \chi ^2}{360}-\frac{\eta_o^2 \chi
		^3}{24}+\frac{73 \eta_o \chi ^4}{2160}-\frac{73 \chi ^5}{10800}\right) \nonumber \\ & \quad
	+ K \left(-\frac{\eta_o^2 \chi
	}{30}+\frac{\eta_o \chi ^2}{6}-\frac{\chi ^3}{18}\right)-\chi; \label{eta2curvedp} \\ 
	(x^i)^{[2]} & = \tilde{n}^i \left[K^2 \left(-\frac{13 \eta_o^4 \chi }{12600}-\frac{\eta_o^3 \chi ^2}{360}+\frac{\eta_o^2 \chi ^3}{20}-\frac{163
		\eta_o \chi ^4}{2160}+\frac{313 \chi ^5}{10800}\right) \right. \nonumber\\ & \quad 
	\left.+K \left(\frac{\eta_o^2 \chi }{30}-\frac{\eta_o \chi
		^2}{6}+\frac{5 \chi ^3}{36}\right)+\chi \right] \;.
	\label{x2curvedpe}
\end{align}
Finally, to get the displacement fields $\Delta x^i$ to second order, we need the second-order $\delta \chi$. We have
\begin{align}
	\nonumber \delta \chi^{[2]} & = \frac{K^2}{604800}\Big(1296 \eta_o^4 \Tilde{\chi} -2940 \eta_o^3 \Tilde{\chi} ^2+1960 \eta_o^2 \Tilde{\chi} ^3-735 \eta_o \Tilde{\chi} ^4+147
	\Tilde{\chi} ^5\Big)\\ & \quad 
	+\frac{K}{360} \Big(18 \eta_o^2 \Tilde{\chi} +15 \eta_o \Tilde{\chi} ^2-5 \Tilde{\chi} ^3\Big) \;,
\end{align}
whence
\begin{equation}
	\begin{aligned}
		\Delta x^i(\Tilde{\chi}, \tilde{n}^i) & = \tilde{n}^i\left[\frac{K^2}{5760} \Big(16 \eta_o^4 \Tilde{\chi} -132 \eta_o^3 \Tilde{\chi} ^2+344 \eta_o^2 \Tilde{\chi} ^3-315 \eta_o \Tilde{\chi} ^4+135 \Tilde{\chi}
		^5\Big) \right.\\ & \quad
		\left.+\frac{K}{24} \Big(2 \eta_o^2 \Tilde{\chi} -3 \eta_o \Tilde{\chi} ^2+3 \Tilde{\chi} ^3\Big) \right] \;.
	\end{aligned}
\end{equation}
Inserting all of these into equations \eqref{Ccurved} and \eqref{Mcurved}, we find 
\begin{equation}\label{eqn:constant curvature C pert}
	\begin{aligned}
		1 - \mathfrak{C} & = \frac{3 K^2}{22400} \Big(16 \eta_o^4-280 \eta_o^3 \Tilde{\chi} +840 \eta_o^2 \Tilde{\chi} ^2-700 \eta_o \Tilde{\chi} ^3+175 \Tilde{\chi}
		^4\Big)\\ & \quad 
		+\frac{K}{40} \Big(2 \eta_o^2-10 \eta_o \Tilde{\chi} +5 \Tilde{\chi} ^2\Big)+1 \;,
	\end{aligned}
\end{equation}
and 
\begin{equation}\label{eqn:constant curvature M pert}
	\begin{aligned}
		2 - \mathfrak{M} & = \frac{K^2}{100800} \Big(432 \eta_o^4-3780 \eta_o^3 \Tilde{\chi} +280 \eta_o^2 \Tilde{\chi} ^2+7875 \eta_o \Tilde{\chi} ^3-1575 \Tilde{\chi}
		^4\Big)\\ & \quad 
		+\frac{K}{20}\Big(2 \eta_o^2-5 \eta_o \Tilde{\chi} -5 \Tilde{\chi} ^2\Big)+2 \;.
	\end{aligned}
\end{equation}

\paragraph{Comparison of the values of $\mathfrak{C}$ and $\mathfrak{M}$ to the expected ones}
As expected, the expressions we just found for $\mathfrak{M}$ and $\mathfrak{C}$, equations \eqref{eqn:constant curvature C pert} and \eqref{eqn:constant curvature M pert}, do not agree with those derived previously, equations \eqref{eqn:constant curvature C expected} and \eqref{eqn:constant curvature M expected}, unless one accounts for the difference between $\chi$ and $\chi_K$. To compare these results, we need to determine the relation between the two parameterisations of the geodesic equation, 
which we obtain by comparing equations \eqref{eta2curved} and \eqref{eta2curvedp} with \eqref{x2curved} and \eqref{x2curvedpe}, \emph{viz.}
\begin{equation}
	\begin{aligned}
		\chi_K &= \chi  - K \left(-\frac{\eta_o^2 \chi}{30}+\frac{\eta_o \chi ^2}{6}-\frac{\chi ^3}{18}\right) \\ & \quad 
		- K^2 \left(\frac{13 \eta_o^4 \chi }{12600}+\frac{\eta_o^3 \chi ^2}{360}-\frac{\eta_o^2 \chi^3}{24}+\frac{73 \eta_o \chi ^4}{2160}-\frac{73 \chi ^5}{10800}\right) \;.
	\end{aligned}
\end{equation}
To express $\chi$ as a function of $\tilde{\chi}$, we rely on equations \eqref{eta2curvedp} and \eqref{x2curvedpe} instead of \eqref{eta2curved} and \eqref{x2curved}, whence
\begin{equation}
	\frac{1}{a_0(\eta_o - \Tilde{\chi})}= \frac{a_K(\eta_o)}{a_K(x^0(\chi))}\;.
\end{equation}
This yields
\begin{equation}
	\begin{aligned}
		\chi & = \frac{K^2}{604800} \Big(147 \Tilde{\chi}^5-735 \eta_o \Tilde{\chi}^4+1960 \eta_o^2 \Tilde{\chi}^3-2940 \eta_o^3
		\Tilde{\chi}^2+1296 \eta_o^4 \Tilde{\chi}\Big)\\ & \quad 
		+\frac{K}{360} \Big(-5 \Tilde{\chi}^3+15 \eta_o
		\Tilde{\chi}^2+18 \eta_o^2 \Tilde{\chi}\Big)+\Tilde{\chi} \;.
	\end{aligned}
\end{equation}
Substituting this expression into equations \eqref{Ccurvedtran} and \eqref{Mcurvedtran}, we get the expected values of $\mathfrak{C}$ and $\mathfrak{M}$,
\begin{equation}
	\begin{aligned}
		1-\mathfrak{C} & = \frac{3 K^2}{22400} \Big(16 \eta_o^4-280 \eta_o^3 \Tilde{\chi} +840 \eta_o^2 \Tilde{\chi} ^2-700 \eta_o \Tilde{\chi} ^3+175 \Tilde{\chi}
		^4\Big)\\ & \quad
		+\frac{K}{40} \Big(2 \eta_o^2-10 \eta_o \Tilde{\chi} +5 \Tilde{\chi} ^2\Big)+1 \;,
	\end{aligned}
\end{equation}
and 
\begin{equation}
	\begin{aligned}
		2-\mathfrak{M} & = \frac{K^2}{100800} \Big(432 \eta_o^4-3780 \eta_o^3 \Tilde{\chi} +280 \eta_o^2 \Tilde{\chi} ^2+7875 \eta_o \Tilde{\chi} ^3-1575 \Tilde{\chi}
		^4\Big) \\ & \quad 
		+\frac{K}{20} \Big(2 \eta_o^2-5 \eta_o \Tilde{\chi} -5 \Tilde{\chi} ^2\Big)+2 \;.
	\end{aligned}
\end{equation}
which, this time, is the result we expected from the second-order formula we derived, i.e.~equations \eqref{eqn:constant curvature C pert} and \eqref{eqn:constant curvature M pert}. 

This concludes the constant-curvature test case. We have verified that, when both the rulers and the co-ordinates in which they are evaluated are expanded to second order in 
$K$ and matched appropriately, our perturbative results for $\mathcal{T}$, $\mathfrak{C}$ and $\mathfrak{M}$ agree with the expected values. As in the other test cases, since $\mathcal{C}$, $\mathcal{B}_I$ and $\mathcal{A}_{IJ}$ are known to satisfy this test at first order, the results of this Appendix imply that $\mathcal{C}$ and $\mathcal{M}$ also do so at second order.


%
%
%


\section{List of Symbols}
\label{appendix:symbol list}

\begin{center}
	\begin{longtable}{|c|l|c|c|}
		\hline
		Symbol & Meaning & Defined & Calculated \\
		\hline
		$a(\eta)$ & background scale factor & \S \ref{sec:geodesic equation} & \\
		$\tilde{a}$ & $1/(1+\tilde{z})$ & & \\
		$\at(\tau)$ & scale factor evaluated at source proper time $\tau$ & \eqref{eqn: a bar definition} & \\
		$a_o$ & $a(\eta_o)$ & \S \ref{sec:geodesic equation} & \eqref{eqn:delta a observer} \\
		$\mathcal{A}_{IJ}$ & transverse cosmic ruler (metric pull-back) & \eqref{eq:gs1}& \eqref{eqn:relation between fraktur and calligraphic rulers} \\
		$\mathfrak{A}_{IJ}$ & transverse cosmic ruler ($1$-form pull-back) & \eqref{eq:frak}& \\
		$\mathcal{B}_I$ & vector cosmic ruler (metric pull-back) & \eqref{eq:gs1}& \\
		$\mathfrak{B}_I$ & $\mathfrak{D}_I + \hat{\mathfrak{D}}_I$ & \eqref{eqn:fraktur B definition}& \\
		$\mathcal{C}$ & longitudinal cosmic ruler (metric pull-back) & \eqref{eq:gs1} & \eqref{eqn:relation between fraktur and calligraphic rulers}\\
		$\mathfrak{C}$ & longitudinal cosmic ruler ($1$-form pull-back) & \eqref{eq:frak} &\eqref{eqn:C 2nd order}\\
		$\chi$ & global co-moving radial co-ordinate & \S \ref{sec:geodesic equation} & \\
		$\chi_e$ & global co-moving radial co-ordinate at source & \S \ref{subsec:frequency shift} & \\
		$\tilde\chi$ & $\chi(\tilde z)$, the inferred co-moving distance to redshift $\tilde z$ & & \\
		$\mathfrak{D}_I$, $\hat{\mathfrak{D}}^I$ & vector cosmic rulers ($1$-form pull-back) & \eqref{eq:frak}& \\
		$\delta a_o$ & $a_o - 1$ & above \eqref{eqn:delta nu at observer} & \eqref{eqn:delta a observer}  \\
		$\delta \chi$ & $\chi_e - \tilde{\chi}$ & \S \ref{subsec:Delta and delta} & \eqref{eqn:deltachi1} \\
		$\delta n^i$ & perturbation to light propagation direction & \eqref{eqn: k mu definition} & \eqref{eqn: delta n i integral expression} \\
		$\delta n^i_o$ & $\delta n^i$ at the observer &  \S \ref{subsec: initial conditions} & \eqref{eqn:delta n at observer} \\
		$\delta \nu$ & perturbation to light propagation direction & \eqref{eqn: k mu definition} & \\
		$\delta \nu_o$ & $\delta \nu$ at the observer & \S \ref{subsec: initial conditions} & \eqref{eqn:delta nu at observer} \\
		$\dgobs$ & observed galaxy over-density & \S \ref{sec:intro}& \eqref{eqn:dgobs} \\
		$\Delta_{a,e}$ & perturbation to $a$ & \eqref{eqn: delta ln a} & \eqref{eqn: delta ln a} \\
		$\Delta_{a,\tau} $ & perturbation to proper age & \eqref{eqn: delta ln alpha} & \eqref{eqn: delta ln alpha calculated} \\
		$\delta\tau$ & $\tau-t$ at the source & \eqref{eqn: delta tau definition} & \eqref{eqn: delta tau calculated} \\
		$\delta \tau_o$ & $\tau_o - t_o$ & above \eqref{eqn: delta tau 0} & \eqref{eqn: delta tau 0}\\
		$\delta u$ & perturbation to time component of $u$ & \eqref{eqn: source 4 velocity} & \eqref{eqn: delta u calculation} \\
		$\delta x^\mu$ & solution to the geodesic equation & \eqref{eqn: lower-case delta x mu definition} & \\
		$\Delta x^\mu$ & difference between actual and inferred co-ordinates & \eqref{eqn: total displacement definition} & (\ref{eqn: total displacement eta}--\ref{eqn: total displacement x i}) \\
		$\delta \vx_s(\eta)$ & perturbation to source's time-like geodesics & \S \ref{subsec: source trajectory} & \eqref{eq:delta_x_s} \\
		$\delta \vx_o(\eta)$ & perturbation to observer's time-like past geodesics & \S \ref{subsec: initial conditions} & \eqref{eq:delta_x_o} \\
		\hline
		$e_{\underline{\alpha}}^\mu$, $e_{\alpha}^\mu$ & observer Fermi tetrad & \S \ref{subsec: initial conditions} & App.~\ref{appendix:observer tetrad} \\
		$\eta$ & conformal time co-ordinate & \eqref{eqn: conformal metric Poisson gauge} & \\
		$\eta_o$ & conformal time at observer & below \eqref{eqn: a bar definition} & \\
		$\tilde{\eta}$ & inferred conformal time & \S \ref{sec:frames} & \eqref{eqn: tilde co-ordinates definition} \\
		$\overline\eta(t)$ & background cosmic-to-conformal-time relation & below \eqref{eqn: a bar definition} & \\
		$F_{[2]}^0$ & $\mathrm{d}\delta \nu/\mathrm{d}\chi$ to $2^{\rm nd}$ order & below \eqref{eqn: kernel for frequency shift} & \eqref{eqn: kernel for frequency shift} \\
		$F_{[2]}^i$ & $\mathrm{d}\delta n^i/\mathrm{d}\chi$ to $2^{\rm nd}$ order & below \eqref{eqn:Geoe} & \eqref{eqn:Geoe} \\
		$g_s$ & inferred metric & below \eqref{eq:gs1} & \\
		$\Gamma^\alpha_{\beta\gamma}$ & Christoffel symbol & \eqref{eqn: light geodesic equation} & App.~\ref{appendix:Christoffel symbols} \\
		$\gamma_1$, $\gamma_2$ & shear (components of $\mathcal{A}_{IJ}$) & \eqref{eqn:A_IJ helicity decomposition} & \\
		$h_{ij}$ & tensor metric component & \eqref{eqn: conformal metric Poisson gauge} & \\
		$\tilde{H}$ & background expansion rate evaluated at $\tilde a$ & & \\
		$\tilde{\mathcal{H}}$ & background conformal Hubble rate evaluated at $\tilde a$ & & \\ 
		$\barH$ & $H\left(\overline{\eta}(\tau)\right)$ & below \eqref{eqn:delta ln H} & \\
		$i$ & Embedding of the observer's light-cone in $\mathbb{M}$ & \S \ref{sec:frames} & \\
		$i,j,\ldots$ & spatial indices $\in \set{1,2,3}$ & & \\
		$I,J,\ldots$ & transverse indices $\in \set{\pm}$ & below \eqref{eq:gs1} & \\
		$K$ & FLRW spatial curvature (for test case) & \S \ref{sec: implementation} & App.~\ref{appendix:test separate universe} \\
		$k^\mu$ & light wave-vector & \eqref{eqn: k mu definition} & \S \ref{subsec:frequency shift} \\
		$\lambda$ & light-ray affine parameter (solves equation \eqref{eqn: light geodesic equation}) & \eqref{eqn: light geodesic equation} & $\lambda = \chi$ \\
		$\mathbb{M}$ & space-time & \S\ref{sec:frames} & \\
		$\mathcal{M}$ & $\tr \mathcal{A}_{IJ}$ & \eqref{eqn:A_IJ helicity decomposition} & \\
		$\mathfrak{M}$ & $\tr \mathfrak{A}^I_{J}$ & & \eqref{eqn:M 2nd order}\\
		$\ntvh$ & observed sky direction & \S \ref{sec:intro} & \\ 
		$\nvh_s$ & ray propagation direction measured at source & \eqref{eqn: n_s hat definition} & \\ 
		$\omega^i$ & vector metric component & \eqref{eqn: conformal metric Poisson gauge} & \\
		$\Psi$ & `temporal' Bardeen potential & \eqref{eqn: conformal metric Poisson gauge} & \\
		$\Phi$ & `spatial' Bardeen potential & \eqref{eqn: conformal metric Poisson gauge} & \\
		$\Sigma_{\tau_o}$ & observer's past light-cone & \S \ref{sec:frames} & \\
		$t$ & cosmic time & \S \ref{sec:geodesic equation}& \\
		$t_o$ & cosmic time co-ordinate of observer & below \eqref{eqn: a bar definition} & \\
		$\mathcal{T}$ & cosmic clock & & \\
		$\tau_o$ & proper age of observer (age of the Universe) & \S \ref{sec:geodesic equation} & \eqref{eqn: delta tau 0} \\
		$\tau$, $\tau_s$ & proper time of the source, as measured by it & & \\
		$\theta$ & eccentric anomaly for separate-universe test & \eqref{eqn:curved universe scale factor solution} & \eqref{eqn:eccentric anomaly solution second order} \\
		$\Theta$ & Ruler-size intrinsic growth rate & \eqref{eqn:ruler growth rate intrinsic} &  \\
		$u^\mu$ & source 4-velocity & \eqref{eqn: source 4 velocity} & \eqref{eqn: source 4 velocity} \\
		$\vv$ & source spatial velocity field & \eqref{eqn: source 4 velocity} & \\
		$\tilde{\vx}$ & inferred position & \S \ref{sec:frames} & \eqref{eqn: tilde co-ordinates definition} \\
		$x^\mu$ & global co-ordinate & \eqref{eqn: conformal metric Poisson gauge} & \\
		$x_e^\mu$ & source co-ordinate position & & \\
		$\xlc^\mu(\chi)$ & solution to geodesic equation & above \eqref{eqn: light geodesic equation} & \S \ref{subsec:frequency shift} \\
		$\tilde{x}^\mu$ & inferred source position & \S \ref{sec:frames} & \eqref{eqn: tilde co-ordinates definition}  \\
		$\tilde{z}$ & observed redshift & \S \ref{sec:intro} & \\
		\hline 
		\newpage
		\hline
		$L^{[n]}$ & evaluation of a quantity $L$ up to order $n$ & \S \ref{sec:geodesic equation} &  \\
		$L^{(n)}$ & order $n$ term of some quantity $L$ & \S \ref{sec:geodesic equation} & \\
		$L'$ & $\mathrm{d}L/\mathrm{d}\eta$ & \S \ref{sec:frames} & \\
		\hline
	\end{longtable}
\end{center}

\bibliographystyle{JHEP}
\bibliography{references}

\label{lastpage}

\end{document}

%% file: defs.tex
\def\lsim{~\rlap{$<$}{\lower 1.0ex\hbox{$\sim$}}}
\def\bsim{~\rlap{$>$}{\lower 1.0ex\hbox{$\sim$}}}

\def\apjl{Astrophys.\ J.\ Lett.}

\def\aap{Astron.\ Astrophys.}

\def\apjs{Astrophys.\ J. Supp.}

\newcommand{\be}{\begin{equation}}
\newcommand{\ee}{\end{equation}}

\let\ln\relax
\DeclareMathOperator{\ln}{ln}
\DeclareMathOperator{\tr}{tr}
\let\det\relax
\DeclareMathOperator{\det}{det}

\def\mathbi#1{\textbf{\em #1}}

\renewcommand{\v}[1]{\bm{#1}}

\def\ntvh{\tilde{\mathbi{n}}}

\def\nvh{\hat{\mathbi{n}}}

\def\nh{\mathit{\hat{n}}}

\def\ve{\mathbi{e}}
\def\vk{\mathbi{k}}

\def\vv{\mathbi{v}}

\def\vx{\mathbi{x}}

\def\dgobs{\delta_g^{\rm obs}}

\def\at{a_t}
\def\xlc{x_{\rm lc}}
\def\barH{\bar{H}}
\def\cH{\mathcal{H}}